\documentclass{LMCS}
\usepackage{amsbsy}
\usepackage[all]{xy}
\usepackage{latexsym,amsfonts}
\usepackage{amssymb}
\usepackage{enumerate,hyperref}

\newcommand{\byre}[1]{\stackrel{ #1 }{\longleftarrow}}
\newcommand{\ex}{\exists}
\newcommand{\ol}{\overline}
\renewcommand{\ni}{\noindent}

\newcommand{\eset}[1]{ \{ #1 \}}

\newcommand{\Nil}{{\bf 0}}
\newcommand{\fa}{\forall}
\newcommand{\G}{\mathsf{G}}
\newcommand{\dom}{\mathrm{dom}}
\newcommand{\app}{@}

\def\doi{5 (3:2) 2009}
\lmcsheading%
{\doi}
{1--52}
{}
{}
{Sep.~20, 2006}
{Jul.~30, 2009}
{} 

\begin{document}
\title[Decidability of Higher-order Matching]{Decidability of Higher-Order Matching}
\author[C.~ Stirling]{Colin Stirling}
\address{School of Informatics \\
University of Edinburgh \\ 
Edinburgh EH8 9AB, UK}
\email{cps@staffmail.ed.ac.uk}

\begin{abstract}
We show that the higher-order matching problem is decidable
using a game-theoretic argument.
\end{abstract}
\maketitle
\tableofcontents

\section{Introduction}

Higher-order unification is the problem given an equation
$t= u$ containing free variables
is there a solution substitution $\theta$ such that
$t \theta$ and $u \theta$ have the same normal form? 
The terms $t$ and $u$ are from the simply typed lambda calculus
and the same normal form is with respect to $\beta \eta$-equivalence.
Higher-order matching is the particular instance when the term
$u$ is closed;   can $t$ be pattern matched to $u$?
Although higher-order unification is undecidable (even if free
variables are only
second-order \cite{gold}), higher-order matching  was conjectured to be decidable 
by Huet  \cite{Hu1}.  
If matching  is decidable then it is known
to have non-elementary complexity \cite{Stat,Wier}.
Decidability has been proved  for the general problem
up to order $4$ (by showing  decidability 
of observational equivalence of lambda terms)
and for various special cases \cite{Pad1,Pad2,Sch,SchS,DW}.
Comon and Jurski define tree automata that characterise
all solutions to a $4$th-order problem, thereby, showing
that they form  a regular set \cite{CJ}.
Loader showed that matching  is undecidable
for the variant  definition of 
the same normal form that only uses  $\beta$-equivalence
by encoding lambda definability as matching \cite{Lo}: 
see \cite{joly} for a proof that uses the halting problem.
An excellent source of information about unification and matching
is \cite{Dow1}.

In this paper, we confirm Huet's conjecture
that higher-order matching is
decidable. The proof first appeals to Padovani's and Schubert's
reduction of matching  to the
(dual) interpolation problem   \cite{Sch,Pad2}
and is then in the tradition
described by Dowek \cite{Dow1}:
``these [decidability] proofs are rather technical $\ldots$ because
they all proceed by transforming potential solutions into
smaller ones cutting and pasting term pieces''.
The proof method is partly inspired by
model-checking games (such as in \cite{Sti})
where a model, a transition graph, is traversed relative
to a property and players make choices at appropriate
positions. Given a (dual) interpolation problem $P$,
we define a game 
where the model is a closed lambda term $t$
in $\eta$-long normal form that is a potential solution to $P$;
game-playing moves around the term  $t$ (dependent on $P$).
The  game captures the dynamics of $\beta$-reduction on  $t$
without changing it (using substitution).
Unlike model-checking games, play may arbitrarily jump around
a term because of binding.

The principal virtue 
of the game is that small pieces
of a solution term can  be understood in terms of sequences of positions in the
game 
and how they, thereby, contribute to solving the problem $P$.
We   identify regions
of a term, ``tiles'',  and classify
them according to these intervals of play.
Two transformations that
preserve solution terms are introduced. With these, we show
that $3$rd-order matching is decidable 
using the small model property:  if there
is a solution to a problem then there  is a small solution
to it. In \cite{St2}, we introduced the game
and two more  transformations that uniformly  solved
known subcases of matching (including $4$th-order).
The key observation for decidability at $3$rd-order is
the tree-model property: each play descends a branch
of a solution term because of the paucity of binding.
For higher-orders, the idea  is to  
induce as far as possible  the tree-model property, to tame play jumping
within a term due to binding. The mechanism for doing this 
involves  unfolding  a lambda term with respect to game
playing which is  
analogous to 
unravelling a transition system in modal logic. Unfolding involves 
``tile lowering'', copying  regions of a term
down its branches.
The proof of decidability at higher-order 
uses  unfolding 
and from  its combinatorial properties
the small  model property follows.

In Section~\ref{section2} we introduce higher-order matching 
and (dual) interpolation and in Section~\ref{prelim}
we define some basic ingredients of the problem.
In Section~\ref{treecheck} we describe 
the tree-checking game that characterises 
(dual) interpolation and in Section~\ref{props} some properties of
the resulting game are highlighted. 
In Section~\ref{tiles}  we  identify tiles as regions
of terms and define their plays and 
in Section~\ref{third} we show that 
$3$rd-order matching is decidable using a tree model property
of game playing; the first step in this argument is to define a 
partition of each play in a game. The definition of partition 
is extended to all orders in Section~\ref{unfold}.
This then forms the basis for 
the notion of tile unfolding that 
is described in Section~\ref{fifth} and how  it leads to decidability
of matching via the small model property.
The complexity analysis is the size of
a smallest term, if there is one,  that solves a problem.
However, the bounds are extremely coarse. 
Finally, we conclude with general remarks
and ideas for future work.

\section{Matching and dual interpolation}
\label{section2}

Simple types are generated from a single base type
$\Nil$ using the binary $\rightarrow$ operator.
For simplicity, we assume only one base type:
everything that is to follow can be extended to arbitrary  
many base types. 
A type is $\Nil$ or $A \rightarrow B$
where $A$ and $B$ are types. 
If $A \not= \Nil$ then it has
the form $A_{1} \rightarrow \ldots \rightarrow A_{n} \rightarrow \Nil$,
assuming $\rightarrow$ associates to the right,
which we abbreviate to  $(A_{1},\ldots, A_{n},\Nil)$ following 
Ong \cite{ong}. 
A standard  definition
of \emph{order} is:  the order of $\Nil$ is $1$ and the order of 
$(A_{1}, \ldots, A_{n}, \Nil)$
is $k +1$ where $k$ is the maximum of the orders of the $A_{i}$s.

Terms of the simply typed lambda calculus 
are built from a countable set of typed
variables $x,y, \ldots$ and constants $a, f, \ldots$
(so, each variable and constant 
has a unique type). 

\begin{defi}
The set of simply typed lambda terms is the smallest set $T$
such that 
\begin{enumerate}[(1)]
\item if $x$ ($f$) has type $A$ then $x : A \in T$ $(f : A \in T)$,
\item if $t : B \in T$ and $x : A \in T$
then 
$\lambda x. t : A \rightarrow B \in T$,
\item if $t : A \rightarrow B \in T$ and $u : A \in T$ 
then $(t u) : B \in T$.
\end{enumerate}
\end{defi} 

\ni
The \emph{order} of a typed term $t : A$ is the order of its type $A$.

In a sequence of unparenthesised applications, 
we adopt the usual convention 
that application associates to the left;  so $t u_{1} \ldots u_{k}$
is  $(( \ldots (t u_{1}) \ldots) u_{k})$. 
The usual definitions of free and bound variable occurrences and when
a  typed term is closed are assumed.
Moreover, we assume the standard 
definitions and properties of $\alpha$-equivalence,
$\beta$-reduction, $\eta$-reduction and 
$\beta\eta$-equivalence, $=_{\beta \eta}$, 
such as strong normalisation of $\beta$-reduction:  
see for instance, Barendregt
\cite{Bar}.

\begin{defi}
A \emph{matching problem} is an equation  $v = u$ where $v, u : A$ 
and $u$ is closed. 
The \emph{order}
of the problem is the maximum of the orders of the free variables
$x_{1}, \ldots, x_{n}$ in $v$.
A \emph{solution} is a sequence
of terms $t_{1}, \ldots, t_{n}$ such that $v \{ t_{1}/x_{1}, \ldots,
t_{n}/x_{n} \} =_{\beta \, \eta} u$
where $v \{ t_{1}/x_{1}, \ldots,
t_{n}/x_{n} \}$ is the simultaneous substitution of $t_{i}$
for each free occurrence of $x_{i}$ in $v$ for each $i: 1 \leq i \leq n$.
\end{defi}

\ni
The decision question is:  given a matching problem, does it have
a solution? 
It suffices to consider the case when $A= \Nil$:
a problem $v = u$ of type $(A_{1},\ldots,A_{n},\Nil)$
reduces to the equivalent problem $v f_{1} \ldots f_{n} = u f_{1} \ldots f_{n}$
of type $\Nil$ where each $f_{i}: A_{i}$ is a fresh constant
which is not allowed in solution terms. 
Equivalent variants of matching include  the ``range question'' 
and the ``pattern matching problem'' \cite{Stat2,joly}.


As described by Huet \cite{Hu1},  
every simply typed lambda calculus term 
is $\beta \eta$-equivalent to a unique term in 
\emph{$\eta$-long normal form},
\begin{enumerate}[(1)]
\item if  $t : \Nil$ then it is
$u : \Nil$ where $u$ is a constant or a variable,
or $u \, t_{1} \ldots  t_{k}$ 
where $u : (B_{1}, \ldots, B_{k}, \Nil)$
is a constant or a variable and each $t_{i} : B_{i}$ is 
in $\eta$-long normal form,
\item if $t : (A_{1},\ldots,A_{n},\Nil)$
then $t$ is
$\lambda y_{1}\ldots y_{n}. t'$ where each $y_{i} : A_{i}$
and $t' : \Nil$ is
in $\eta$-long normal form.
\end{enumerate}
Throughout, we write $\lambda z_{1} \ldots z_{n}$
for $\lambda z_{1} \ldots \lambda z_{n}$.
A term is \emph{well-named} if each occurrence
of a variable $y$ within a lambda abstraction is unique.
In the following,  we assume that a term  in \emph{normal form}
is always in  $\eta$-long normal form; consequently, 
$\beta$-equality and 
$\beta \, \eta$-equality 
coincide (for instance, see \cite{Sto}).

\begin{defi}
Assume $u : \Nil$ and each $v_{i} : A_{i}$, $1 \leq i \leq n$, is a 
closed term in normal form and $x : (A_{1},\ldots,A_{n},\Nil)$.
\begin{enumerate}[(1)]
\item $x \, v_{1} \ldots v_{n} = u$ is an \emph{interpolation equation}.
\item $x \, v_{1} \ldots  v_{n} \not= u$ is an 
\emph{interpolation disequation}.
\item A finite
family of interpolation equations
$x \, v^{i}_{1}  \ldots  v^{i}_{n} = u_{i}$, where $i: 1 \leq i \leq m$, 
with the same free variable $x$, is an \emph{interpolation problem} $P$.
\item A  finite family 
of interpolation equations and disequations 
$x \, v^{i}_{1}  \ldots  v^{i}_{n} \approx_{i} u_{i}$, 
when $i: 1 \leq i \leq m$, 
with the same free variable $x$ and where  each 
$\approx_{i} \, \in \eset{=, \not=}$, is a  \emph{dual interpolation problem} 
$P$.
\item The type of problem $P$ is that of $x$
and the order of $P$ is the order of $x$.
\item  A \emph{solution} of $P$ of type $A$
is a closed term $t : A$ in normal form 
such that for each equation
$t \, v^{i}_{1} \ldots  v^{i}_{n}  =_{\beta} u_{i}$
and, in the case of dual interpolation,  for each disequation
$t \, v^{i}_{1}  \ldots  v^{i}_{n}  \not=_{\beta} u_{i}$.
We write $t \models P$ if $t$ is a solution of $P$.
\end{enumerate}
\end{defi}

Conceptually, (dual) interpolation is simpler than 
matching because there is a single variable $x$ that appears
at the head of each  (dis)equation.
However, Schubert shows
that a  matching problem of order $n$  
reduces to an interpolation problem of order at most $n+2$ and
Padovani shows it reduces to a
dual interpolation problem of order $n$,  \cite{Sch,Pad2}.
Consequently, the higher-order matching problem reduces to the
following decision question.
\vspace{2mm}

\ni
{\bf Decision Question}
Given a (dual) interpolation problem $P$,
is there a term $t \models P$?
\vspace{2mm}

\ni
It is this question that is solved positively  in the rest of the paper.
Throughout,  we assume a fixed 
dual interpolation problem $P$ of type $A$ whose order is greater
than $1$ (as  an order $1$ problem is easily decided).
A  problem $P$ has  the form 
$x \, v^{i}_{1} \ldots  v^{i}_{n}  \approx_{i} u_{i}$, $1 \leq i \leq m$,
where the normal form terms 
$v^{i}_{j}$ and $u_{i}$ are well-named
and no pair share  bound variables.

\section{Preliminaries}
\label{prelim}
We start with  some examples of interpolation problems.

\begin{exa}\label{examp1} The following is a $4$th-order problem
\[ \begin{array}{lcl}
x \, \lambda y_{1} y_{2}.y_{1}y_{2}  & = & f a \\
x \, \lambda y_{3} y_{4}.y_{3}(y_{3} y_{4}) & = & f(f a)
\end{array} \]
with   
$x : (((\Nil, \Nil),\Nil , \Nil),\Nil)$ and $f : (\Nil, \Nil)$.
\qed \end{exa}

\begin{exa}\label{examp2} The problem  $x(\lambda z. z) = 
f(\lambda x_{1} x_{2} x_{3}.x_{1} x_{3}) a$ has order $3$
where $x$ has type  $((\Nil, \Nil), \Nil)$
and $f :
(((\Nil,\Nil),\Nil,\Nil,\Nil),\Nil,\Nil)$ assuming $x_{2}:\Nil$.
\qed \end{exa}

\begin{exa}\label{examp3} The next  equation, 
due to Luke Ong,  is $5$th-order 
\[ x(\lambda y_{1} y_{2}.y_{1}(\lambda y_{3}.y_{2}(y_{1}(\lambda y_{4}.y_{3}))) = h(g(h(h a )))
\]
with $x : ((((\Nil, \Nil), \Nil),(\Nil, \Nil ), \Nil),\Nil)$
and $g,h : (\Nil,\Nil)$.
\qed \end{exa}

A  right term $u$ of an interpolation (dis)equation
$x v_{1} \ldots v_{n}  \approx u$
may contain bound variables, such as 
$f(\lambda x_{1}x_{2}x_{3}.x_{1} x_{3}) a$ of Example~\ref{examp2}.
Let $X = \eset{ x_{1}, \ldots, x_{k} }$ be the set
of bound variables in $u$ and let
$C = \eset{c_{1},\ldots,c_{k}}$ be a fresh set of constants
where  each $c_{i}$ has the same
type as $x_{i}$. 

\begin{defi} The \emph{ground closure} of
a closed term $w$, whose bound variables belong to
$X$, with respect to $C$, written Cl$(w,X,C)$,
is defined inductively:
\begin{enumerate}[(1)]
\item if $w = a: \Nil$ then Cl$(w, X,C) = \{ a \}$,
\item if $w = f \, w_{1} \ldots w_{n}$ then Cl$(w, X,C)
= \eset{w} \cup  \bigcup \eset{  {\rm Cl}(w_{i},X,C) \, | \, 
1 \leq i \leq n}$,
\item if $w = \lambda x_{j_{1}} \ldots x_{j_{n}}.u$ 
then Cl$(w,X,C)= 
{\rm  Cl}(u \eset{ c_{j_{1}}/x_{j_{1}}, \ldots,
c_{j_{n}}/x_{j_{n}}}, X,C)$.
\end{enumerate}
\end{defi}

\ni
If $u = f(\lambda x_{1}x_{2}x_{3}.x_{1}x_{3}) a$
then its ground closure 
with respect to $\eset{c_{1},c_{2},c_{3}}$
is the set of terms 
$\eset{u, c_{1} c_{3},c_{3},a}$.
The ground closure of 
$h(g(h(h a)))$ of Example~\ref{examp3}
with respect to the empty set is
its subterms $\eset{h(g(h(ha))),g(h(ha)),h(h a),h a,a}$.
An element of a ground closure always has base type;
in the case  of a right term $u$ of an interpolation
(dis)equation, its ground closure  
with respect to a set of constants consists of 
all  subterms of type $\Nil$
when free variables $x_{i}$ are replaced with constants $c_{i}$
of the same type.

We also  identify subterms of left terms $v_{j}$
of a (dis)equation $x \, v_{1} \ldots v_{n}  \approx u$
relative to the finite  set $C$ of constants for $u$.
Such a subterm may contain free variables
and may  have a higher-order type.

\begin{defi} The \emph{subterms} of $w$  relative
to $C$, written Sub$(w,C)$,
is defined inductively using an auxiliary set
Sub$'(w,C)$: 
\begin{enumerate}[(1)]
\item if $w$ is a variable
or a constant then Sub$(w,C)$ $=$ Sub$'(w,C)$ $=$ $\eset{w}$,
\item  if $w$ is
$x \, w_{1} \ldots  w_{n}$ then Sub$(w,C)$ $=$ Sub$'(w,C)$ $=$
$\eset{w} \cup \bigcup \eset{ 
{\rm Sub}(w_{i},C) \, | \, 1 \leq i \leq n}$,
\item if $w$ is $f \, w_{1} \ldots w_{n}$ then Sub$(w,C)$ $=$ Sub$'(w,C)$
$=$ $\eset{w} \cup \bigcup \eset{ {\rm Sub}'(w_{i},C) \, | \, 1 \leq i 
\leq n}$,
\item if $w$ is $\lambda y_{1} \ldots
y_{n}. v$ then Sub$(w,C)$ $=$ $\eset{w} \cup 
{\rm Sub}(v,C)$ and 
Sub$'(w,C)$ $=$ \\ $\bigcup\eset{ {\rm Sub}(v\eset{
c_{i_{1}}/y_{1}, \ldots, c_{i_{n}}/y_{n}},C) \, | \,  
c_{i_{j}} \in C \mbox{ has the same type as } y_{j}}$.
\end{enumerate}
\end{defi}

\ni
The subterms of $\lambda z.z$ relative
to $\eset{c_{1},c_{2},c_{3}})$ of 
Example~\ref{examp2} is $\eset{\lambda z.z,z}$. If instead
of $\lambda z.z$ the left term of this example is 
$v = \lambda z. f(\lambda z_{1} z_{2} z_{3}.z_{1} z_{2}) z$,
then Sub$(v,\eset{c_{1},c_{2},c_{3}})$ is
$\eset{v, f(\lambda z_{1}z_{2} z_{3}.z_{1} z_{2}) z,
c_{1} c_{2},c_{1} c_{3},c_{2},c_{3},z}$: bound variables directly beneath
a constant are replaced in their body by constants in  $C$
with the same type.

Given the problem $P$ with (dis)equations 
$x \, v^{i}_{1} \ldots  v^{i}_{n} \approx_{i} u_{i}$,  
$i: 1 \leq i \leq m$, for each $i$
let $X_{i}$ be the (possibly empty) set of bound variables in
$u_{i}$ and $C_{i}$ be a corresponding set of new
constants (that do not occur in $P$), the \emph{forbidden} constants. 
We are interested in closed terms $t$ in normal form 
where  $t \models P$ and $t$ does not contain
forbidden constants.

\begin{defi}\label{def3}
Assume $P$ is the fixed dual interpolation problem of 
type $A$.
\begin{enumerate}[(1)]
\item $\mathsf{T}$ is the set of
\emph{subtypes} of $A$ including $A$ and the subtypes of subterms of 
$u_{i}$.
\item For each $i$, the \emph{right subterms} are 
$\mathsf{R}_{i}$ $=$ Cl$(u_{i},X_{i},C_{i})$ and 
$\mathsf{R} = \bigcup \eset{ \mathsf{R}_{i} \, | \, 1\leq i \leq m}$.
\item For each $i$, the \emph{left subterms} are $\mathsf{L}_{i}$ $=$
$C_{i} \cup \bigcup \eset{ {\rm Sub}(v^{i}_{j},C_{i}) \, | \, 1 \leq 
j \leq n}$ and 
$\mathsf{L} = \bigcup \eset{ \mathsf{L}_{i} \, | \, 1 \leq i \leq m}$.
\item The \emph{arity} of $P$, $\alpha$,  is the largest width $k$ of any
type $(A_{1},\ldots,A_{k},\Nil) \in \mathsf{T}$.
\end{enumerate}
\end{defi}

\ni
Clearly, the  sets $\mathsf{T}$, $\mathsf{R}$ and
$\mathsf{L}$ are each  finite and $\alpha$ is bounded 
with respect to a given problem $P$.
In  Example~\ref{examp1}, the set of forbidden constants $C_{1} \cup C_{2}$
is empty and $\mathsf{R}$ is
$\eset{f(f a),f a,a}$, $\mathsf{L}_{1}$ is
$\eset{\lambda y_{1}y_{2}.y_{1} y_{2} ,y_{1} y_{2},y_{2}}$
and its arity is $2$.

\begin{defi}
The \emph{right size}, $\delta(u)$, of a right term $u$ relative
to its constants $C$
is defined inductively:
\begin{enumerate}[(1)]
\item if $u = a: \Nil$ then $\delta(u) = 0$,
\item  if $u = f w_{1} \ldots  w_{k}$
then $\delta(u) = 1 + \sum \eset{ \delta(w_{i}) \, | \, 1 \leq i \leq k}$,
\item if $u = \lambda x_{i_{1}} \ldots x_{i_{k}}.w$, then 
$\delta(u) = \delta(w\eset{c_{i_{1}}/x_{i_{1}}, \ldots, c_{i_{k}}/x_{i_{k}}})$.
\end{enumerate}
\end{defi}

\begin{defi} \label{delta}
The \emph{right size} for $P$, $\delta$, is 
$\sum \eset{ \delta(u_{i}) \, | \, 1 \leq i \leq m}$ of its right terms.
\end{defi}

\ni
For instance, $\delta(h(g(h(h a)))) = 4$
and $\delta(f(\lambda x_{1} x_{2} x_{3}.x_{1} x_{3}) a)
= 2$. If the right size $\delta$ for $P$ is $0$,
then each (dis)equation in $P$ has restricted form 
$x \, v_{1} \ldots v_{n}  \approx a$ where 
$a :\Nil$. 
Padovani proved 
that dual interpolation is decidable for this special case,
the \emph{atoms} case,  by showing  decidability of
observational equivalence within  minimal
models \cite{Pad1}.

\section{Tree-checking games}
\label{treecheck}

We introduce a game-theoretic characterisation of dual interpolation
inspired by model-checking games (such as in \cite{Sti}) 
where a model, a transition graph, is traversed relative
to a property and players make choices at appropriate
positions. Similarly, in the following game the model
is a putative solution term $t$ that is traversed relative
to the dual interpolation problem. The central motivation
is to model the dynamics, $\beta$-reduction, without changing $t$
by substituting into it.
Because of binding play may jump around $t$.

A potential solution term $t$ for $P$ has the right type,
is in normal form, is well-named (with variables that are disjoint
from variables in $P$) and does not contain forbidden constants.
Term $t$ is represented as a tree, tree$(t)$.
If $t$ is $y : \Nil$ or $a: \Nil$ then tree$(t)$ is the single
node labelled with $t$. In the case of  $u \, v_{1} \ldots v_{k}$
when $u$ is a variable or a constant, we assume that a dummy
lambda with the empty sequence of variables
is placed directly above any subterm $v_{i} : \Nil$ in its
tree representation. With this understanding, if $t$ is
$u \, v_{1} \ldots v_{k}$  then tree$(t)$ consists of the root
node labelled $u$ and $k$-successors, tree$(v_{1})$,$\ldots$,tree$(v_{k})$. 
We use the notation $t \downarrow_{i} t'$
to represent that tree $t'$ is the $i$th successor of the root
node of $t$. We also use the standard abbreviation 
$\lambda \ol{y}$ for $\lambda y_{1} \ldots y_{n}$
for some  $n \geq 0$, so  $\ol{y}$ is possibly
the empty sequence of variables.
If $t$ is $\lambda \ol{y}.v$  then  tree$(t)$
consists of the root node labelled $\lambda \ol{y}$ and
a single successor  tree$(v)$, so
$t \downarrow_{1} {\rm tree}(v)$.

For ease of exposition, we allow  $t$  to range over  lambda terms,
their
trees and their root nodes: the context will make it clear which is meant. 
The introduction of dummy lambdas is a slight extension to
$\eta$-long normal form;
they make term trees more homogeneous, allow for an easier
analysis of game playing and, 
as we shall see in later sections, 
they  are   useful for individuating regions of a term 
and for defining region transformations.
\begin{figure}
\begin{center}
\xymatrix{
& & & (1) {\lambda} z \ar[d]& & & \\
& & & \ar[dll](2) z \ar[drr] &  & & \\
& (3) \ar[d] \lambda x & & & & \ar[d] (11) \lambda & \\
& \ar[d] (4) f & & & & \ar[dl] (12) z  \ar[dr] & \\
& \ar[d] (5) \lambda & & & \ar[d] (13) \lambda y  & &  \ar[d] (19) \lambda \\
& \ar[dl](6) z \ar[dr] & & & \ar[dl] (14) z \ar[dr] & &  (20) a  \\
\ar[d] (7) \lambda u  & & \ar[d] (9) \lambda   & \ar[d] (15)  \lambda s  & & 
\ar[d] (17) \lambda  &   \\
(8) x  & & (10) b   & (16) s&  & (18) y &
} 
\end{center}
\caption{A term tree that solves  Example~\ref{examp1}}
\label{ex1}
\end{figure}

\begin{exa}\label{examp41} A solution $t$ from \cite{CJ} for the 
problem of Example~\ref{examp1} is the following term
$\lambda z.z(\lambda x.f(z(\lambda u.x) b)) 
(z(\lambda y.z(\lambda s.s) y) a)$. For instance,
if  $v = \lambda y_{1} y_{2}.y_{1}y_{2}$ then the normal
form of $tv$ is $fa$.
\[ \begin{array}{lcl}
t \, v & =_{\beta} & v (\lambda x.f(v(\lambda u.x) b)) 
(v(\lambda y.v(\lambda s.s) y) a) \\
& =_{\beta} & \lambda x.f(v(\lambda u.x) b) 
(v(\lambda y.v(\lambda s.s) y) a) \\
& =_{\beta} & f( v (\lambda u.v (\lambda y.v (\lambda s.s)y a )) b) \\
& =_{\beta} & f( \lambda u.(v(\lambda y.v(\lambda s.s) y) a) b) \\
& =_{\beta} & f( v(\lambda y.v(\lambda s.s) y) a) \\
& =_{\beta} & f( \lambda y.(v (\lambda s.s) y) a) \\
& =_{\beta} & f( v (\lambda s.s) a) \\
& =_{\beta} & f( (\lambda s.s)  a) \\
& =_{\beta} & f a
\end{array} \]
The tree for $t$ (without  indices on edges)
is depicted in Figure~\ref{ex1}.
For instance, in this tree $(6) \downarrow_{1} (7)$
and $(6) \downarrow_{2} (9)$.
Each node (which we have uniquely identified with a natural number)
is labelled with a $\lambda \ol{z}$, a variable or a constant.
A node labelled with a constant or variable of type $\Nil$ is a
leaf of the tree (such as node $(10)$).
A node labelled with a higher-order constant or variable 
of type $(B_{1},\ldots,B_{k},\Nil)$ has precisely $k$-successor
nodes each  labelled with a $\lambda \ol{y}$ (which may be dummy).
A node labelled with a $\lambda \ol{z}$ has a single successor
which is labelled with a constant or a variable. 
Therefore,  every even level of the tree
(when the root is at level $0$) is a lambda node.
\qed \end{exa}

Innocent game semantics following Ong in \cite{ong}
provides a possible game-theoretic foundation
for (dual) interpolation.
Given  a potential solution term $t$ and a 
(dis)equation $x v^{i}_{1} \ldots v^{i}_{n}  \approx_{i} u_{i}$ 
there is the game board in Figure~\ref{ex32}.
\begin{figure}
\begin{center}

\ \ \ \ \ \ \ \ \  \xymatrix{
& & \ar[dl] \app \ar[d]  \ar[dr] \ar[drr] & & &  \\
&t \ar@{.>}[d]& v^{i}_{1} & \ldots  v^{i}_{j} \ldots \ar@{.>}[d] &  v^{i}_{n}
& u_{i} \\
& \ar@{-}[dl] y_{j}  \ar@{-}[dr] \ar@{.>}[urr]  & &  \ar@{-}[dl] x_{k}
\ar@{-}[dr] \ar@{.>}[dll] & \\ 
\ar@{-}[r] & \lambda \ol{z} \ar@{.>}[d] \ar@{-}[r] & \ar@{-}[r] & 
\lambda \ol{w} \ar@{-}[r] \ar@{.>}[d] & \\
& \ar@{-}[dl] z_{l} \ar@{-}[dr] \ar@{.>}[urr] & &  w_{m} \ar@{.>}[dll]& \\
\ar@{-}[r] & \lambda  \ar@{.>}[d] \ar@{-}[r] &   & 
 & \\
& & &  &
} 

\end{center}
\caption{Illustrating game semantics}
\label{ex32}
\end{figure}
Player Opponent  chooses a branch of $u_{i}$.
Then,   there is
a finite play that starts at the root of $t$ and
may repeatedly  jump in and out of $t$ and in and out of
the $v^{i}_{j}$'s. At a constant $a : \Nil$
play ends. At other constants  $f$, player Proponent tries
to match  Opponent's choice of branch. Proponent
wins, when the play finishes,  if the sequence of constants
encountered matches the branch
chosen by Opponent (assuming a mechanism for forbidden constants).
Play, for example,  may reach $y_{j}$ in $t$
and then jump to $v^{i}_{j}$, as it is this subtree 
that is applied to $\lambda \ol{y}$ at the root of $t$,
and then when at $x_{k}$ in $v^{i}_{j}$ play may return to $t$ to an immediate
successor of $y_{j}$
labelled $\lambda \ol{z}$; play then may proceed
to $z_{l}$ and return to a successor of $x_{k}$ in $v^{i}_{j}$,
and so on. Game semantics models $\beta$-reduction on the fixed
structure of Figure~\ref{ex32} without changing it 
using  substitution. This is the rationale for the
tree-checking game\footnote{I am indebted
to Luke Ong for pointing out there is
a  close formal relationship between the tree-checking
game and game semantics.}. However, the game that we now define
starts from the assumption
that only $t$ is the common structure for the problem $P$.
Moreover, in later sections, we shall define transformations
on $t$ justified by game playing
which introduces an asymmetry between
it and the argument
terms $v^{i}_{j}$ of $P$ which are fixed.  
Consequently, we insist that play is  always in the term $t$. 
Jumping in and out
of the $v^{i}_{j}$'s is coded using states, as play traverses
$t$. The game also 
avoids the justification pointers of game
semantics by appealing to  iteratively defined look-up tables.

The tree-checking game $\mathsf{G}(t,P)$
is played by one participant, player
$\forall$, the \emph{refuter} who
attempts to show that $t$ is not a solution of $P$.
It appeals to a finite set of states involving
elements of $\mathsf{L}$ and $\mathsf{R}$ from Definition~\ref{def3}:
$\mathsf{L}$ are the subterms of the $v^{i}_{j}$'s and
$\mathsf{R}$ are those of the $u_{i}$'s, both modulo the forbidden
constants. 
There are four  kinds of state, as follows.
\begin{enumerate}[$\bullet$]
\item An \emph{argument} state has the form $q[(l_{1},\ldots, l_{k}), r]$
where each $l_{j} \in \mathsf{L}$ (and $k$ can be $0$)
and $r \in \mathsf{R}$. Such a state will only occur at a node in $t$
labelled
$\lambda z_{1} \ldots z_{k}$  where each $l_{j}$ has
the same type as $z_{j}$: $l_{1} \ldots l_{k}$ are the subterms
that are applied to the subterm rooted at
$\lambda z_{1}\ldots z_{k}$. 
A state $q[ ( \ ),r]$ occurs at a node of $t$ labelled with
a dummy lambda. 
\item A \emph{value} state has the form $q[l,r]$ where $l \in \mathsf{L}$
and $r \in \mathsf{R}$. This state can only occur at 
a node of $t$ labelled with a variable $y$ which has the same type as  $l$: 
$l$ is the subterm of some $v^{i}_{j}$
that play at $y$ would jump
to in game semantics.
\item An \emph{empty} state  has the form $q[- ,r]$ where $r \in \mathsf{R}$
and can only occur at a node of $t$ labelled 
with  a higher-order constant $f : (B_{1},\ldots,B_{k},\Nil)$
when  $r$ has the 
form $f u_{1} \ldots u_{k}$.   
\item A \emph{final} state is either 
$q[\, \fa \, ]$, winning for the refuter, or $q[\, \ex \,]$, 
losing for the refuter.
\end{enumerate}

As play traverses  $t$, there are two kinds
of free variables: those in the current subtree
of $t$ (such as $y_{j}$ in Figure~\ref{ex32})
and those
in the left terms of a  current argument or value state
$q[(l_{1},\ldots,l_{k}),r]$ or $q[l,r]$  (such as $x_{k}$ 
in Figure~\ref{ex32}).
A free variable in a subtree of $t$
is  associated with a single
left subterm (an element of $\mathsf{L}$) and 
a free variable in a left subterm $l$ of an argument or value
state is  associated with a unique subtree  of $t$. 
So, the game appeals to look-up tables or nested environments
$\theta \in \Theta_{k}$ and $\xi \in \Xi_{k}$ at 
a position $k \geq 1$
that are defined iteratively.

\begin{defi} \label{defi42} The sets of partial mappings $\Theta_{k}$ and
$\Xi_{k}$ are defined iteratively as follows.
\begin{enumerate}[(1)] 
\item $\Theta_{1} = \eset{\theta_{1}}$ and $\Xi_{1} = \eset{\xi_{1}}$
where both $\theta_{1}$ and $\xi_{1}$ are empty (that is, have no entries).
\item For $k > 1$,  $\theta \in \Theta_{k}$ iff
$\theta$ is a partial
map from variables that are labels of nodes in
$t$ to triples   $l \xi j$ where  
$l \in \mathsf{L}$, $j < k$ and $\xi \in \Xi_{j}$.
 \item For $k > 1$, $\xi \in \Xi_{k}$ iff $\xi$  is a partial
map from variables that can occur in terms of 
$\mathsf{L}$ to triples  $t' \theta j$
where $t'$ is a subtree of $t$, $j < k$ and $\theta \in \Theta_{j}$.
\end{enumerate}
\end{defi}

\ni
A variable $y$
in $t$ may be associated with a left subterm $l \in \mathsf{L}$
which contains free variables: hence, the need for $\theta(y)$
to be a triple  $l \xi j$ as $\xi$ records
the values of the free variables in $l$ as determined at the earlier position
$j$: we include the position $j$ which is crucial
to the  understanding of game playing in later sections,
when relationships between positions are analysed. 
Similarly, a variable $z$ in a left subterm may
be associated with a subtree of $t$ which contains free variables;
so, $\xi(z) = t'\theta j$ where $\theta$ has entries for the
free variables in $t'$ at the earlier position $j$.
Initially, at the beginning of play when there are no free variables
and no previous moves, 
$\theta \in \Theta_{1}$ and $\xi \in \Xi_{1}$ are both
empty. 
The look-up tables play the same role as environments for abstract
machines of the lambda calculus (such as a Krivine machine).
So the game will simulate the 
evaluation of a branch of the normal form in the same way
abstract machines compute closures.
However, unlike these abstract machines, the game here
essentially depends on $\eta$-long normal forms.

\begin{defi} A \emph{play} of $\mathsf{G}(t,P)$  is a   sequence of positions
$t_{1} q_{1} \theta_{1} \xi_{1}, \ldots, t_{n} q_{n} \theta_{n} \xi_{n}$ 
where 
\begin{enumerate}[(1)]
\item each $t_{i}$ is a
node of $t$ and  $t_{1}$ is the root node of $t$, 
\item each $q_{i}$ is a state,  $q_{n}$ is a final state
and $q_{1}$ is decided as follows:  $\fa$ chooses
a (dis)equation $x \, v^{i}_{1} \ldots v^{i}_{n} \approx_{i} u_{i}$ 
in  $P$ and
$q_{1} = q[(v^{i}_{1}, \ldots, v^{i}_{n}),u_{i}]$,
\item for each $i$, $\theta_{i} \in \Theta_{i}$ and $\xi_{i} \in \Xi_{i}$,
\item position  $t_{m+1} q_{m+1} \theta_{m+1} \xi_{m+1}$, $m < n$, 
is determined by a single move
in  Figure~\ref{game} from position $t_{m}q_{m} \theta_{m} \xi_{m}$
according to the label at $t_{m}$.
\end{enumerate}
\end{defi}

\begin{figure}

\begin{enumerate}[A.]

\item $t_{m}$ is labelled $\lambda y_{1} \ldots y_{j}$ and $j \geq 0$.
Assume $q_{m} =
q[(l_{1},\ldots, l_{j}),r]$.

\ni
Then, $t_{m+1} = t'$ such that $t_{m} \downarrow_{1} t'$, 
$\theta_{m+1} = \theta_{m} \eset{l_{1}\xi_{m} m/y_{1}, 
\ldots, l_{j}\xi_{m} m/y_{j}}$
and $q_{m+1}$, $\xi_{m+1}$ are defined 
by cases on the label at $t_{m+1}$.

\begin{enumerate}[1.]
\item $a : \Nil$. 
Then, $\xi_{m+1} = \xi_{m}$. If $r = a$ then 
$q_{m+1} = q[ \, \ex \, ]$
else  $q_{m + 1} = q[ \, \fa \, ]$.

\item $f: (B_{1}, \ldots, B_{k},\Nil)$.  
Then, $\xi_{m+1} = \xi_{m}$.
If $r = f s_{1} \ldots s_{k}$ then
$q_{m+1} = q[ -, r]$ else $q_{m+1}
= q[\,  \fa \, ]$.

\item $y : B$. 
If $\theta_{m+1}(y) = l \xi i$,
then $\xi_{m+1} = \xi$ and $q_{m+1} = q[l,r]$.

\end{enumerate}

\item $t_{m}$ is labelled $f : (B_{1}, \ldots, B_{k},\Nil)$.
Assume $q_{m} = q[ -, f s_{1} \ldots s_{k} ]$.

\begin{enumerate}[1.]
\item Then, $\theta_{m+1} = \theta_{m}$, $\xi_{m+1} = \xi_{m}$
and $\fa$ chooses $d : 1 \leq d \leq k$
and  $t_{m+1} = t'$ such that $t_{m} \downarrow_{d} t'$
and $q_{m+1}$ is by cases on $s_{d}$.

\ni
$s_{d} : \Nil$.  Then
$q_{m+1} = q[( \ ), s_{d}]$. 

\ni
$s_{d} =
\lambda x_{i_{1}} \ldots x_{i_{n}}.s$.
Then  $q_{m+1}$ $=$ $q[(c_{i_{1}},\ldots,c_{i_{n}}), 
s \eset{c_{i_{1}}/x_{i_{1}}, \ldots, c_{i_{n}}/x_{i_{n}}}]$.
\end{enumerate}

\item $t_{m}$ is labelled $y$. Assume 
$q_{m} = q[l,r]$.

\noindent
If $l : \Nil$ then $\xi_{m+1} = \xi_{m}$
else  $l = \lambda z_{1} \ldots z_{j}.w$ and
$t_{m} \downarrow_{i} t'_{i}$, $1 \leq i \leq j$,  and
$\xi_{m+1} = \xi_{m}\eset{t'_{1} \theta_{m} m/z_{1}, \ldots, t'_{j}
\theta_{m} m/z_{j}}$.
Elements
$t_{m+1}$, $q_{m+1}$ and $\theta_{m+1}$
are  by cases on $l$. 
\begin{enumerate}[1.]
\item $a$ or $\lambda z_{1} \ldots z_{j}. a$ where $a : \Nil$.
Then, $t_{m+1} = t_{m}$ and $\theta_{m+1} = \theta_{m}$.

\ni
If $r =a$ then $q_{m+1} = q[ \, \ex \,]$
else $q_{m+1} = q[ \, \fa \, ]$. 

\item $c : (B_{1}, \ldots, B_{k},\Nil)$.
Then $\theta_{m+1} = \theta_{m}$. 

\ni
If $r \not= c s_{1}  \ldots  s_{k}$ then $t_{m+1} = t_{m}$
and $q_{m+1} = q [ \, \fa \, ]$
else  $r = c s_{1}  \ldots  s_{k}$ and
$\fa$ chooses  $d: 1 \leq d \leq k$, 
and $t_{m+1} = t'$ such that  $t_{m}
\downarrow_{d} t'$ and $q_{m+1}$ is by cases on $s_{d}$.

\ni
$s_{d} : \Nil$. Then $q_{m+1} = q[( \ ), s_{d}]$.

\ni
$s_{d} = \lambda x_{i_{1}} \ldots x_{i_{n}}.s$. Then
$q_{m+1}$ $=$ $q[(c_{i_{1}},\ldots,c_{i_{n}}), 
s \eset{c_{i_{1}}/x_{i_{1}}, \ldots, c_{i_{n}}/x_{i_{n}}}]$.

\item $f w_{1} \ldots w_{k}$ or $\lambda z_{1} \ldots z_{j}.
f w_{1} \ldots  w_{k}$.  
Then  $t_{m+1} = t_{m}$ and  $\theta_{m+1} 
=\theta_{m}$.

\ni
If $r \not= f s_{1} \ldots  s_{k}$, then $q_{m+1} = q [ \, \fa \, ]$
else 
$r = f s_{1} \ldots  s_{k}$ and
$\fa$ chooses $d : 1 \leq d \leq k$ and $q_{m+1}$ is by cases on $s_{d}$.

\ni
$s_{d} : \Nil$. Then $q_{m+1} = q[ w_{d}, s_{d}]$.

\ni
$s_{d} = \lambda x_{i_{1}} \ldots x_{i_{n}}. s$
and $w_{d} = \lambda y_{1} \ldots y_{n}. w'$.
Then $q_{m+1}$ $=$ $q[w'
\eset{c_{i_{1}}/y_{1},\ldots,c_{i_{n}}/y_{n}}, 
s \eset{c_{i_{1}}/x_{i_{1}}, \ldots, c_{i_{n}}/x_{i_{n}}}]$.

\item $x \, l_{1} \ldots l_{k}$ or $\lambda  z_{1} \ldots z_{j}.
x \, l_{1} \ldots l_{k}$ where $k \geq 0$.

\ni
If $\xi_{m+1}(x) = t' \theta i$
then  $t_{m+1} = t'$, $\theta_{m+1} = \theta$
and  $q_{m+1} = q[(l_{1},\ldots,l_{k}),r]$.

\end{enumerate}

\end{enumerate}
\caption{Game moves}
\label{game}
\end{figure}

The initial position of a play of $\mathsf{G}(t,P)$
is $t_{1} q[(l_{1},\ldots,l_{n}),r] \theta_{1} \xi_{1}$
where $t_{1}$ is the initial node of $t$ labelled $\lambda y_{1} \ldots y_{n}$
for some $y_{1}, \ldots, y_{n}$
and $x l_{1} \ldots l_{n} \approx r$ is a (dis)equation in $P$.
At this position we are interested whether $(t_{1} \theta_{1})
(l_{1} \xi_{1}) \ldots (l_{n} \xi_{1}) =_{\beta} r$ where the look-up tables
are viewed as substitutions: initially, they are empty because there are 
no free variables in the terms $t, l_{1}, \ldots, l_{n}$
. The initial position is the same as an initial position
in game semantics, except the terms $l_{j}$  and $r$  in the game here are
part of the state (and the choice of branch in $r$ will take place as play
proceeds).

The possible moves from position $m$ to $m+1$ in a play
are listed in Figure~\ref{game} and
are divided into three groups that depend on the label at 
$t_{m}$ of position $m$. 
Group A covers the case when it is a $\lambda \ol{y}$,
group B  a constant $f$ (whose type is not $\Nil$)
and group C a variable $y$.
For look-up tables $\theta_{m+1}$
and $\xi_{m+1}$, we assume that  the substitution notation 
also stands  for
function updating:
$\mu' = \mu\eset{\nu_{1}/y_{1},\ldots, \nu_{k}/y_{k}}$
means that the entries in $\mu'$ are  the same as
in $\mu$ except for the $y_{i}$'s, as $\mu'(y_{i}) = \nu_{i}$
for each $y_{i}$, $1 \leq i \leq k$.
This notation is permitted even if $k =0$.

Consider group A moves when position 
$m$ is at the node $t_{m}$ labelled $\lambda y_{1} \ldots y_{j}$, $j \geq 0$, 
so  $q_{m}$ has the form $q[(l_{1},\ldots,l_{j}),r]$ where
each $l_{i}$ has the same type as $y_{i}$. 
(If $j = 0$ then the position is at a dummy lambda.)
The ``interpretation'' of position $m$ in terms of
$\beta$-reduction
(see  the proof of Theorem~\ref{imp})
is whether $(t_{m} \theta_{m}) (l_{1} \xi_{m}) \ldots (l_{j} \xi_{m})$
$=_{\beta} r$.
Node $t_{m}$ has a single successor  $t_{m+1}$
and so play descends to it. 
However, the subtree at $t_{m+1}$ may contain 
free occurrences of the $y_{i}$'s (when
$j > 0$): the interpretation of each such occurrence is
$l_{i} \xi_{m}$ because
$t_{m}  = \lambda y_{1} \ldots y_{j}. t_{m+1}$
and so $\theta_{m+1}$ is
an updated version  of $\theta_{m}$ reflecting
this association of the $y_{i}$'s with the $l_{i}$'s
and the look-up table $\xi_{m}$ at position $m$. 
Move A1 is when
$t_{m+1}$ is labelled with $a : \Nil$.
We can now immediately decide whether $(t_{m} \theta_{m})
(l_{1} \xi_{m}) \ldots (l_{j} \xi_{m}) =_{\beta} r$ just
by comparing $r$ and $a$;
so position $m+1$ is final (and won by the refuter if $r$ is different
from $a$). 
For move A2, if $t_{m+1}$ is labelled with $f : (B_{1},\ldots,B_{k},\Nil)$
and $r$ does not have the form $f s_{1} \ldots s_{k}$
then $q_{m+1} = q[ \, \fa \, ]$ as we now know that 
$(t_{m} \theta_{m}) (l_{1} \xi_{m}) \ldots (l_{j} \xi_{m})$
$\not=_{\beta} r$. If $r = f s_{1} \ldots s_{k}$ then
position $m+1$ is $t_{m+1} q[-, r] \theta_{m+1} \xi_{m+1}$
where $\xi_{m+1} = \xi_{m}$. 
Move A3 is when $t_{m+1}$ is 
labelled with a  variable $y$. Thus, $y$ is a free variable occurrence
in the tree $t_{m+1}$ whose interpretation 
is $\theta_{m+1}(y) = l \xi i$ decided at the earlier position $i \leq m$.
So, $\xi_{m+1}$ is set to $\xi$ (as it interprets the free variables
in $l$ in $q_{m+1}= q[l,r]$).

The  $B$ move covers the case when $t_{m}$ is labelled with a constant
$f : (B_{1},\ldots,B_{k},\Nil)$;
because of move A2, we only need to consider it
when the state is $q_{m} = q[-,fs_{1} \ldots s_{k}]$
for some $s_{1},\ldots,s_{k}$.
The ``interpretation'' of such a position $m$ is
whether $(t_{m} \theta_{m}) =_{\beta} f s_{1} \ldots s_{k}$.
In which case $t_{m} = f t'_{1} \ldots t'_{k}$.  
So, $(t_{m} \theta_{m}) =_{\beta} f s_{1} \ldots s_{k}$
if, and only if, for each $d: 1 \leq d \leq k$, 
$t'_{d} \theta_{m} =_{\beta} s_{d}$.
The refuter $\fa$ chooses such a $d$. 
The delicacy is that $s_{d}$ may be of higher type,
of the form $\lambda x_{i_{1}} \ldots x_{i_{n}}.s$
and so, therefore, $t'_{d}$ is labelled $\lambda y'_{1} \ldots y'_{n}$
for some $y'_{1},\ldots y'_{n}$
because it has the same type: position $m+1$
is then $t'_{d} q[(c_{i_{1}},\ldots,c_{i_{n}}), s\eset{c_{i_{1}}/x_{i_{1}},
\ldots, c_{i_{n}}/x_{i_{n}}}] \theta_{m+1} \xi_{m+1}$ where
$\theta_{m+1} = \theta_{m}$,
and $\xi_{m+1} = \xi_{m}$:
here we are making use of the forbidden constants $c_{i_{j}}$.  
If $s_{d}$ has ground type then $q_{m+1} = q[( \ ), s_{d}]$
(as the label at $t'_{d}$ is a dummy lambda).

Group C moves cover the case where $t_{m}$ is labelled with a variable
$y$.  The ``interpretation'' of position $m$, 
$t_{m} q[l,r] \theta_{m} \xi_{m}$, is whether 
$(l \xi_{m}) (t'_{1} \theta_{m}) \ldots (t'_{j} \theta_{m})
=_{\beta} r$ where for $j \geq 0$, $t_{m} \downarrow_{i} t'_{i}$ 
when $1 \leq i \leq j$. 
If $j > 0$ then $l = \lambda z_{1} \ldots z_{j}.w$
for some $z_{1},\ldots, z_{j}$ and the free 
occurrences of $z_{i}$ in $w$ are associated
with  $t'_{i} \theta_{m}$ at position $m$;
so, $\xi_{m+1}$
is an updated version  of $\xi_{m}$ 
reflecting this association.
When $j =0$, $l =w$ and the interpretation is whether $l \xi_{m} =_{\beta} r$.
For both cases $l = \lambda z_{1} \ldots z_{j}.w$ and $l = w$
play proceeds by examining  the ``head'' of $w$. 
Move C1 covers the case
where it is $a : \Nil$ (which is possibly a forbidden constant). 
We can now immediately decide whether
$(l \xi_{m}) (t'_{1} \theta_{m}) \ldots (t'_{j} \theta_{m})
=_{\beta} r$ just by comparing $l$ and $r$.
Move C2 covers the case where $l$ is a forbidden constant $c$ at higher type
$(B_{1},\ldots,B_{j},\Nil)$.
If $r \not= c s_{1} \ldots s_{j}$ then we 
know that $c (t'_{1} \theta_{m}) \ldots (t'_{j} \theta_{m})
\not=_{\beta} r$ and, so $q_{m+1} = q[\,  \fa \, ]$. 
If $r = c s_{1} \ldots s_{j}$ then
$c (t'_{1} \theta_{m}) \ldots (t'_{j} \theta_{m})
=_{\beta} r$ if, and only if, 
for each $d$, $(t'_{d} \theta_{m})  =_{\beta} s_{d}$. 
As with B1, the refuter $\fa$ chooses such a $d$. 
Again we need to specially deal with the case that
$s_{d}$ has  higher type.
Move C3 deals with the case that $w$ is 
$f w_{1} \ldots w_{k}$. 
Again, if $r$ does not have the form $f s_{1} \ldots s_{k}$
then $q_{m+1} = q[\, \fa \, ]$ as we now know that 
$(l \xi_{m}) (t'_{1} \theta_{m}) \ldots (t'_{j} \theta_{m})
\not=_{\beta} r$.
If $r = f s_{1} \ldots s_{k}$ then
$(l \xi_{m}) (t'_{1} \theta_{m}) \ldots (t'_{j} \theta_{m})
=_{\beta} r$ if, and only if, for each $d$,
$(w_{d} \xi_{m+1}) =_{\beta} s_{d}$. Therefore,
$\fa$ chooses such a $d$.
For the next position, again we need to deal with the possibility
that $s_{d}$ is of higher type: if $s_{d} =
\lambda x_{i_{1}} \ldots x_{i_{n}}.s$
then $w_{d}$ which has the same type must be of the form
$\lambda y_{1} \ldots y_{n}.w'$ for some $y_{1}, \ldots,
y_{n}$; so we must substitute the same forbidden constant
for each $x_{i_{j}}$ and $y_{j}$;
so $q_{m+1}$ is $q[w'
\eset{c_{i_{1}}/y_{1},\ldots,c_{i_{n}}/y_{n}}, 
s \eset{c_{i_{1}}/x_{i_{1}}, \ldots, c_{i_{n}}/x_{i_{n}}}]$.
If $s_{d} : \Nil$
then the  next position
$m+1$ is  $t_{m+1} q[w_{d},s_{d}] \theta_{m+1} \xi_{m+1}$
where $t_{m+1}  = t_{m}$ and $\theta_{m+1} =\theta_{m}$.
Move C4 covers the case when $w = x l_{1} \ldots l_{k}$,
$k \geq 0$. Therefore,
$x$ is a free  variable in $l$ whose interpretation 
$\xi_{m+1}(x) = t' \theta i$ which was determined
at the earlier position $i \leq m$; so, $x$ is associated with
the subtree $t'$ of $t$ and play therefore jumps to it.
\begin{figure}
\[
\xymatrix{
& (1) \lambda y \ar[d] & \\
& (2) y \ar[d] & \\
& (3) \lambda \ar[d] & \\
& (4) y \ar[d] & \\
 & (5) \lambda \ar[d]& \\
 & \ar[dl] (6) f \ar[dr] &  \\
\ar[d] (7) \lambda x z_{1}z_{2} & & (13) \lambda \ar[d] \\
\ar[d] (8) x & & (14) y \ar[d] \\
\ar[d] (9) \lambda & & (15) \lambda \ar[d] \\
\ar[d] (10) y & & (16) a \\
\ar[d] (11) \lambda & & \\
 (12) z_{2} & & 
} 
\]
\caption{A solution term with order $3$ for Example~\ref{examp42}}
\label{picture2}
\end{figure}

$\fa$ can exercise choice, by carving out a branch of a right term
of a (dis)equation modulo forbidden constants, with 
moves  B1, C2 and C3.
The look-up tables are used in earnest with moves  A3 and C4
to interpret the two kinds of free variable.
Move  C4 allows play to jump elsewhere 
in  the term tree (always to a node
labelled with a lambda): it also opens the possibility that
play can repeatedly be at the same node of $t$. 
With moves   A1--A3, B1 and C2 (unless play finishes)
control passes down  the term tree 
while it remains stationary in the case of C1 and
C3.

\begin{defi}
If $t_{1} q_{1} \theta_{1} \xi_{1}, \ldots, t_{n} q_{n} \theta_{n}
\xi_{n}$ is a play of  $\mathsf{G}(t,P)$ 
then player $\fa$ \emph{wins the play} if the final state is $q[ \, \fa \, ]$
and she \emph{loses} it otherwise (if the final state is
$q[ \, \ex \,]$).
\end{defi}

\begin{exa}\label{examp42} Let  $P$ be the problem 
$x(\lambda z. z) = u$ where $u = f(\lambda x_{1} x_{2} x_{3}.x_{1} x_{3}) a$
of Example~\ref{examp2}.
The bound variables in the right term $u$ are
$\eset{x_{1},x_{2},x_{3}}$; so, assume  
corresponding forbidden constants $C = \eset{c_{1},c_{2},c_{3}}$. 
Let $t = \lambda y. y(y(f(\lambda x z_{1} z_{2}.x (y z_{2})) (y a) ))$:
tree$(t)$ is depicted in  Figure~\ref{picture2}.
$\mathsf{G}(t,p)$ consists of two plays that descend $t$.
(For $3$rd-order problems $P$ play cannot  jump around a term tree,
as we observe in Section~\ref{third}.) 
Both plays start as follows where we have supplied which move is
applied to produce the next position: the initial component
of each position is a node of $t$.
\[ \begin{array}{llll}
(1) \, q[(\lambda z.z), u]\,  \theta_{1} \, \xi_{1} & & &\\
(2) \, q[ \lambda z.z, u] \, \theta_{2} \xi_{2} &
\theta_{2} = \theta_{1} \eset{(\lambda z.z) \xi_{1} 1/y} 
& \xi_{2} = \xi_{1} & A3\\
(3) \, q[( \ ), u] \, \theta_{3} \xi_{3}
& \theta_{3} = \theta_{2} &
\xi_{3} = \xi_{1}\eset{(3)\theta_{2} 2/z} & C4\\
(4) \, q[\lambda z. z, u] \, \theta_{4} \xi_{4} &
\theta_{4} = \theta_{2} & \xi_{4} = \xi_{1} & A3\\
(5) \, q[( \ ), u] \, \theta_{5} \, \xi_{5} &
\theta_{5} = \theta_{2} & \xi_{5} = \xi_{1}\eset{(5) \theta_{2} 4/z}
& C4\\
(6) \, q[-, u] \, \theta_{6} \, \xi_{6} &
\theta_{6} = \theta_{2} & \xi_{6} = \xi_{5} & A2
\end{array} \]
The initial state is an argument state
$q[(\lambda z.z),u]$ and control is at node (1).
Play descends from node (1) to (2) calling the value
$\lambda z.z$ by A3.
Next, by C4, because $z$ is the head variable
in the body of $\lambda z.z$, has no arguments
and is  associated with node (3),
the next state is the argument state
$q[ ( \ ),u]$ and control is at (3).
Play descends from (3) to (4) calling the value
$\lambda z.z$ by A3. Again by  C4, $z$ is the head variable
in the body of $\lambda z.z$, has no arguments and is now
associated with (5), the next state is the argument
state $q[( \ ), u]$ and control is at (5).
A2 is then applied (because the right term $u$ in the state
has $f$ as head constant) and control passes from (5) to (6).
Move B1 is now applied and there is a $\fa$ choice as to
which branch of $u$ to take.
If direction $1$ is chosen then
play continues as follows.
\[ \begin{array}{llll}
(7) \, q[(c_{1},c_{2},c_{3}), c_{1} c_{3} ] \, \theta_{7} \, \xi_{7} & \theta_{7} = \theta_{2} &
\xi_{7} = \xi_{5} & B1\\
(8) \, q[c_{1}, c_{1} c_{3} ] \, \theta_{8} \, \xi_{8} & \theta_{8} = 
\theta_{2}\eset{c_{1}\xi_{5} 7\!/\!x,c_{2} \xi_{5} 7\!/\!z_{1},
c_{3}\xi_{5} 7\!/\!z_{2}} & \xi_{8} = \xi_{5} & A3\\
(9) \, q[( \ ), c_{3}] \, \theta_{9} \, \xi_{9} & \theta_{9} = \theta_{8}
& \xi_{9} = \xi_{5} & C2 \\
(10) \, q[\lambda z. z, c_{3}] \, \theta_{10} \xi_{10} &
\theta_{10} = \theta_{8} & \xi_{10} = \xi_{5} & A3\\
(11) \, q[( \ ), c_{3}] \, \theta_{11} \, \xi_{11} &
\theta_{11} = \theta_{8} & \xi_{11} = \xi_{5}\eset{(11) \theta_{8} 10\!/\!z}
& C4\\
(12) \, q[ c_{3},c_{3}] \, \theta_{12} \, \xi_{12} & \theta_{12}
= \theta_{8} & \xi_{12} = \xi_{5} & A3 \\
(12) \, q[ \ \ex \ ] \, \theta_{13} \, \xi_{13} & \theta_{13} = \theta_{8}
& \xi_{13} = \xi_{5} & C1
\end{array} \]
Forbidden constants are introduced for replacing
$x_{1},x_{2},x_{3}$ in the body of $u$ to give the right term
$c_{1} c_{3}$ and as arguments $(c_{1},c_{2},c_{3})$ 
for the variables $x,z_{1},z_{2}$ bound at (7); see the updated
look-up table $\theta_{8}$.
At (8) the value $c_{1}$ is called using move A3 and 
then by C2, control
proceeds to (9) and  the right term becomes the argument
$c_{3}$ of $c_{1}$. 
At (10) the value $\lambda z.z$ is called again
by A3 and by C4 as $z$ is the head variable and is associated
with node (11) control passes to it (with the empty sequence
of arguments). Finally, at (12), the value $c_{3}$
is called by A3, and then by C1, $\fa$ loses the play.
She also loses if direction $2$ is chosen
at position $7$ as the reader can verify. 
\qed \end{exa}

\begin{exa}\label{examp43}
Let  $P$ be the first equation of Example~\ref{examp1}
$x\, v = f a$  where $v = \lambda y_{1}y_{2}.y_{1}y_{2}$
 (which is a $4$th-order problem). 
The tree  $t$ 
of  Figure~\ref{ex1} solves $P$ as was demonstrated
in Example~\ref{examp41} through $\beta$-reduction.
The single play in $\mathsf{G}(t,P)$ is 
presented in Figure~\ref{ex2}.
\begin{figure}
\[ \begin{array}{llll}
(1) \, q[(v), fa]\,  \theta_{1} \, \xi_{1} & & &\\
(2) \, q[ v, fa] \, \theta_{2} \xi_{2} &
\theta_{2} = \theta_{1} \eset{v \xi_{1} 1/z} 
& \xi_{2} = \xi_{1} & A3\\
(3) \, q[(y_{2}), fa] \, \theta_{3} \xi_{3}
& \theta_{3} = \theta_{2} &
\xi_{3} = \xi_{1}\eset{(3)\theta_{2} 2/y_{1},(11)\theta_{2} 2/y_{2}} & C4\\
(4) \, q[-, fa] \, \theta_{4} \xi_{4} &
\theta_{4} = \theta_{2}\eset{y_{2}\xi_{3} 3/x} & \xi_{4} = \xi_{3} & A2\\
(5) \, q[( \ ), a] \, \theta_{5} \, \xi_{5} &
\theta_{5} = \theta_{4} & \xi_{5} = \xi_{4}
& B1\\
(6) \, q[v, a] \, \theta_{6} \, \xi_{6} &
\theta_{6} = \theta_{4} & \xi_{6} = \xi_{1} & A3\\
(7) \, q[(y_{2}), a ] \, \theta_{7} \, \xi_{7} & \theta_{7} = \theta_{4} &
\xi_{7} = \xi_{1}\eset{(7)\theta_{4} 6/y_{1},(9)\theta_{4} 6/y_{2}} & C4\\
(8) \, q[y_{2}, a] \, \theta_{8} \, \xi_{8} & \theta_{8} = 
\theta_{4}\eset{y_{2} \xi_{7}7/u} & \xi_{8} = \xi_{3} & A3\\
(11) \, q[ ( \ ), a] \, \theta_{9} \, \xi_{9} & \theta_{9} = \theta_{2}
& \xi_{9} = \xi_{3} & C4 \\
(12) \, q[ v, a] \, \theta_{10} \xi_{10} &
\theta_{10} = \theta_{2} 
& \xi_{10} = \xi_{1} & A3\\
(13) \, q[(y_{2}), a] \, \theta_{11} \xi_{11}
& \theta_{11} = \theta_{2} &
\xi_{11} = \xi_{1}\eset{(13)\theta_{2} 10/y_{1},(19)\theta_{2} 10/y_{2}} & C4\\
(14) \, q[v, a] \, \theta_{12} \xi_{12} &
\theta_{12} = \theta_{2}\eset{y_{2}\xi_{11} 11/y} & \xi_{12} = \xi_{1} & A3\\
(15) \, q[(y_{2}), a] \, \theta_{13} \, \xi_{13} &
\theta_{13} = \theta_{12} & \xi_{13} = \xi_{1}\eset{(15)\theta_{12} 12/y_{1},
(17)\theta_{12} 12/y_{2}} & C4\\
(16) \, q[y_{2}, a] \, \theta_{14} \, \xi_{14} &
\theta_{14} = \theta_{12}\eset{y_{2}\xi_{13} 13/s} & \xi_{14} = \xi_{13} & A3\\
(17) \, q[( \ ), a ] \, \theta_{15} \, \xi_{15} & \theta_{15} = \theta_{12} &
\xi_{15} = \xi_{13} & C4\\
(18) \, q[y_{2}, a] \, \theta_{16} \, \xi_{16} & \theta_{16} = 
\theta_{12} & \xi_{16} = \xi_{11} & A3\\
(19) \, q[( \ ), a ] \, \theta_{17} \, \xi_{17} & \theta_{17} = \theta_{2} &
\xi_{17} = \xi_{11} & C4\\
(20) \, q[ \ \ex \ ] \, \theta_{18} \, \xi_{18} & \theta_{18} = 
\theta_{2} & \xi_{18} = \xi_{11} & A1\\
\end{array} \]
\caption{The play of Example~\ref{examp43} on the tree in Figure~\ref{ex1}}
\label{ex2}
\end{figure}
The initial state is $q[(v),fa]$ at the root of $t$.
By move A3, play descends to node $(2)$ calling the value
$v$. Next by C4 because $y_{1}$ is the head variable in the body
of $v$, is associated with the subtree
at $(3)$ and has argument $y_{2}$ play moves to $(3)$ with argument
state $q[(y_{2}),fa]$. 
By A2 play descends to node $(4)$ labelled $f$ and then move B1
is applied without a choice for $\fa$ because the type of $f : (\Nil,\Nil)$
has arity one, and so the play descends to node $(5)$ with a change
in the right term of the state from $f a$ to $a$. 
By move A3 play descends to node (6) calling the value $v$ again.
By C4 because the head variable $y_{1}$ is associated with 
the subtree at $(7)$ and has argument $y_{2}$ play moves to
$(7)$. By move A3, play descends to $(8)$. The entry $\theta_{8}(x)$
is $y_{2} \xi_{3} 3$, so the state is $q[y_{2},a]$. By move C4
because $\xi_{3}(y_{2}) = (11) \theta_{2} 2$, that is, 
$y_{2}$ is associated with the subtree rooted at $(11)$, 
play jumps from node $(8)$ to node $(11)$.
If node $(8)$ were  labelled $u$ then because the entry
$\theta_{8}(u) = y_{2} \xi_{7} 7$ the state would
again be $q[y_{2},a]$; play would then jump to node $(9)$
because $\xi_{7}(y_{2}) = (9) \theta_{4} 6$.
Play descends from node $(11)$ to $(12)$, $(13)$, $(14)$,
$(15)$ and $(16)$ and jumps to node $(17)$ and descends
to $(18)$ and then jumps to node $(19)$ before descending
to node $(20)$ by move A1,  where the refuter  loses the play.
\qed \end{exa}
 
\begin{figure}
\[
\xymatrix{
& &  (1) \lambda z \ar[d]& \\
& & \ar[dl](2) z \ar[dr] &  \\
& (3) \ar[d] \lambda z_{1} & & \ar[d] (17) \lambda z_{2} & \\
& \ar[dl] (4) z \ar[dr] & &  \ar[d] (18) h  \\
\ar[d] (5) \lambda x_{1}& & \ar[d] (13) \lambda x_{2}  & \ar[d] (19) \lambda \\
\ar[d](6) z_{1} & &  \ar[d] (14) g  &  (20) z_{2}  \\
\ar[d] (7) \lambda  & & \ar[d] (15) \lambda   & \\
\ar[d] (8) x_{1}  & & (16) x_{2} &  \\
\ar[d] (9) \lambda   & & & \\
\ar[d] (10) z_{1}   & & & \\
\ar[d] (11) \lambda &  & & \\
(12) a & & & 
} 
\]
\caption{A $5$th order term tree for Example~\ref{examp44}}
\label{ex14}
\end{figure}

\begin{exa}\label{examp44}
We now examine the equation of Example~\ref{examp3} which 
illustrates play jumping in more detail
and how the game moves, especially A3 and C4, 
essentially depend on 
$\eta$-long normal forms
Let $P$ be the equation 
\[ x(\lambda y_{1} y_{2}.y_{1}(\lambda y_{3}.y_{2}(y_{1}(\lambda y_{4}.y_{3}))) = h(g(h(h a)))
\]
and let $t$ be 
$\lambda z.z (\lambda z_{1}.z (\lambda x_{1}.z_{1}(x_{1}(z_{1} a))) 
\lambda x_{2}. g x_{2}) \lambda z_{2}.h z_{2}$ which is a solution; $t$ as a 
tree is depicted in  Figure~\ref{ex14}. 
The single play for $\mathsf{G}(t,P)$ is 
presented in Figure~\ref{ex15} where the following abbreviations 
for left and right subterms are employed.
\[ \begin{array}{lcl}
v & = & \lambda y_{1} y_{2}.y_{1}(\lambda y_{3}.y_{2}(y_{1}(\lambda y_{4}.y_{3}))) \\
v_{1} & = & \lambda y_{3}.y_{2}(y_{1}(\lambda y_{4}.y_{3})) \\
v_{2} & = & y_{1}(\lambda y_{4}.y_{3}) \\
v_{3} & = & \lambda y_{4}.y_{3} \\
u & = & h(g(h(h a))) \\
u_{1} & = & g(h(h a)) \\
u_{2} & = & h(h a)
\end{array} \]
\begin{figure}
\[ \begin{array}{llll}
(1) \, \, q[(v), u]\,  \theta_{1} \, \xi_{1} & & &\\
(2) \, \, q[ v, u] \, \theta_{2} \xi_{2} &
\theta_{2} = \theta_{1} \eset{v \xi_{1}1/z} 
& \xi_{2} = \xi_{1} & A3\\
(3) \, \, q[(v_{1}), u] \, \theta_{3} \xi_{3}
& \theta_{3} = \theta_{2} &
\xi_{3} = \xi_{1}\eset{(3)\theta_{2} 2/y_{1},(17)\theta_{2} 2/y_{2}} & C4\\
(4) \, \, q[v, u] \, \theta_{4} \xi_{4} &
\theta_{4} = \theta_{2}\eset{v_{1}\xi_{3} 3/z_{1}} & \xi_{4} = \xi_{1} & A3\\
(5) \, \, q[(v_{1}), u] \, \theta_{5} \, \xi_{5} &
\theta_{5} = \theta_{4} & \xi_{5} = 
\xi_{1}\eset{(5)\theta_{4} 4/y_{1}, (13)\theta_{4} 4/y_{2}} & C4\\
(6) \, \, q[v_{1}, u] \, \theta_{6} \, \xi_{6} &
\theta_{6} = \theta_{4}\eset{v_{1}\xi_{5} 5/x_{1}} & \xi_{6} = \xi_{3}&A3 \\
(17) \, q[(v_{2}), u ] \, \theta_{7} \, \xi_{7} & 
\theta_{7} = \theta_{2} &
\xi_{7} = \xi_{3}\eset{(7)\theta_{6} 6/y_{3}} & C4\\
(18) \, q[-, u] \, \theta_{8} \, \xi_{8} & 
\theta_{8} = 
\theta_{2}\eset{v_{2} \xi_{7} 7/z_{2}} & \xi_{8} = \xi_{7} & A2\\
(19) \, q[ ( \ ), u_{1}] \, \theta_{9} \, \xi_{9} & \theta_{9} = \theta_{8}
& \xi_{9} = \xi_{7} & B1\\
(20) \, q[ v_{2}, u_{1}] \, \theta_{10} \xi_{10} &
\theta_{10} = \theta_{8} 
& \xi_{10} = \xi_{7} & A3\\
(3) \, \, q[(v_{3}), u_{1}] \, \theta_{11} \xi_{11}
& \theta_{11} = \theta_{2} &
\xi_{11} = \xi_{7} & C4\\
(4) \, \, q[v, u_{1}] \, \theta_{12} \xi_{12} &
\theta_{12} = \theta_{2}\eset{v_{3} \xi_{7} 11/z_{1}} 
& \xi_{12} = \xi_{1} & A3\\
(5) \,\, q[(v_{1}), u_{1}] \, \theta_{13} \, \xi_{13} &
\theta_{13} = \theta_{12} & \xi_{13} = \xi_{1}\eset{(5)\theta_{12} 12/y_{1},
(13)\theta_{12} 12/y_{2}} & C4\\
(6) \, \, q[v_{3}, u_{1}] \, \theta_{14} \, \xi_{14} &
\theta_{14} = \theta_{12}\eset{v_{1}\xi_{13} 13/x_{1}} & \xi_{14} = \xi_{7} &
A3 \\
(7) \, \,  q[( \ ), u_{1}] \, \theta_{15} \, \xi_{15} & 
\theta_{15} = \theta_{6} &
\xi_{15} = \xi_{14}\eset{(7)\theta_{14} 14/y_{4}} & C4\\
(8) \, \, q[v_{1}, u_{1}] \, \theta_{16} \, \xi_{16} & \theta_{16} = 
\theta_{6} & \xi_{16} = \xi_{5} & A3\\
(13) \, q[(v_{2}), u_{1} ] \, \theta_{17} \, \xi_{17} & \theta_{17} = \theta_{4} &
\xi_{17} = \xi_{5}\eset{(9)\theta_{6} 16/y_{3}} & C4\\
(14) \, q[-, u_{1} ] \, \theta_{18} \, \xi_{18} & \theta_{18} = \theta_{4}\eset{v_{2} \xi_{17} 17/x_{2}} &
\xi_{18} = \xi_{17} & A2\\
(15) \, q[( \, ), u_{2} ] \, \theta_{19} \, \xi_{19} & \theta_{19} = 
\theta_{18} &
\xi_{19} = \xi_{17} & B1\\
(16) \, q[v_{2}, u_{2} ] \, \theta_{20} \, \xi_{20} & \theta_{20} = 
\theta_{18} &
\xi_{20} = \xi_{17} & A3\\
(5) \, \, q[(v_{3}), u_{2}] \, \theta_{21} \, \xi_{21} &
\theta_{21} = \theta_{4} & \xi_{21} = \xi_{17} & C4\\
(6) \, \, q[v_{1}, u_{2}] \, \theta_{22} \, \xi_{22} &
\theta_{22} = \theta_{4}\eset{v_{3}\xi_{17} 21/x_{1}} & \xi_{22} = \xi_{3} &
A3\\
(17) \, q[(v_{2}), u_{2} ] \, \theta_{23} \, \xi_{23} & 
\theta_{23} = \theta_{2} &
\xi_{23} = \xi_{3}\eset{(7)\theta_{22} 22/y_{3}} & C4\\
(18) \, q[-, u_{2}] \, \theta_{24} \, \xi_{24} & 
\theta_{24} = 
\theta_{2}\eset{v_{2} \xi_{23} 23/z_{2}} & \xi_{24} = \xi_{23} & A2\\
(19) \, q[ ( \ ), h a] \, \theta_{25} \, \xi_{25} & \theta_{25} = 
\theta_{24} & \xi_{25} = \xi_{23} & B1\\
(20) \, q[ v_{2}, h a] \, \theta_{26} \xi_{26} &
\theta_{26} = \theta_{24} 
& \xi_{26} = \xi_{23} & A3\\
(3) \, \, q[(v_{3}), h a] \, \theta_{27} \xi_{27}
& \theta_{27} = \theta_{2} &
\xi_{27} = \xi_{23} & C4\\
(4) \, \, q[v, h a] \, \theta_{28} \xi_{28} &
\theta_{28} = \theta_{2}\eset{v_{3} \xi_{23} 27/z_{1}} 
& \xi_{28} = \xi_{1} & A3\\
(5) \,\, q[(v_{1}), h a] \, \theta_{29} \, \xi_{29} &
\theta_{29} = \theta_{28} & \xi_{29} = \xi_{1}\eset{(5)\theta_{28} 28/y_{1},
(13)\theta_{28} 28/y_{2}} & C4\\
(6) \, \, q[v_{3}, h a] \, \theta_{30} \, \xi_{30} &
\theta_{30} = \theta_{28}\eset{v_{1}\xi_{29} 29/x_{1}} & \xi_{30} = \xi_{23} 
& A3\\
(7) \, \,  q[( \ ), h a] \, \theta_{31} \, \xi_{31} & 
\theta_{31} = \theta_{22} &
\xi_{31} = \xi_{23}\eset{(7)\theta_{30} 30/y_{4}} & C4\\
(8) \, \, q[v_{3}, h a] \, \theta_{32} \, \xi_{32} & \theta_{32} = 
\theta_{22} & \xi_{32} = \xi_{17} & A3\\
(9) \, \,  q[( \ ), h a ] \, \theta_{33} \, \xi_{33} & 
\theta_{33} = \theta_{6} &
\xi_{33} = \xi_{32}\eset{(9)\theta_{22} 32/y_{4}} & C4 \\
(10) \, q[v_{1}, h a ] \, \theta_{34} \, \xi_{34} & \theta_{34} = 
\theta_{6} & \xi_{34} = \xi_{3} & A3\\
(17) \, q[(v_{2}), h a ] \, \theta_{35} \, \xi_{35} & 
\theta_{35} = \theta_{2} &
\xi_{35} = \xi_{3}\eset{(11)\theta_{6} 34/y_{3}} & C4\\
(18) \, q[-, h a] \, \theta_{36} \, \xi_{36} & 
\theta_{36} = 
\theta_{2}\eset{v_{2} \xi_{35} 35/z_{2}} & \xi_{36} = \xi_{35} & A2\\
(19) \, q[ ( \ ), a] \, \theta_{37} \, \xi_{37} & \theta_{37} = 
\theta_{36} & \xi_{37} = \xi_{35} & B1\\
(20) \, q[ v_{2}, a] \, \theta_{38} \xi_{38} &
\theta_{38} = \theta_{36} 
& \xi_{38} = \xi_{35} & A3\\
(3) \, \, q[(v_{3}), a] \, \theta_{39} \xi_{39}
& \theta_{39} = \theta_{2} &
\xi_{39} = \xi_{35} & C4\\
(4) \, \, q[v, a] \, \theta_{40} \xi_{40} &
\theta_{40} = \theta_{2}\eset{v_{3} \xi_{35} 39/z_{1}} 
& \xi_{40} = \xi_{1} & A3\\
(5) \,\, q[(v_{1}), a] \, \theta_{41} \, \xi_{41} &
\theta_{41} = \theta_{40} & \xi_{41} = \xi_{1}\eset{(5)\theta_{40} 40/y_{1},
(13)\theta_{40} 40/y_{2}} & C4\\
(6) \, \, q[v_{3}, a] \, \theta_{42} \, \xi_{42} &
\theta_{42} = \theta_{40}\eset{v_{1}\xi_{41} 41/x_{1}} & \xi_{42} = \xi_{35}
& A3 \\
(11)  \,  q[( \ ), a] \, \theta_{43} \, \xi_{43} & 
\theta_{43} = \theta_{6} &
\xi_{43} = \xi_{35}\eset{(7)\theta_{42}42/y_{4}}& C4 \\
(12) \, q[ \ \ex \ ] \, \theta_{44} \, \xi_{44} & \theta_{44} = 
\theta_{6} & \xi_{44} = \xi_{43} & A1\\
\end{array} \]
\caption{A $5$th-order play}
\label{ex15}
\end{figure}
Play starts at node $(1)$ with state $q[(v),u]$ and 
by A3 descends to $(2)$. The head variable of the body
of $v$ is $y_{1}$  which has argument $v_{1}$ and so play
moves to $(3)$ by C4 with state
$q[(v_{1}),u]$. Similarly, it descends to $(4)$ by A3 and then
to $(5)$ by C4 with state $q[(v_{1}),u]$ at which point it
moves to $(6)$ which is labelled $z_{1}$; $\theta_{6}(z_{1}) =
v_{1} \eta_{3} 3$ and the head variable of the body of $v_{1}$
is $y_{2}$ and its argument is $v_{2}$; so, play jumps to 
$(17)$ with state $q[(v_{2}),u]$. It then descends
to $(18)$, $(19)$ and $(20)$ with state $q[v_{2},u_{1}]$;
the head variable of $v_{2}$ is $y_{1}$ and it has argument
$v_{3}$; so play returns to $(3)$ with state $q[(v_{3}),u_{1}]$.
It then descends to $(4)$ and $(5)$ and moves to $(6)$;
the head variable of the body of $v_{3}$ is $y_{3}$
which is associated with node $(7)$, as $\xi_{14}(y_{3})
= (7) \theta_{6} 6$.
Play proceeds to (8), jumps to (13), reaches (16) and
then returns to (5) and so on.
Player $\fa$ eventually loses when play reaches (12) as the reader can
check.
\qed \end{exa}

\begin{defi} \label{def45}
If $P$ is a (dual) interpolation problem then  
$\fa$ \emph{loses the
game}  $\mathsf{G}(t,P)$ if and only if
\begin{enumerate}[(1)]
\item for every equation in $P$,
$\fa$ loses  every play 
whose initial state is given from it,    
\item for each  disequation in $P$, 
$\fa$ wins some play whose
initial state is given from it.
\end{enumerate}
\end{defi}

\ni 
The game \emph{characterises} dual interpolation.

\begin{thm} \label{thm1}
$\fa$ loses $\mathsf{G}(t,P)$ if, and only if,
$t \models P$.
\end{thm}
\proof
The simply typed $\lambda$-calculus is strongly normalising
and is Church-Rosser modulo $\alpha$-equivalence.
For every term $t$ there is an $m$ such that $t$ reduces to
normal form using at most $m$ $\beta$-reductions (whatever the reduction
strategy). 
Therefore, for any position $t_{i} q_{i} \theta_{i} \xi_{i}$ of a play of
$\G(t,P)$ we say that it $m$-holds ($m$-fails) if $q_{i} = q[\, \ex \,]$
($q_{i} = q[\, \fa \,]$) and when $q_{i}$ is not final, by cases on $t_{i}$
and $q_{i}$ (and look-up tables become delayed substitutions)
\begin{enumerate}[$\bullet$]
\item if $t_{i} = \lambda \ol{y}$, $q_{i} = q[(l_{1},\ldots,l_{k}),r]$
and $t'$ is $(t_{i}\theta_{i}) l_{1}\xi_{i} \ldots l_{k}\xi_{i}$
then $t' =_{\beta}r$ 
($t' \not=_{\beta}r$) and
$t'$ reduces to normal form with at most
$m$ $\beta$-reductions,
\item if $t_{i} = f$, $q_{i} = q[-,r]$ and $t'$ is
$t_{i} \theta_{i}$ then $t' =_{\beta} r$ ($t' \not=_{\beta} r$) and
$t'$ reduces to normal form  with at most m $\beta$-reductions,
\item if $t_{i} = z$, $q_{i} = q[l,r]$ and $t_{i} \downarrow_{j} t'_{j}$,
$1 \leq j \leq k$ for $k \geq 0$,
and $t'$ is $(l \xi_{i}) t'_{1} \theta_{i} \ldots t'_{k} \theta_{i}$ then
$t' =_{\beta} r$ ($t' \not=_{\beta} r$)
and $t'$  reduces to normal form 
with at most $m$ $\beta$-reductions.
\end{enumerate}

\ni
The proof is done by invoking as a measure  a pair of integers,
first the   
size of right term in $q_{i}$ and second
the  maximal number of $\beta$-reductions needed
for $t'$ to reduce to normal form; the pair is ordered with the lexicographic
ordering. 
The following properties are easy to show by case analysis.
\begin{enumerate}[(1)]
\item If $t_{i} q_{i} \theta_{i} \xi_{i}$ $m$-holds then 
$q_{i} = q[\, \ex \,]$ or for any next position 
$t_{i+1} q_{i+1} \theta_{i+1} \xi_{i+1}$
there is an $m'$ such that it $m'$ holds and either
$m' \leq m$ or the size of the right term
in $q_{i+1}$ is strictly smaller than in $q_{i}$.
\item If $t_{i} q_{i} \theta_{i} \xi_{i}$ $m$-fails 
then $q_{i} = q[\, \fa \,]$ or there is an $m'$ and a   next position 
$t_{i+1} q_{i+1} \theta_{i+1} \xi_{i+1}$ that 
$m'$-fails and either $m' \leq m$ or the size of the right term
in $q_{i+1}$ is strictly smaller than in $q_{i}$.
\end{enumerate}

\ni
For instance, assume $t_{i} q_{i} \theta_{i} \xi_{i}$ $m$-holds,
$t_{i} = \lambda y_{1}\ldots y_{k}$, $t_{i} \downarrow_{1}
t_{i+1} = y$, $t_{i+1} \downarrow_{j} t'_{j}$ for $1 \leq j \leq p$
and $q_{i} =
q[(l_{1},\ldots,l_{k}),r]$. Therefore,
$\theta_{i+1} = \theta_{i}\eset{l_{1}\xi_{i} i/y_{1}, \ldots,
l_{k} \xi_{i} i / y_{k}}$, $q_{i+1} = q[l,r]$ and  
$\xi_{i+1} = \xi'$
when   $\theta_{i+1}(y) = l \xi' n$.
So, $t_{i} = \lambda y_{1} \ldots y_{k}.y \, t'_{1}
\ldots t'_{p}$ and by assumption $(t_{i} \theta_{i})l_{1} \xi_{i}
\ldots  l_{k}\xi_{i} =_{\beta} r$. With $k$ $\beta$-reductions
we obtain  $(l \xi_{i+1}) t'_{1} \theta_{i+1} \ldots t'_{p}\theta_{i+1}$
and  position $t_{i+1}q_{i+1} \theta_{i+1} \xi_{i+1}$, therefore,
$(m-k)$-holds.
Next, assume $t_{i} q_{i} \theta_{i} \xi_{i}$ $m$-holds,
$t_{i} = f$,  $q_{i} =
q[-,f s_{1} \ldots s_{k} ]$
and $t_{i} \downarrow_{j} t'_{j}$ for $1 \leq j \leq k$.
By assumption, $(f t'_{1} \ldots t'_{k})  \theta_{i} =_{\beta}
f s_{1} \ldots s_{k}$. So, $t'_{j} \theta_{i} =_{\beta} s_{j}$.
Consider any choice of next position. If $s_{j} : \Nil$
then $q_{i+1} = q[( \ ),s_{j}]$, 
$t_{i+1} = t'_{j}$
and $\theta_{i+1} = \theta_{i}$. 
Therefore,
$t'_{j} \theta_{i+1} =_{\beta} s_{j}$ and so this next position
$m'$-holds for some $m'$ and $s_{j}$
is strictly smaller than $f s_{1} \ldots s_{k}$.
Alternatively, $s_{j} = \lambda x_{i_{1}} \ldots x_{i_{n}}. s$.
Therefore, $t'_{j} = \lambda z_{1} \ldots z_{n}. t'$
and $t'\theta_{i}\eset{c_{i_{1}} /z_{1}, \ldots, c_{i_{n}} / z_{n}}
=_{\beta} 
s\eset{c_{i_{1}} /x_{i_{1}},\ldots, c_{i_{n}}/x_{i_{n}}}$ 
$m'$-holds for some $m'$ provided
that the $c_{i_{j}}$'s are new (which is guaranteed as they
are forbidden constants). So the next position $m'$-holds and
the right term of $q_{i+1} = q[(c_{i_{1}},\ldots,c_{i_{n}}),
s\eset{c_{i_{1}} /x_{i_{1}},\ldots, c_{i_{n}}/x_{i_{n}}}]$
is strictly smaller than $f s_{1} \ldots s_{k}$. 
Assume $t_{i} q_{i} \theta_{i} \xi_{i}$ $m$-holds
and $t_{i} = y$,  $q_{i} = q[l,r]$, $l = \lambda z_{1} \ldots z_{k}.w$,
$w = z \, l_{1} \ldots l_{p}$, $t_{i} \downarrow_{j} t'_{j}$ for
$1 \leq j \leq k$ and  $\xi_{i+1}(z) = t'' \theta' n$; so, $t_{i+1} = t''$
and $\theta_{i+1} = \theta'$.
By assumption, $((\lambda z_{1} \ldots z_{k}.w)\xi_{i}) (t'_{1}\theta_{i})
\ldots (t'_{k}\theta_{i}) =_{\beta} r$. With $k$ $\beta$-reductions 
the left term in this equation becomes $(t_{i+1} \theta_{i+1}) l_{1}\xi_{i+1}
\ldots l_{p}\xi_{i+1}$ 
and so the next position $(m-k)$-holds. 
All other cases of (1) are close to one of these three, and the proof
of (2) is also very similar.
The only cases where the measure, size of right term in state
and the number of $\beta$-reductions to normal form, does not decrease
are applications of A3 when $t_{m}$ is labelled with a dummy
lambda and C4 when $l = w$ of Figure~\ref{game}. As a supplementary argument
we show that there cannot be an indefinite sequence
of such applications of A3 followed by C4 by examining the index $j$ 
that is called at these positions; namely, $\theta_{i+1}(y) = l \xi j$
in the case of A3 and $\xi_{i+1}(x) = t' \theta' j$ in the case of 
C4;  this index must be strictly decreasing in such repeated
sequence of applications of A3 followed by C4.

The result  now follows: if $t \models P$ then
for each initial position  that starts from an equation 
there is an $m$ such that it
$m$-holds and for each disequation there is an $m$ such that it
$m$-fails. Conversely, if $t \not \models P$ then
there is an initial position for some equation that $m$-fails
or for some  disequation there is an $m$ such that it $m$-holds.
\qed

The game is analogous to a model-checking game 
in the sense that deciding a possibly complex temporal
property of a transition
graph can be formulated as a game
whose arena is the  graph and where the moves are locally  small steps 
that traverse it; similarly, the complex property
whether $t$ solves $P$ is here formalised as a game whose arena
is $t$  involving 
locally small steps
and local moves. 
In both cases, play proceeds until 
one definitely knows an  outcome.

\section{Properties of game playing}
\label{props}

In the following we let $\pi, \sigma, \ldots$ range over plays
in a game $\G(t,P)$. 
The total number of  different plays is at most the sum of
the number of branches
in the right terms $u$ of $P$.
For instance, in the case of Example~\ref{examp44} whose right term
is $h(g(h(ha)))$ there is a single play.
 We now examine some properties
of plays and introduce relationships between
play positions that uses the play indices in the look-up tables.

\begin{rem}
\label{rem1} Theorem~\ref{thm1} allows one to restrict the set of constants
that can appear in a potential solution term $t$ for $P$.
Let $d : \Nil$ be a new constant that does not occur in any right term 
$u$ of a (dis)equation in $P$
(and which is also not a forbidden constant). 
Without loss of generality, we can  assume that 
any potential solution term $t$ to $P$ only  contains  the constant
$d$ and constants that occur in the right terms of  $P$:
a  similar  observation is made  in \cite{Pad2}. 
The justification appeals to  moves A1 and  A2 of Figure~\ref{game}.
Assume $t \models P$.
Control in a play associated with an interpolation  equation in $P$
can never be at  a node in $t$ labelled with a constant 
that does not occur
in a right term (as $\fa$ would win the play).
If  control
in a play associated with an interpolation disequation in $P$
is at a node $t'$ labelled with a constant that does not occur in
a right term then  replacing $t'$ in $t$ with a single node labelled
$d$ preserves $\fa$'s win.
\qed
\end{rem}

\begin{defi} \hfill
\begin{enumerate}[(1)]
\item The \emph{length} of $\pi$, 
$|\pi|$, is the number of positions $t_{i} q_{i} \theta_{i} \eta_{i}$
 in $\pi$.
\item The \emph{$i$th position} of $\pi$  is 
$\pi(i)$ where  $1 \leq i \leq |\pi|$.
\item The sequence of positions   
$\pi(i), \ldots, \pi(j)$, $i \leq j$,
is written  $\pi(i,j)$.
\end{enumerate}
\end{defi}

We write $t \in \pi(i)$, $q \in \pi(i)$,
$\theta \in \pi(i)$ and $\xi \in \pi(i)$ when $\pi(i) = t q \theta \xi$
and $t \not\in \pi(i)$ means that $\pi(i) = t' q \theta \xi$ and $t \not= t'$.
We shall describe  a sequence
of positions $\pi(i,j)$ as an \emph{interval}.

\begin{defi} The \emph{right term}
of state $q[(l_{1},\ldots,l_{k}),r]$, $q[-,r]$
or  $q[l,r]$ is the term  $r$. Each $l_{i}$
is a \emph{left term} of $q[(l_{1},\ldots,l_{k}),r]$
and $l$ is the \emph{left term} of $q[l,r]$.
\end{defi}

\begin{defi}\label{ri} \label{defi48}
The interval   $\pi(i,j)$ is \emph{ri}, \emph{right term invariant}, 
if
$q \in \pi(i)$ and $q' \in \pi(j)$ share the same right term $r$.
It is \emph{nri} if it is not ri and $q' \in \pi(j)$ is not a 
final state. 
\end{defi}

\ni
Clearly, if $\pi(i,j)$ is ri then \emph{every}  state
at every position  in this interval shares the
same right term. For instance,
when  $\pi$ is the play of Figure~\ref{ex2}, the interval
$\pi(5,17)$ is ri as all its states share the right term $a$; also,
each position in $\pi(6,17)$ is the result of  moves A3 or C4.
The outcome of the other moves
in Figure~\ref{game}, A1, A2, B1, C1, C2 and C3, depend on the right term
of the state. 

\begin{fact} \label{fact55}
If $\pi(i,j)$ is ri and $t_{j} \in \pi(j)$ is labelled
$\lambda \ol{y}$ for some $\ol{y}$ then each position
in $\pi(i+1,j)$ is the result of move A3 or C4 of
Figure~\ref{game}.
\end{fact}

\ni
Intervals  that are ri do not directly  contribute to the solution of $P$.

\begin{fact}\label{factri}
If  $t_{i}q_{i} \theta_{i}\xi_{i}, \ldots,
t_{n}q_{n}\theta_{n}\xi_{n}$ is a sequence of positions that
is ri, $t_{n}$ is labelled  $\lambda \ol{y}$  
and $q\{ r'/r \}$ is state q
with right term $r'$ instead of $r$, then 
$t_{i}q_{i}\{ r'/r \} \theta_{i}\xi_{i}, \ldots,
t_{n}q_{n}\{ r'/r \}\theta_{n}\xi_{n}$ is also a sequence of positions
that is  ri.
\end{fact}

Consider a position $t q \theta \xi$ of a play.
If there is a free occurrence of $y$
in the subtree $t$ then $\theta(y)$ is defined; similarly,
if there is a free occurrence of a variable $z$ in a left term
of $q$ then $\xi(z)$ is defined. In contrast, if there is a bound
occurrence of $y$, a node labelled $\lambda y_{1} \ldots y_{k}$
with $y = y_{j}$,  in the subtree $t$ then $\theta(y)$
is not defined and similarly, if there is a bound occurrence of
$z$ in a left term of $q$ then $\xi(z)$ is not defined.

\begin{prop} \label{imp}
Assume $\pi(i) = t q  \theta  \xi$.
\begin{enumerate}[\em(1)]
\item If $\lambda y_{1} \ldots y_{k}$ labels a node in the subtree rooted
at $t$ then for each $j: 1 \leq j \leq k$,  $\theta(y_{j})$ is undefined.
\item If $y$ occurs free in the subtree rooted at $t$ then $\theta(y)$
is defined.
\item If $q = q[l,r]$ or $q[(l_{1},\ldots,l_{k}),r]$
and $\lambda z_{1} \ldots z_{m}$ occurs in $l$ or in some
$l_{j}$, $1\leq j \leq k$, then for any $n: 1 \leq n \leq m$, 
$\xi(z_{n})$ is not defined.
\item If $q = q[l,r]$ or $q[(l_{1},\ldots,l_{k}),r]$
and $z$ occurs free in $l$ or in some $l_{j}$, $j: 1 \leq j \leq k$,
then $\xi(z)$ is defined.
\end{enumerate}
\end{prop}
\proof 
We prove this by induction on the position $i$ in a play $\pi$.
We also show by induction on $i$
that if 
$\pi(i+1)$ $=$ $t' \, q' \, \theta' \,
\xi'$ then the following additional four properties hold.
\begin{enumerate}[$\bullet$]
\item For any $y$, if $\theta'(y) = l \xi''j$ and 
$\lambda z_{1} \ldots z_{m}$ occurs in $l$, 
then for any $n: 1 \leq n \leq m$, 
$\xi''(z_{n})$ is not defined.
\item For any $y$, if $\theta'(y) = l \xi''j$ and $z$ occurs free in $l$,
then $\xi''(z)$ is defined.
\item If $\xi'(z) = t'' \theta''j$ and 
$\lambda y_{1} \ldots y_{k}$ labels a node in the subtree rooted
at $t''$ then for $i: 1 \leq i \leq k$,  $\theta''(y_{i})$ is undefined.
\item If $\xi'(z) = t'' \theta''j$ and 
$y$ occurs free in the subtree rooted at $t''$ then $\theta''(y)$
is defined.
\end{enumerate}
For the base case, 
consider an initial position $\pi(1)$ $=$ $t_{1}
\, q[(v^{i}_{1},\ldots,v^{i}_{n}), u_{i}] \, \theta_{1} \xi_{1}$
where $t_{1}$ is the root node of $t$ labelled
$\lambda z_{1} \ldots z_{n}$. There are no free
variables in $t$ or in the $v^{i}_{j}$'s.
Moreover, both $\theta_{1}$ and $\xi_{1}$ are empty;
therefore (1) to (4) hold. Now we need to show the
additional properties for $\pi(2) = t' q' \theta' \xi'$:
by definition $\theta' = \theta_{1}\eset{v^{i}_{1} \xi_{1} 1/z_{1},
\ldots, v^{i}_{n} \xi_{1} 1/z_{n}}$ and $\xi' = \xi_{1}$.
Therefore, these properties hold.
Consider next the general case for position $\pi(i)$.
If $\pi(i)$ is the result of moves A1-A3 of Figure~\ref{game}
applied to $\pi(i-1)$ then $(1)$ and $(2)$ follow because
they are true  at $\pi(i-1)$
and  $\theta \in \pi(i)$ is an update of $\theta' \in \pi(i-1)$ 
with respect to  the (potentially) free variables $y_{1},\ldots,y_{j}$ such
that $t_{i-1} \in \pi(i-1)$ is labelled
$\lambda y_{1} \ldots y_{j}$.
Parts $(3)$ and $(4)$ trivially hold for A1 and A2.
In the case of A3, if $t_{i} \in \pi(i)$ is labelled
$y \in \eset{y_{1},\ldots,
y_{j}}$ when $t_{i-1} \in \pi(i-1)$
is labelled $\lambda y_{1} \ldots y_{j}$ then $(3)$ and
$(4)$ follow from the induction hypothesis
that they hold at $\pi(i-1)$; otherwise
they  follow from the induction hypothesis for
the first two additional properties at $\pi(i-1)$.
If $\pi(i)$ is the result of move B1 of Figure~\ref{game}
to $\pi(i-1)$ then as $\theta$, $\xi$ are unchanged $(1)-(4)$
remain true.
Finally, we examine the case when $\pi(i)$ is the result of
moves C1-C4 to $\pi(i-1)$. Cases $(3)$ and $(4)$ hold because
$\xi \in \pi(i)$ is a simple updating of $\xi' \in \pi(i-1)$
where they hold. Cases $(1)$ and $(2)$ hold for C1-C3
because $\theta \in \pi(i)$ and  $\theta' \in \pi(i-1)$
are the same.
In the case of move C4, $(1)$ and $(2)$ either follow
from the induction hypothesis that they
 hold at $\pi(i-1)$ or from the induction hypothesis
for the final two additional properties
at $\pi(i-1)$. Using that $(1)-(4)$ are true at $\pi(i)$,
the argument is similar for showing that the four 
additional properties hold 
at $\pi(i+1) = t' q' \theta' \xi'$.
\qed

Now we examine some simple relationships between look-up tables.
We allow $\mu$, $\nu$ to range over both kinds of look-up
tables.

\begin{defi}
Two look-up tables $\mu$, $\mu'$ are equal, $\mu = \mu'$,
if, and only if, $\dom(\mu) = \dom(\mu')$ and
for all $x \in \dom(\mu)$,  $\mu(x) = \mu'(x)$; that is, 
if $\mu(x) = s \nu i$
and $\mu'(x) = s' \nu' i'$ then $s = s'$, 
$\nu = \nu'$ and $i=i'$.
\end{defi}

\ni
This is well defined because the definitions of  look-up tables
$\theta \in \Theta_{k}$ and $\xi \in \Xi_{k}$ 
in Definition~\ref{defi42} are well-founded with respect to
the embedding of look-up tables. 


\begin{defi}
A look-up table $\mu$ \emph{extends} $\mu'$ 
if  for all $x \in \dom(\mu')$,
$\mu(x) = \mu'(x)$.
\end{defi}

\begin{exa} Assume $\pi$ is the play in Figure~\ref{ex15}
that operates on the term tree in Figure~\ref{ex14}.
The look-up table $\theta \in \pi(18)$ consists of three entries
$\eset{v \xi_{1}1/z, v_{1} \xi_{3}3/z_{1},v_{2} \xi_{17} 17/x_{2}}$;
it therefore extends $\theta' \in \pi(11)$ which is just
the single entry $\eset{v \xi_{1} 1/z}$.
On the other hand, although $\theta'' \in \pi(36)$
extends $\theta'$ it does not extend $\theta$;
it consists of the entries $\eset{v \xi_{1} 1/z,v_{2} \xi_{35} 35/z_{2}}$.
In similar fashion, $\xi \in \pi(35)$ which is
$\eset{(3)\theta_{2} 2/y_{1}, (17)\theta_{2} 2/y_{2}, (11)\theta_{6} 34/y_{3}}$
extends $\xi_{6} \in \pi(6)$ which only contains the first two
of these entries.
\qed
\end{exa}

Let $\pi(i) = t q \theta \xi$ and let $\pi(j) = t' q' \theta' \xi'$
be a later position. If $t'$ is a subtree of $t$ and $\theta'$ extends 
$\theta$ then the free variable occurrences that are common to both
$t$ and $t'$ have the same interpretation; their meaning is preserved
at position $j$. Similarly, if $\eta'$ extends $\eta$ then
the  free variable occurrences that are
common to  the left terms of $q$ and $q'$ have the same interpretation.

Both  look-up tables of positions that are the result
of moves A1, A2, B1, C1-C3 of Figure~\ref{game} extend (or are equal to) 
those of the previous position. 
In the case of A3 the $\theta$ look-up table extends the one from the previous 
position but this is not true, in general,  for the $\xi$ table.
Dually, in the case of C4 the $\xi$ table extends the one 
of the previous position but this may not hold
for the  $\theta$ look-up table. 
We want to restore when \emph{both} look-up tables are extensions of
an earlier position.
For this we introduce a similar notion to that in game semantics
that later positions  
are justified by earlier positions \cite{ong}. 
We define
when  a  later position
is  a child
of an earlier position. It is at this point that we appeal to 
the third component of an entry
in a  look-up table.

\begin{defi} \label{def410} 
Position $\pi(j)=
t q \theta \xi$ is a \emph{child} of position $\pi(i)$ if
$i < j < |\pi|$ and the following by  cases of which move $\pi(j)$
is the result of

\begin{enumerate}[(1)]
\item A2, B1, C2 or C3:  then  $i = j-1$, 
\item A3:   then $t$ is labelled $y$ and $\theta(y) = l \xi' i$,
\item C4:  then $q[l,r] \in \pi(j-1)$,
the head variable in $l$ is  $x$ and $\xi(x)  = t \theta i$.
\end{enumerate}
\end{defi}

Assume  
$\pi(j) = t q \theta \xi$ is the child of $\pi(i) = t' q' \theta' \xi'$;
if $\pi(j)$ is the result of A3 
then $t'$ is the binder of  $t$
and if it is the result of C4 then
$t$  is a successor of $t'$.

\begin{fact} \label{fact411}
Assume $\pi(j)$ is a child of
$\pi(i)$. 
\begin{enumerate}[(1)]
\item If $\pi(j) = t' q[l',r'] \, \theta' \xi'$ is the result
of A3 and $\pi(i) = t \, q[(l_{1},\ldots,l_{m}),
r] \theta \, \xi$, then $t$ binds $t'$, $\xi' = \xi$
and for some $k: 1 \leq k \leq m$, $l' = l_{k}$.
\item If $\pi(j) = t'  q[(l_{1},\ldots,l_{n}),r'] \theta' 
\xi'$ is the result of C4, $q[l',r'] \in \pi(j-1)$
 and  $\pi(i) = t q[l,r] \theta \xi$
then $\theta' = \theta$, $l = \lambda z_{1} \ldots z_{m}.w$ and
for some $k:1 \leq k \leq m$, $t \downarrow_{k} t'$
and the head variable of $l'$ is $z_{k}$.
\end{enumerate}
\end{fact}

\begin{fact}\label{child} If $1 < j < |\pi|$ then there is a unique $i <j$
such that $\pi(j)$ is a child of $\pi(i)$.
\end{fact}

\ni
For this reason, we also say that $\pi(i)$ is \emph{the parent} of $\pi(j)$ 
instead of $\pi(j)$ is a child of $\pi(i)$. 

\begin{exa}
Let $\pi$ be the play in Figure~\ref{ex15} which is on the tree in
Figure~\ref{ex14}.
Every position that occurs at nodes $(2)$ and $(4)$
labelled with $z$ is a child of $\pi(1)$; examples include 
$\pi(2)$, $\pi(12)$ and $\pi(40)$.
Not every position that occurs at the nodes $(6)$ and $(10)$
labelled $z_{1}$ is a child of $\pi(3)$;
positions $\pi(6)$, $\pi(22)$ and $\pi(34)$ are
whereas $\pi(14)$, $\pi(30)$ and $\pi(35)$ are not.
Nodes $(5)$ and $(6)$ are the successors of
$(4)$; children of position $\pi(4)$ at $(4)$
are $\pi(5)$, $\pi(17)$ and $\pi(21)$
while $\pi(13)$, $\pi(29)$ and $\pi(41)$ are not.
Position $\pi(24)$ is a child of $\pi(23)$ through A2 and $\pi(9)$
is a child of $\pi(8)$ through B1. 
\qed
\end{exa}

\begin{fact}
If $\pi(j)$ is a child of
$\pi(i)$, $t \in \pi(i)$ and $t' \in \pi(j)$ then $t= t'$ or
there is a path of successors from $t$ to $t'$.
\end{fact}

The following is a critical consequence of the definition of a child position
that \emph{both} its look-up tables extend those of its parent.

\begin{prop}\label{child1}
Assume  $\pi(j)$ is a child of $\pi(i)$.
\begin{enumerate}[\em(1)]
\item $\theta'\in \pi(j)$
extends $\theta \in \pi(i)$,
\item $\xi'  \in \pi(j)$ extends
$\xi \in \pi$.
\end{enumerate}
\end{prop}
\proof
Assume that $\pi(j)$ is a child of $\pi(i)$
and $\theta', \xi' \in \pi(j)$ and $\theta, \xi \in \pi(i)$.
If $\pi(j)$ is a child of $\pi(i)$ as a result of
B1 or  C2 of Figure~\ref{game} then the result is true
because the look-up tables of $\pi(i)$ and $\pi(j)$
are the same. 
In the case of C3, $\theta' = \theta$ and 
$\xi'= \xi$ or $\xi' = \xi\eset{t'_{1}\theta i /z_{1},\ldots,t'_{k}
\theta i/z_{k}}$ for some $t'_{i}$, $z_{i}$, $1 \leq i \leq k$.
Therefore,  $\xi'$ extends $\xi$ as the $z_{i}$'s are not defined 
in $\xi$
by  Proposition~\ref{imp}. A similar argument applies when $\pi(j)$
is the result of A2; 
now  $\xi' = \xi$ and $\theta' = \theta$
or $\theta' = \theta\eset{l'_{1}\xi i /y_{1},\ldots,l'_{k}
\xi i/y_{k}}$ for some $l'_{i}$ and $y_{i}$, $1 \leq i \leq k$,
and $\theta'$ extends $\theta$ as the $y_{i}$'s are  not defined
in $\theta$  using Proposition~\ref{imp}.
Next we examine  the cases for A3 and C4. By definition of A3,
the look-up table $\xi' = \xi$ and by definition of
C4, $\theta' = \theta$. Therefore, we just need to prove
the result for the other look-up tables. In both cases the proof proceeds
by case analysis  of $j-i$. The initial  case is when $j = i+1$.
Therefore, for   A3 this means that
$t' \in \pi(j)$ is labelled with a variable $y$
which is bound by $\lambda \ol{y}$ which labels $t  \in \pi(j-1)$;
so $\theta'$ $=$ $\theta \eset{l_{1} \xi i/y_{1},\ldots,
l_{m} \xi i/y_{m}}$ for some $m$ and $l_{k}$, $1 \leq k \leq m$. 
Consequently,
$\theta'$ extends $\theta$ using Proposition~\ref{imp}
(since $\theta$ does not have an entry for $y$). Similarly, 
in the case of C4 when $j = i+1$, 
$q[\lambda \ol{z}.z_{p} l_{1} \ldots l_{k},r] \in
\pi(i)$
for some $k \geq 0$ and 
$\xi'$ $=$ $\xi\eset{t'_{1}\theta i/z_{1},
\ldots,t'_{m}\theta i/z_{m}}$ and $\xi$ does not have
entries for the $z_{i}$'s using  Proposition~\ref{imp}.
For the general  case for A3, we examine the branch  between 
$t \in \pi(i)$ and $t' \in \pi(j)$ where $t$ is labelled $\lambda \ol{y}$
and $t'$ is labelled $y$; 
which is the sequence of nodes with labels
$u_{1}, \lambda \ol{y}_{1}, \ldots, u_{n}, \lambda \ol{y}_{n}$
where each $u_{i}$ is a constant or variable and $n \geq 0$. Consider
position $\lambda \ol{y}_{n} \in \pi(j-1)$. Clearly,
$\theta'$ extends $\theta_{j-1} \in \pi(j-1)$ by a
similar
argument to the base case.
Position $\pi(j-1)$ is a child of $\pi(j_{n})$ for some $j_{n}$ with
$t_{n} \in \pi(j_{n})$ which is labelled $u_{n}$
by C4, B1 or  C2: in all cases $\theta_{j-1}$ extends $\theta_{j_{n}} \in
\pi(j_{n})$
by definition of these moves. The argument continues
for position $\pi(j_{n} -1)$. So, we reach a position
$\pi(i_{1})$ with $t_{1} \in \pi(i_{1})$ labelled $u_{1}$. By assumption,
$\theta' \in \pi(j)$ extends
$\theta_{i_{1}} \in \pi(i_{1})$: this means they have the same
entry for the $y_{i}$'s  in $\theta(j)$ when they are bound by 
$\lambda \ol{y}$
which is the label of 
$t \in \pi(i)$.  
Consider the relationship between $\pi(i_{1})$ and $\pi(i)$.
Clearly, it cannot be the case that $i > i_{1}$ because
this would contradict the entries in $\theta_{i_{1}}$ for the 
variables in $\ol{y}$. Moreover, by definition of the moves in A,
$\theta_{i+1} \in \pi(i+1)$ has the same entries for the $y$'s in 
$\ol{y}$ as $\pi(i_{1})$ and $t_{1} \in \pi(i+1)$.
A small argument shows that  $\pi(i_{1})$ must, therefore, be $\pi(i +1)$:
otherwise, $\pi(j_{1})$ would not be a child of
$\pi(i_{1})$.
Consequently, $\theta'$ extends $\theta$.
We now examine the general case for C4.  One possibility is that
$\pi(j)$ is the result of a sequence of C3 moves followed by C4:
clearly, in this case  $\xi'$ extends $\xi$. 
Otherwise,  $q[\lambda \ol{z}.w,r]
\in \pi(i)$ and $\xi_{i+1} \in \pi(i+1)$
is  $\xi\eset{t'_{1}\theta i/z_{1},
\ldots,t'_{m}\theta i /z_{m}}$ for some $m$;  so $\xi_{i+1}$ extends $\xi$.
Therefore, $t_{j-1} \in \pi(j-1)$ is 
labelled with some $y'$ and $q \in \pi(j-1)$ has the form $q[l,r']$
where $l$ is $z_{k} l'_{1} \ldots l'_{m'}$ or 
$\lambda \ol{x}.z_{k} l'_{1} \ldots l'_{m'}$ for some $k$:
we know that  $\xi'$ extends $\xi_{j-1} \in \pi(j-1)$.
There may be a sequence of positions $\pi(j_{n},j-1)$
where each $\pi(j')$, $j_{n} < j' \leq j-1$
is the result of C3 (and, therefore, $\xi'$ extends $\xi_{j_{n}}
\in \pi(j_{n})$).
Otherwise $j_{n} = j-1$. 
Position $\pi(j_{n})$ is a child of
a unique position $t_{j_{n-1}} \in \pi(j_{n-1})$ labelled
$\lambda \ol{y}'$
by A3 and so $\xi_{j_{n}} \in \pi(j_{n})$ extends $\xi_{j_{n-1}} \in
\pi(j_{n-1})$.
This argument is now repeated: consider position $t'' \in \pi(j_{n-1}-1)$
labelled with $y''$.
Again, there may be a sequence $\pi(j_{n-2},j_{n-1}-1)$
where the moves are the result of C3 and either $j_{n-2} = i$
or $\pi(j_{n-2})$ is a child of $\pi(j_{n-3})$. 
Eventually, for some $k$, $j_{n-k} = i$
as each $\xi_{j_{n}} \in \pi(j_{n})$ has the entries $t'_{l} \theta i$
for the  $z_{l}$'s.
\qed

\begin{defi} \label{def416}
The binary relation  \emph{is a descendent of} on positions 
is the reflexive and transitive closure of ``is a child of''.
\end{defi}

\section{Tiles and their plays}
\label{tiles}

In this section, we connect  the static structure, regions 
of a potential solution term of a (dual) interpolation problem, 
with the dynamics of game playing. 
To this end, 
partial subtrees of a term tree  are introduced.

\begin{defi}
Assume
$B  = (B_{1},\ldots,B_{k},\Nil)$.
\begin{enumerate}[(1)]
\item Node $t'$ labelled $\lambda$ is an \emph{atomic leaf}
of type $\Nil$. 
\item Node $t'$ labelled $\lambda x_{1} \ldots x_{k}$ is an \emph{atomic leaf}  of type
$B$ when  each $x_{j} : B_{j}$
\item  If $t'$ is labelled $u:\Nil$ then  
$t'$ is a \emph{simple tile}. 
\item If node $t'$ is labelled $u :B$ and each node $t_{j}$ is an  atomic leaf
of type $B_{j}$ and   $t' \downarrow_{j} t_{j}$, $1 \leq j \leq k$, 
then $t'(t_{1}, \ldots, t_{k})$
is a \emph{simple tile}.
\end{enumerate}
\end{defi}

\ni
A potential solution tree  without its initial lambda is a tree  of
simple tiles.
For instance, the region 
$(2)((3), (11))$ of Figure~\ref{ex1}
is  a  simple tile labelled $z(\lambda x,\lambda)$ with atomic leaves
$(3)$ and $(11)$ labelled $\lambda x$ and $\lambda$;
the region $(12)((13), (19))$ is labelled  $z(\lambda y,\lambda)$;
the region $(4)((5))$  is a simple  tile labelled $f(\lambda)$.
Single  nodes such as (16) and (20) are also simple 
tiles but without atomic leaves.
The definition precludes node $(2)$ by itself or  $(2)((3))$
as simple tiles.

In the following, as it makes the presentation cleaner,
we describe tiles directly through their labelling.
For example, $z(\lambda x,\lambda)$ identifies 
$(2)((3), (11))$ of Figure~\ref{ex1}. When there is
ambiguity, such as with $z_{1}(\lambda)$
of Figure~\ref{ex14}, we disambiguate by
describing the root node;  $z_{1}(\lambda)$
at $(6)$ versus $z_{1}(\lambda)$ at $(10)$. 

Tiles can be composed to form composite tiles.
A  (possibly composite)
tile is a partial tree which can be extended at any of its atomic leaves.
If $\tau(\lambda \ol{x})$ is a tile with leaf 
$\lambda \ol{x}$ and $\tau'$ is a simple tile,
then $\tau(\lambda \ol{x}.\tau')$ is the composite tile that is the result
of placing $\tau'$
directly beneath $\lambda \ol{x}$ in $\tau$.
For instance, we can compose the tile $z(\lambda y,\lambda)$
of Figure~\ref{ex1} with the tile $z(\lambda s,\lambda)$ beneath
$\lambda y$ and
produce the composite tile $z(\lambda y. z(\lambda s,\lambda), \lambda)$
which has three atomic leaves: in Figure~\ref{ex1} this tile is the region
$(12)((13) (14)((15),(17)),(19))$.  
We write $\tau(\lambda \ol{x}_{1},\ldots,\lambda \ol{x}_{k})$
if $\tau$ is a (composite)  
tile with atomic leaves $\lambda \ol{x}_{1}, \ldots,
\lambda \ol{x}_{k}$.

A tile $\tau(\lambda \ol{x}_{1},\ldots,\lambda \ol{x}_{k})$ is a multi-holed
context. It is also a subregion of a term and  
we  assume that the usual definitions of free and bound variable occurrences
apply: for instance, 
the free variables in
$z(\lambda y. z(\lambda s. s,\lambda), \lambda)$
are the two occurrences
of $z$. Later we shall   manipulate tiles and, therefore, we have
given them an 
independent existence.

\begin{defi}\label{basic}
A (composite) tile $\tau$ is said to be \emph{basic} if
$\tau$ contains
\begin{enumerate}[(1)]
\item exactly one occurrence of a free variable 
and no occurrences of constants, or  
\item exactly one occurrence of a constant 
and no occurrences of free variables.
\end{enumerate}
\end{defi}

\ni
By definition, simple tiles are basic. 
The single occurrence of the free variable or constant in a basic tile 
is its ``head'' element.
Particular contiguous regions of a term tree  are basic tiles.
In Figure~\ref{ex1} the region 
$z(\lambda s.s,\lambda)$ is a basic tile rooted at (14) with
the single atomic leaf $\lambda$.
However, if we also included  node (18) then it would 
be a composite tile $z(\lambda s.s,\lambda. y)$
without atomic leaves, but not a basic tile.

\begin{defi} \label{defi53}
Assume $\tau$ and $\gamma$ are basic tiles in a tree.
\begin{enumerate}[(1)]
\item $\gamma$ is \emph{$j$-below}
$\tau(\lambda \ol{x}_{1},\ldots,\lambda \ol{x}_{k})$
if there is a path of successors from $\lambda \ol{x}_{j}$ to
$\gamma$. 
\item $\gamma$ is \emph{below} $\tau$ 
(or $\tau$ is \emph{above} $\gamma$)   if $\gamma$ is $j$-below
$\tau$ for some $j$. 
\end{enumerate}
\end{defi}

\ni
In Figure~\ref{ex14}, the tile $x_{1}(\lambda)$, rooted at (8),
is $1$-below $z(\lambda z_{1},\lambda z_{2})$ 
and $z_{2}$ at (20) is $2$-below the same tile.

\begin{defi} \label{defdepend}
Assume $\tau$ and $\gamma$ are basic tiles in a tree.
\begin{enumerate}[(1)]
\item $\gamma$ is an \emph{immediate $j$-dependent}
of tile $\tau$ if $\gamma$ is $j$-below $\tau$
and $\gamma$ contains a free variable that is bound in
$\tau$.
\item $\gamma$ is  a \emph{$j$-dependent} of $\tau$ 
if it  is an immediate
$j$-dependent of $\tau$ 
or there is a $\tau'$ that is an immediate
$j$-dependent  of $\tau$ 
and $\gamma$ is a $k$-dependent  of $\tau'$
for some $k$. 
\item $\gamma$ is a (\emph{immediate})
\emph{dependent} 
of $\tau$ if $\gamma$ is a 
(immediate) $j$-dependent of $\tau$ for some $j$. 
\end{enumerate}
\end{defi}

\ni
The tile $z_{1}(\lambda)$ rooted at (6) of  Figure~\ref{ex14}
is an immediate $1$-dependent
of  $z(\lambda z_{1},\lambda z_{2})$ rooted at (2)
and $z_{2}$ rooted at (20) is a $2$-dependent.
Given a tile $\tau$, its set of dependents are all the tiles below
it whose free variables are either bound within $\tau$ or are bound within
a dependent of $\tau$.
For instance, the dependents of $z(\lambda x_{1},\lambda x_{2})$
of Figure~\ref{ex14} are $x_{1}(\lambda)$ rooted at (8)
and $x_{2}$ at (16).

\begin{defi} \label{family}
Tiles $\tau$ and $\gamma$ belong to the \emph{same family} in a
tree  if
one is a dependent of the other, or there is a $\tau'$ such that
both are dependents of $\tau'$. The \emph{family}
of tiles associated with $\tau$ in a tree  consists of $\tau$ and
each  tile $\gamma$ that belongs  to the same family
as $\tau$. 
\end{defi}

In Figure~\ref{ex14} the family associated with $x_{2}$ at (16)
is the set of tiles containing $z(\lambda x_{1},\lambda x_{2})$
at (4), $x_{1}(\lambda)$ at (8) and $x_{2}$.

\begin{rem}
Assume basic tiles $\tau$ and $\gamma$.  If  position $\pi(j)$
is the result of move A3 of Figure~\ref{game} and 
is a child of $\pi(i)$  
and $t' \in \pi(j)$ is in $\gamma$ and $t \in \pi(i)$ is
in $\tau$  then $\gamma$ is a dependent of
$\tau$. 
If $\pi(j)$ is the result of  move C4  
and $t' \in \pi(j)$ is in $\gamma$ and $t \in \pi(j-1)$
is in $\tau$, then these two tiles
belong to the same family: this important property is proved
in Section~\ref{unfold}.
In the play of  Figure~\ref{ex15}, play at $x_{1}$ at position 16
jumps to $\lambda x_{2}$ of $z(\lambda x_{1},\lambda x_{2})$
and play at $x_{2}$ at position 20 jumps to $\lambda x_{1}$.
\qed
\end{rem}

\begin{defi} \label{defi54}
Assume $\tau$ and $\gamma$ are basic tiles
that each contain
an occurrence of a  free variable.
\begin{enumerate}[(1)]
\item $\tau$ and $\gamma$ 
are \emph{equivalent}, written $\tau \equiv \gamma$, 
if they are $\alpha$-equivalent;
that is, they are the same basic tiles with the same number and type of atomic leaves and
with the same single free variable occurrence $y$. 
\item If $\tau \equiv \gamma$
and $t_{1}$ is a node of $\tau$ and  
$t'_{1}$ is the corresponding equivalent node of $\gamma$
then we write $t_{1} \equiv^{\tau}_{\gamma} t'_{1}$.
\item If $\tau \equiv \gamma$ and $\gamma$ is below $\tau$ 
then  $\gamma$ is said to be an  \emph{embedded} tile.
\end{enumerate}
\end{defi}

\ni
The reader can verify that ``tile equivalence'' is indeed
an equivalence relation.
In Figure~\ref{ex1}, each pair of tiles $z(\lambda x,\lambda)$,
$z(\lambda u,\lambda)$, $z(\lambda y,\lambda)$ and
$z(\lambda s,\lambda)$ is equivalent: nodes such as
those with labels $\lambda u$ and $\lambda s$ correspond.  Each of these tiles
except the first is 
also an embedded tile.
The tile $z_{1}(\lambda)$ rooted at (6) is equivalent
to $z_{1}(\lambda)$ rooted at (10) of Figure~\ref{ex14}:
however, neither is equivalent to $x_{1}(\lambda)$
rooted at (8).

\begin{defi}\label{static}
Assume 
$\tau = \tau(\lambda \ol{x}_{1},\ldots, \lambda \ol{x}_{k})$ is a
basic tile.
\begin{enumerate}[(1)]
\item $\tau$ is a \emph{top} tile if it contains a  free variable occurrence
that is 
is bound by the initial lambda of the term tree.
\item $\tau$ is \emph{$j$-end}  if 
$\tau$ has no immediate $j$-dependents.
It is an \emph{end} tile if it is $j$-end for \emph{all}  
$j : 1 \leq j \leq k$.
\item $\tau$ is a \emph{constant} tile 
if it contains an occurrence of a  constant
or it is  a dependent  of a constant tile.
\end{enumerate} 
\end{defi}

\ni
The tile $z(\lambda x, \lambda)$ rooted at (2)
in Figure~\ref{ex1}
is a top tile as  $z$ is bound by the initial lambda  at node (1).
It  is also $2$-end because no tile
beneath node (11) contains a free variable occurrence
that is bound within it:  however, it is not $1$-end because
of the tile occurrence
$x$ at node (8).
Tile
$z(\lambda u,\lambda)$ rooted at (6)
is a top tile and also an end tile. 
Tiles $f(\lambda x z_{1} z_{2},\lambda)$ rooted at
(6), $x(\lambda)$ at (8) and $z_{2}$ at (12)
in Figure~\ref{picture2} are all constant tiles.   

The previous definitions provide a classification
of basic tiles within a tree  that only appeals to
the static structure of the tree. Tiles  can also be categorised
in terms of 
dynamic properties of game playing.

\begin{defi}
The  interval  $\pi(i,j)$ is a \emph{play} on the simple tile
$u(\lambda \ol{x}_{1},\ldots,\lambda \ol{x}_{k})$
if $u \in \pi(i)$, $\lambda \ol{x}_{m} \in \pi(j)$ for some $m : 1 \leq m \leq k$ and $\pi(j)$ is a child  of $\pi(i)$. 
It is an \emph{m-play} if  $\lambda \ol{x}_{m} \in \pi(j)$.
\end{defi}

A simple tile has the form $y(\lambda \ol{x}_{1},\ldots, \lambda \ol{x}_{k})$
or $f(\lambda \ol{x}_{1},\ldots, \lambda \ol{x}_{k})$.
A play $\pi(i,j)$ on such a tile starts at the head of the  tile and
ends at one of its atomic leaves; importantly, $\pi(j)$ must be a child
of $\pi(i)$. 
A  play on  a simple constant tile 
$u(\lambda \ol{x}_{1},\ldots, \lambda \ol{x}_{k})$
is a consecutive pair of positions $\pi(i,i+1)$ 
with $u \in \pi(i)$ and $\lambda \ol{x}_{m} \in \pi(i+1)$ for some $m$
(by moves B1 or  C2 of Figure~\ref{game}).  

\begin{fact} \label{fact57}
If $\pi(i,j)$ is a play on a simple constant tile then $j = i+1$.
\end{fact}

\ni
For instance, if  $\pi$ be the play in Example~\ref{examp42}
whose term tree is depicted in Figure~\ref{picture2}
then $\pi(6,7)$ is a play on the simple
constant tile $f(\lambda x z_{1}z_{2},\lambda)$
and $\pi(8,9)$ is a play on the simple constant tile $x(\lambda)$.

A play $\pi(i,j)$ on a simple non-constant tile
$y(\lambda \ol{x}_{1},\ldots, \lambda \ol{x}_{k})$ 
can have  arbitrary length.
It  starts at the node labelled with  $y$ and 
finishes at a node labelling an atomic  leaf $\lambda \ol{x}_{m}$.
In between, control  can be almost anywhere in the tree
(including the node labelling $y$).
However, because $\pi(j)$ is a child of $\pi(i)$
the look-up tables of $\pi(j)$ extend
those of  $\pi(i)$ as shown in 
Proposition~\ref{child1}.

\begin{exa}
Let $\pi$ be the play in Figure~\ref{ex15} on the tree
in Figure~\ref{ex14}.
The simple tile $\tau = z(\lambda z_{1},\lambda z_{2})$
is rooted at (2). There are various plays on $\tau$:
$1$-plays include $\pi(2,3)$ and $\pi(2,39)$; 
$\pi(2,23)$ and $\pi(2,35)$ are $2$-plays.
If $\gamma = z(\lambda x_{1},\lambda x_{2})$
rooted at (4) then $\pi(4,5)$ and $\pi(12,13)$ are $1$-plays
on $\gamma$:
however, the interval  $\pi(4,13)$ is not 
a play on $\gamma$ because $\pi(13)$ is a child of $\pi(12)$.
If $\tau$ is the end tile $x_{1}(\lambda)$ rooted at (8)
then there is just one play $\pi(16,33)$ on it;
the interval  $\pi(32,33)$ is not  a play on
$\tau$. \qed
\end{exa}

The definition of  play on a simple tile can be  extended to arbitrary
composite tiles by composing consecutive
plays on the simple tiles from its root to one of its  atomic leaves.

\begin{defi} \label{def612}
The  interval 
$\pi(i,j)$ is a play on the composite tile
$\tau = \tau(\lambda \ol{x}_{1},\ldots,\lambda \ol{x}_{k})$
if there is a path of successor nodes with labels
$u_{1}, \lambda \ol{y}_{1},
\ldots u_{n}, \lambda \ol{y}_{n}$
from the root of $\tau$ to an atomic leaf 
$\lambda \ol{x}_{m} =\lambda \ol{y}_{n}$ such that 
$\pi(i,j) = \pi(i_{1},j_{1}), \ldots, \pi(i_{n},j_{n})$
and $\pi(i_{l},j_{l})$ is a play on the simple tile
$u_{l}(\ldots  \lambda \ol{y}_{l} \ldots)$
with $\lambda \ol{y}_{l} \in \pi(j_{l})$
for $1 \leq l \leq n$.
It is an \emph{$m$-play} if $\lambda \ol{x}_{m} \in \pi(j)$.
\end{defi}

If  $t \in \pi(j)$, $j >1$,  and $t$ is labelled with a lambda then  
there is a unique partition of
$\pi(2,j)$ into plays on the simple
tiles that occur on the branch from the node 
directly beneath the initial $\lambda$ of the tree  to
$t$. The partition also preserves
children.

\begin{prop} \label{prop511}
Assume $t \in \pi(j)$ and $t$ is labelled with $\lambda \ol{y}_{n}$
and 
$\lambda \ol{y}_{0}, u_{1}, \lambda \ol{y}_{1},
\ldots u_{n}, \lambda \ol{y}_{n}$ is the labels of the sequence 
of nodes from the root of the tree to $t$.
Then there is a unique partition of $\pi(1,j)
= \pi(j_{0}),\pi(i_{1},j_{1}), \ldots, \pi(i_{n},j_{n})$
such that for $1 \leq m \leq n$
\begin{enumerate}[\em(1)]
\item $j_{0} =1$ and $\pi(i_{m},j_{m})$ is a play
on the simple tile $u_{m}(\ldots\lambda \ol{y}_{m} \ldots)$
with $\lambda \ol{y}_{m} \in \pi(j_{m})$, 
\item if $u_{m}$ is a variable bound by $\lambda \ol{y}_{k}$
then $\pi(i_{m})$ is a child of $\pi(j_{k})$.
\end{enumerate} 
\end{prop} 
\proof
Assume $t \in \pi(j)$ and $t$ is labelled with $\lambda \ol{y}_{n}$
and $\lambda \ol{y}_{0}, u_{1}, \lambda \ol{y}_{1},
\ldots, u_{n}, \lambda \ol{y}_{n}$ are the labels of
the sequence of nodes from
the root of the term tree to $\lambda \ol{y}_{n}$.
Assume $j_{n} = j$. 
Let $\pi(i_{n},j_{n})$ be the play
on the simple tile $t_{n} = u_{n}(\ldots \lambda \ol{y}_{n} \ldots)$
such that $\pi(j_{n})$ is a child of $\pi(i_{n})$: 
move $\pi(j_{n})$
is the result of B1, C2 or C4 of Figure~\ref{game}
(and, therefore, $\pi(i_{n})$ is uniquely defined 
from Definition~\ref{def410}).
Let $j_{n-1} = i_{n}-1$,  the argument is now repeated for
$\pi(i_{n-1},j_{n-1})$ as a play on $u_{n-1}(\ldots
\lambda \ol{y}_{n-1}\ldots)$ and so on for subsequent tiles in the branch
from the node labelled $\lambda \ol{y}_{n}$ to the root. Clearly,
this will define a partition 
of $\pi(2,j$) into $\pi(i_{1},j_{1}), \ldots, \pi(i_{n},j_{n})$
with $i_{1} = 2$: hence, we can add $\pi(j_{0})$ with $j_{0} = 1$
at the beginning.
Next assume that $u_{m}$ is a variable bound 
by $\lambda \ol{y}_{k}$.
It is straightforward to show that the look-up tables
in $\pi(i_{m})$ extend those in $\pi(j_{k}+1)$,
which, therefore, implies that $\pi(i_{m})$ is a child of $\pi(j_{k})$;
this follows from repeated application of Proposition~\ref{child1}.
\qed 

\begin{defi} \label{def614}
If $t \in \pi(j)$ and $t$ is labelled with $\lambda \ol{y}_{n}$
and 
$\lambda \ol{y}_{0}, u_{1}, \lambda \ol{y}_{1},
\ldots u_{n}, \lambda \ol{y}_{n}$ is the labels of the sequence 
of nodes from the root of the tree to $t$
then the unique partition $\pi(1,j)
= \pi(j_{0}),\pi(i_{1},j_{1}), \ldots, \pi(i_{n},j_{n})$
such that for $1 \leq m \leq n$
where each $\pi(i_{m},j_{m})$ is 
the play on $u_{m}(\ldots \lambda \ol{y}_{m}
\ldots)$ 
of  Proposition~\ref{prop511} is called the \emph{b-partition for
position $\pi(j)$}.
\end{defi}

\begin{exa} \label{example512}
Let $\pi$ be the play in Figure~\ref{ex15}
on the tree in Figure~\ref{ex14}. Consider the two
positions $\lambda  \in \pi(15)$ and $\lambda  \in \pi(31)$
where $\lambda$ is at node (7).
Below we illustrate  the b-partitions for $\pi(15)$
and $\pi(31)$. The branch from the root to $\lambda$
is presented horizontally.
\[ \begin{array}{llll}
\lambda z \ \ \ & z \ \ \ \lambda z_{1} \ \ \ \ \  & z \ \ \ \lambda x_{1} 
\ \ \ \ \ & z_{1} \ \ \ \lambda  \\
\pi(1) & \pi(2,3) & \pi(4,5) & \pi(6,15)  \\
\pi(1) & \pi(2,3) & \pi(4,21) & \pi(22,31) 
\end{array} \]
The b-partitions capture children along their branches:
$\pi(2)$ and $\pi(4)$ are children of $\pi(1)$; 
$\pi(3)$ is a child of $\pi(2)$; $\pi(5)$ and $\pi(21)$
are children  of $\pi(4)$;
$\pi(6)$ and $\pi(22)$ are children
of $\pi(3)$.  Plays within these b-partitions
may jump around the term tree: for instance, the play $\pi(22,31)$
on $z_{1}(\lambda)$ rooted at (10) includes a sequence
of moves that descends   the branch below (17). 
\qed \end{exa}

We now exhibit some useful  uniformities of plays on tiles.
Given a play $\pi$
and tile $\tau$ we examine  three aspects.
First, there can be  multiple plays
$\pi(i,n_{1})$, $\ldots$, $\pi(i,n_{m})$ on $\tau$
from the same initial position. Second, there can be
multiple plays from different  starting positions $\pi(i_{1},n_{1})$,
$\ldots$, $\pi(i_{m},n_{m})$ on $\tau$. A third consideration
is the relationship between plays on $\tau'$ and $\tau$ when 
these tiles are equivalent.

We start with a  pertinent
property of a simple non-constant tile.

\begin{prop}\label{plays}
If 
$\pi(i,m)$ and $\pi(i,n)$ are plays  on 
the simple tile
$y(\lambda \ol{x}_{1},\ldots,\lambda \ol{x}_{k})$ and $i < m < n$
then there is a position $\pi(m')$, $m' < n$, that
is a child  of $\pi(m)$.
\end{prop}
\proof
Assume  $\pi(i) = t \, q[\lambda z_{1} \ldots z_{k}.w,r] \, \theta \, 
\xi$, $t$ is labelled $y$ and  
$\pi(i,m)$, $\pi(i,n)$, $n > m$, are  plays
on the simple tile $y(\lambda \ol{x}_{1},\ldots,\lambda \ol{x}_{k})$;
so,  $\lambda \ol{x}_{j} \in \pi(m)$ for some $j$.
The look-up table $\xi_{i+1} \in \pi(i+1)$ is
$\xi\eset{
\lambda \ol{x}_{1} \theta i /
z_{1},\ldots,\lambda \ol{x}_{k}\theta i /z_{k}}$ 
and
the $\xi$  look-up tables in  $\pi(m-1)$ and $\pi(n-1)$ both extend 
$\xi_{i+1}$ 
because $\pi(m)$, $\pi(n)$ are children of $\pi(i)$. 
Clearly, no look-up table  $\xi_{l} \in \pi(l)$, $l <i+1$,
has entries $\xi(z_{i'}) = \lambda \ol{x}_{i'} \theta i$
because for any $z$, if $\xi_{l}(z) = t' \theta' j$ then $j < i$.
Consider the first position $\pi(m_{1})$ after
$\pi(m)$ that is at a node labelled with a  variable $y_{1}$.
Clearly, this node is below the node labelled  $\lambda \ol{x}_{j}$
in the tree. If $y_{1}$ is bound by $\lambda \ol{x}_{j}$
then $\pi(m_{1})$ is a child of $\pi(m)$ and the
result is proved. 
Otherwise $\pi(m_{1})$ is a child of an earlier position
$\pi(l)$. Either $l <i$ or $i <l$.
Assume the former; so, via move A3 
the look-up table $\xi_{l} \in \pi(l)$ 
cannot have  an entry for any $z$ of the form $t' \theta'' i$
because $l < i$.
Play may then 
jump anywhere in the term tree  using  move C4. If there is not
a play $\pi(m_{1},n_{1})$ on the simple tile whose head node is
labelled $y_{1}$ then for all later positions
$\pi(m_{2})$, $m_{2} > m_{1}$, $\xi_{m_{2}} \in \pi(m_{2})$
cannot include an entry for 
any $z$ of the form $t' \theta'' i$;
this is a contradiction because of the position $\pi(n)$.
Therefore, play must reach a position 
$\pi(n_{1})$ that is a child of $\pi(m_{1})$; so
$t'' \in \pi(n_{1})$ is a successor of the node labelled with $y_{1}$
and is therefore below (the node labelled)
$\lambda \ol{x}_{j}$.
Next assume that $i < l$ so $y_{1}$ is bound by a $\lambda \ol{y}$
that is below (the node labelled) $\lambda \ol{x}_{j}$. 
But then $y_{1}$ is bound to a leaf of a constant tile
that occurs between the node labelled $\lambda \ol{x}_{j}$ and 
the node labelled $y_{1}$
and so move C2 must apply and so play descends to
a successor with  position $\pi(n_{1})$ that is a child of $\pi(m_{1})$
where $n_{1} = m_{1}+1$.
This argument is now repeated for the next
position after $\pi(n_{1})$ that is at a node labelled with a   variable $y_{2}
\in \pi(m_{2})$: this node must be below  the node labelled
$\lambda \ol{x}_{j}$.
The argument proceeds as above, except there is the new case
that $\pi(m_{2})$ is a child of $\pi(n_{1})$. However,
by move A3, $\xi_{m_{2}} \in \pi(m_{2})$ cannot 
include an entry for some $z$ of the form $t' \theta' i$.
Therefore, play must eventually reach a child of $\pi(m)$.
\qed

\ni
By iteration, this property extends  to basic tiles.

\begin{cor}\label{plays1}
If $\pi(i,m)$ and $\pi(i,n)$ are plays  on 
the basic tile $\tau(\lambda \ol{x}_{1},\ldots,\lambda \ol{x}_{k})$,
$i < m < n$, and $\pi(i,m) = \pi(i_{1},j_{1}),
\ldots, \pi(i_{n},j_{n})$ where $u_{1}, \lambda \ol{y}_{1},
\ldots,u_{n},\lambda \ol{y}_{n}$ is the path of (labels of)
nodes from the root of $\tau$ to its atomic leaf
$\lambda \ol{x}_{j} = \lambda \ol{y}_{n} \in \pi(m)$
and $\pi(i_{l},j_{l})$ is a play on the simple tile
$u_{l}(\ldots \lambda \ol{y}_{l} \ldots)$ with $\lambda \ol{y}_{l} \in
\pi(j_{l})$ for $1 \leq l \leq n$ then there is a position
$m'$ such that $m < m' < n$ and $\pi(m')$ is a child of $\pi(j_{l'})$
for some $l'$.
\end{cor}

One  consequence of Corollary~\ref{plays1} is that there cannot
be more than one play on a basic end tile that starts from the same 
position. More precisely, we have the following.

\begin{prop}\label{end1}
If $\pi(i,m)$ is a $j$-play on the basic tile
$\tau$ which is $j$-end,  $\pi(i,n)$ is a play on
$\tau$  and $n \geq m$ then $n = m$.
\end{prop}
\proof 
Assume $\tau = \tau(\lambda \ol{x}_{1},\ldots,\ol{x}_{k})$
is $j$-end. So  $\tau$ has no dependents
below the node labelled $\lambda \ol{x}_{j}$;
therefore, the binder of every free variable
that occurs in the subtree 
below this node occurs above the tile $\tau$
in the tree.
Assume $\pi(i,m)$ is a $j$-play  on $\tau$ 
and $\pi(i,n)$, $n > m$, is also a play
on $\tau$. 
The play $\pi(i,m)$ can be partitioned into
plays on the simple tiles of $\tau$ from its root to its
atomic leaf $\lambda \ol{x}_{j} \in \pi(m)$.
Therefore, by Corollary~\ref{plays1} there is
a position $\pi(m')$, $m < m' < n$, that is a child
of one of the final positions of a simple tile
in the partition of $\pi(i,m)$.
But then, by the definition of child, see Fact~\ref{fact411},  
there must be a free variable
below the node labelled $\lambda \ol{x}_{j}$ that is bound
within $\tau$ which contradicts the assumption that
$\tau$ is $j$-end. \qed

Special restricted plays of basic tiles are defined.

\begin{defi} \label{defi516}
Assume $\pi(i,j)$ is a play on  basic tile $\tau$.
\begin{enumerate}[(1)]
\item It is a \emph{shortest
play} on $\tau$  if no proper prefix $\pi(i,k)$, $k <j$, is also a  
play on $\tau$.
\item It is a \emph{shortest
$m$-play} on $\tau$ if $\pi(i,j)$ is a $m$-play
on $\tau$ and no proper prefix $\pi(i,k)$, $k < j$, is also
an  $m$-play on $\tau$.
\item It is also an \emph{internal} play ($m$-play)
on $\tau$ provided that    for any $n:i \leq n \leq j$,
if  $t \in \pi(n)$ then 
$t$ is a node of $\tau$.
\item It is also an \emph{ri} play ($m$-play)
on $\tau$ if $\pi(i,j)$ is ri (see, Definition~\ref{ri}).
\end{enumerate}
\end{defi}

\ni
Let $\pi$ be the play in Figure~\ref{ex2}.
The interval  $\pi(2,3)$ is a shortest play 
on tile $z(\lambda x,\lambda)$ rooted at (2) of Figure~\ref{ex1}: 
this play is also ri,
internal and 
a shortest $1$-play. Although $\pi(2,9)$ is a shortest $2$-play,
it is neither a shortest play nor an internal play.

Next,  we define   a uniformity condition concerning 
multiple plays on a  basic tile.

\begin{defi}
Assume $\tau$
is a basic tile whose root node is $t'$.
Tile $\tau$ is \emph{$j$-directed with respect to the interval
$\pi(i,|\pi|)$} if 
\begin{enumerate}[(1)]
\item $t' \not\in \pi(m)$ for all $m: i \leq m \leq |\pi|$, or
\item $\pi(m)$ is the first position $m \geq i$
with $t' \in \pi(m)$
and 
there is a shortest $j$-play $\pi(m,n)$ on  $\tau$,
$\pi(m,n)$ is ri and
$\tau$ is $j$-directed with respect to  $\pi(n+1, |\pi|)$.
\end{enumerate}
\end{defi}

\begin{defi}\label{directed}
The basic tile  $\tau$  is
\emph{$j$-directed} with respect to the game $\mathrm{G}(t,P)$
if $\tau$ is $j$-directed with respect to  $\pi(1,|\pi|)$ for every 
play $\pi  \in \G(t,P)$.
\end{defi}

If $\tau$ is $j$-directed with respect to a game
then every play $\pi$ 
contains  a (unique) sequence of ri intervals
$\pi(i_{k},m_{k})$, for some $k \geq 0$, which are shortest $j$-plays on 
$\tau$ as follows (assuming $t'$ is the root of $\tau$ and $\lambda \ol{x}_{j}$
is its $j$th atomic leaf).
\[ \begin{array}{ccccccccccccccc}
& & t' & & \lambda \ol{x}_{j} & & t' & & \lambda \ol{x}_{j} & & \\
\pi(1) & \dots & \pi(i_{1}) & \ldots & \pi(m_{1}) & \ldots 
& \pi(i_{n}) & \ldots &\pi(m_{n})& 
\ldots &
\pi(|\pi|) 
\end{array} \]
By definition,  $t'$ cannot occur in a position
that is outside of these intervals $\pi(i_{k},m_{k})$.
A tile $\tau$  can be $j$-directed with respect to a game
for   multiple  $j$. If $\pi$ is the play
of Figure~\ref{ex15} on the tree in Figure~\ref{ex14} then
tile $z(\lambda x_{1},\lambda x_{2})$ is $1$-directed
with respect to  $\pi(1,|\pi|)$ because of the following sequence:
\[ \begin{array}{lllllllllll}
& & z \ \ \ \ \ \lambda x_{1} & & z \ \ \ \ \ \ \ \lambda x_{1} & & 
z \ \ \ \ \ \ \ \lambda x_{1} 
& & 
z \ \ \ \ \ \ \ \lambda x_{1} & & \\
\pi(1) & \ldots & \pi(4) \, \pi(5)&  \ldots & \pi(12)\,  \pi(13) & \ldots &
\pi(28) \, \pi(29) & \ldots&  \pi(40) \, \pi(41)&  \ldots&  \pi(44)
\end{array} \]
however, it is not $2$-directed with respect to  $\pi(1,|\pi|)$ 
because $\pi(4,17)$
is nri. 

Recall that a basic top tile, see Definition~\ref{static},
has a single variable occurrence that
is bound by the initial lambda of the term tree.
Equivalent
basic top tiles are subject to the following local uniformity 
properties.

\begin{prop} \label{prop515}
Assume $\tau$, $\gamma$ are basic top tiles, 
$\tau \equiv \gamma$ and $\pi(i,i+m)$ is a shortest play on $\tau$.
\begin{enumerate}[\em(1)]
\item $\pi(i,i+m)$ is an internal play on $\tau$.
\item If $\pi(i,i+m)$ is ri and $\pi(i',m')$ is a shortest
play on $\gamma$ then $\pi(i',m')$ is ri, $m' = i'+m$
and for all $j:1 \leq j \leq m$, $t_{1} \in \pi(i+j)$ implies
$t'_{1} \in \pi(i'+j)$ where $t_{1} \equiv^{\tau}_{\gamma} t'_{1}$.
\item If $\pi(i,i+m)$ is ri and a $j$-play
on $\tau$ 
then $\tau$ and $\gamma$ are $j$-directed with respect to  $\pi(1,|\pi|)$.
\item If $\pi(i,i+m)$ is nri and $t \in \pi(i')$ where $t$ is the root node
of $\gamma$ 
then there is an interval
$\pi(i',i'+m')$ which  is internal to $\gamma$
that is either a shortest  play
on $\gamma$ and nri  or $|\pi| = i'+m'$.
\end{enumerate}
\end{prop}
\proof 
Let $t$ be the root node of top tile $\tau$
and assume  $t \in \pi(i)$. The state $q \in \pi(i)$ has the form
$q[v,r]$ where
$v$ is a closed left term, a $v^{l}_{j}$ from a (dis)equation of $P$.
Therefore, a shortest play $\pi(i, i+m)$ on $\tau$ is internal
(as a jump outside $\tau$ requires there to be a free variable
in $v$ via move C4 of Figure~\ref{game}) which shows (1).
If $\tau \equiv \gamma$, $t'$ is the root of $\gamma$ 
and $t' \in \pi(i')$ then the  
state $q' \in \pi(i')$ has the same left term
$q[v,r']$ as  $q[v,r] \in \pi(i)$.
If the play $\pi(i,i+m)$ is ri then there is a corresponding ri
play $\pi(i',i'+m)$ on $\gamma$ consisting of the same sequence
of corresponding positions in $\tau$  and states 
(except for their right terms $r$ and $r'$) which shows (2).
Furthermore, tile $\tau$ is $j$-directed with
respect to  $\pi$ when $\pi(i+m)$ is a $j$-play on $\tau$
because each time play is at $t$
there is the same repeated shortest play on it (and similarly for $\gamma$). 
For (4), if the shortest play $\pi(i,i+m)$ on $\tau$ is nri
and $t' \in \pi(i')$ then either there is a shortest play
on $\gamma$ that is nri (but may involve different $\fa$-choices)
or play remains within $\gamma$ and the final state is reached.
\qed

Tiles  are  equivalent to themselves, $\tau \equiv \tau$;
so Proposition~\ref{prop515} also applies to repeated shortest plays
on a single top tile $\tau$. 
Consider the tree in Figure~\ref{ex14} 
and the play in Figure~\ref{ex15}. 
The equivalent
tiles $z(\lambda x_{1},\lambda x_{2}) \equiv z(\lambda z_{1},\lambda z_{2})$
are top tiles.
Shortest $\pi$ plays on the first of these tiles 
$\pi(4,5)$, $\pi(12,13)$,
$\pi(28,29)$ and $\pi(40,41)$  are $1$-plays that are ri and
each corresponds to the single shortest
play $\pi(2,3)$ on $z(\lambda z_{1},\lambda z_{2})$.

Top basic tiles are distinguished because their only
free variable is bound by the initial lambda. 
We now   show that there are also
play uniformities on other equivalent tiles, 
in the case  of \emph{embedded} tiles
(as in Definition~\ref{defi54}). 
If $\tau \equiv \gamma$
and $\gamma$ is below $\tau$ then shortest
plays on the embedded tile $\gamma$ are constrained by earlier plays
on $\tau$.

\begin{prop} \label{prop516} 
Assume  $\tau$, $\gamma$  are basic tiles, $\tau \equiv \gamma$
and $\gamma$ is below $\tau$.
\begin{enumerate}[\em(1)]
\item If $\pi(j,j +m)$ is a shortest play
on $\gamma$ that is ri and a $k$-play
then there is a shortest play $\pi(i,i+m)$ on
$\tau$, $i < j$, that is ri and a $k$-play.
\item If $\pi(j,j+m)$ is a shortest
play on $\gamma$ that is nri then 
there is a shortest play $\pi(i,i+n)$ on $\tau$, $i < j$,
that is nri.
\end{enumerate}
\end{prop}
\proof 
Assume $\tau$ and $\gamma$ are basic equivalent tiles,
$\tau \equiv \gamma$, $\gamma$ is below $\tau$
and $\pi(j,j+m)$ is a shortest play on $\gamma$ that is ri
and a $k$-play. 
Consider the unique b-partition of position
$\pi(j+m)$ from the root of the term tree to the atomic leaf of $\gamma$,
see Definition~\ref{def614}.
Because $\gamma$ is below $\tau$, this b-partition
contains a play $\pi(i,i +m')$ on $\tau$.
Because $\tau$ and $\gamma$ are equivalent and $\pi(j,j+m)$
is ri, these tiles contain the same single free variable $y$
at their root; therefore, 
positions $\pi(i)$ and $\pi(j)$ share a common
parent. Therefore,   $\xi \in \pi(i)$ is the same look-up table
as $\xi \in \pi(j)$, $\theta' \in \pi(j)$ extends $\theta \in \pi(i)$
and if $q[l,r'] \in \pi(j)$ then $q[l,r] \in \pi(i)$ for some $r$.
Now, it is an easy argument that $\pi(i,i+m)$  is ri and 
a $k$-play on $\tau$.  If instead $\pi(j,j+m)$
is nri then some prefix  of $\pi(i,i+m')$ 
is a shortest nri play 
on $\tau$ that may involve different $\fa$ choices than in $\pi(j,j+m)$.
\qed

\ni
In the case of the tree in Figure~\ref{ex14} 
and the play in Figure~\ref{ex15}, 
tile $z_{1}(\lambda)$ rooted at (10) is 
an embedded (end) tile: its single play $\pi(34,43)$
which is nri corresponds to the earlier play
$\pi(22,31)$ on $z_{1}(\lambda)$ rooted at (6).

There is an even stronger property of embedded end tiles:
an embedded end tile is either $j$-directed with respect to
a game for some $j$, 
has at least one nri play or play finishes within it.

\begin{prop}\label{end} 
Assume $\tau$, $\gamma$ are basic end tiles
in a term tree $t$, $\gamma$ is $j$-below $\tau$,
$\tau \equiv \gamma$ and $ \pi \in \G(t,P)$.
Then either
\begin{enumerate}[\em(1)]
\item $\gamma$ is $j$-directed with respect to $\pi(1,|\pi|)$, or
\item there exist $m, n, m', n'$
such that $m < n < m' \leq n'$, $\pi(m,n)$ is a  nri 
$j$-play on $\tau$, $t' \in \pi(m')$ is the root node of $\gamma$
and $\pi(m',n')$
is a  nri play on $\gamma$ or $n' = |\pi|$ and no prefix of $\pi(m',n')$
is a play on $\gamma$. 
\end{enumerate}
\end{prop}
\proof
Assume $\tau$, $\gamma$ are basic end tiles
in term tree $t$, $\tau \equiv \gamma$ and 
and $\gamma$ is $j$-below
$\tau$. We follow the argument in the proof of Proposition~\ref{prop516}.
Both $\tau$ and $\gamma$ have the same head variable $y$
that is bound to the same $\lambda \ol{y}$ above $\tau$  in $t$. 
Let $\pi \in \G(t,P)$. Consider the first position $t' \in \pi(m)$
where $t'$ is the root of $\gamma$
and the b-partition for $\pi(m-1)$.
This partition must contain a $j$-play on $\tau$,
$\pi(i,i+k)$ (because $\gamma$ is $j$-below $\tau$);
positions $\pi(i)$ and $\pi(m)$
share a common parent. If $\pi(m,m')$ is play on
$\gamma$ then the b-partition for $\pi(m')$ is a simple extension
of that for $\pi(m-1)$; the play on $\tau$ is still $\pi(i,i+k)$.
By Proposition~\ref{end1}, $\pi(i,i+k)$ is a $j$-play and
so it is  a corresponding play to $\pi(m,m')$ by
Proposition~\ref{prop516}; so, $\pi(m,m')$ is ri and is a $j$-play
on $\gamma$. 
This argument is repeated for the next position, $m'' > m'$, such
that $t' \in \pi(m'')$  or until
the $j$-play on $\tau$ in the b-partition is nri;
in which case, there is either an nri play on $\gamma$
or a final state is reached.
\qed

\section{Two transformations}
\label{third}

Given an interpolation  problem $P$, 
the aim   is to prove decidability
of higher-order matching by establishing a small model property:
if $t \models P$ then there is a \emph{small} 
term $t'\models P$ (whose size is determined from $P$).
As we saw in Section~\ref{props},
the number of plays in the game $\mathsf{G}(t,P)$
is bounded by the number of branches
in the right terms $u$ of the (dis)equations of $P$.
However,  there is no upper bound  on the length of a play 
that is independent of the size 
and order of $t$.
Nevertheless, a long play must contain significant ri intervals
that do not directly contribute to solving $P$:
the number of times the right term in a state changes within
all plays 
is bounded by $\delta$, the right size of $P$,
Definition~\ref{delta}. 
Game-theoretically, what will underpin the conversion
of an arbitrary solution term $t$ to a small solution term $t'$
is manipulation of tiles and their ri plays (using omission, repetition
and permutation). 
The proof for the general case is quite intricate.
So, we proceed in stages. In this section we provide two
proofs  of  decidability
of  $3$rd-order matching, one  using a tree model property
of game playing 
and the other using properties of  embedded end tiles.
Both proofs appeal to omission  of tiles and their ri intervals.

As a first step, we introduce two transformations on term trees
(for terms of any order).
A transformation ${\bf T}$
converts   a term tree $t$ into a 
term tree $t'$, written $t \, ${\bf T}$\, t'$.

\begin{defi}
Assume  $t'$ is a subtree of $t$ whose root 
is labelled with a variable $y$ or a constant $f:B \not= \Nil$.
The game $\G(t,P)$ \emph{avoids} $t'$ if for every play 
$\pi \in \G(t,P)$, $t' \not\in \pi(i)$ for 
all  $i : 1 \leq i \leq |\pi|$.
\end{defi}

\begin{defi}
Assume  $t'$ and $t''$ are trees whose roots are labelled 
with a constant or a variable.
Let $t[t''/t']$ be  the result of replacing the subtree $t'$ of $t$ 
with the tree $t''$.
\end{defi}

The first transformation is straightforward: if no play enters
a subtree of $t$ then it can be replaced with the 
single node labelled with constant $d: \Nil$
(introduced in Remark~\ref{rem1}).

\begin{enumerate}[{\bf T1}]
\item If $\G(t,P)$ avoids $t'$ then 
transform $t$ to $t[d/t']$
\end{enumerate}


\ni
The second transformation removes inner tiles from $t$:
if a basic tile is both $j$-end,
Definition~\ref{static}, and $j$-directed with respect
to the game $\mathsf{G}(t,P)$, Definition~\ref{directed},  then it
is redundant.

\begin{enumerate}[{\bf T1}]
\item[{\bf T2}] If $\tau(\lambda \ol{x}_{1},\ldots,\lambda \ol{x}_{k})$ is 
a  $j$-end basic tile and $j$-directed with respect to $\G(t,P)$,
$t'$ is the subtree of $t$
rooted at $\tau$ and  $t_{j}$ is the subtree directly beneath $\lambda \ol{x}_{j}$ of $\tau$
then transform $t$ to $t[t_{j}/t']$. 
\end{enumerate}

\ni
An application of {\bf T2} not only removes the  tile 
$\tau(\lambda \ol{x}_{1},\ldots,\lambda \ol{x}_{k})$ from $t$
but also all subtrees that occur directly
beneath any atomic leaf $\lambda \ol{x}_{i}$, $i \not= j$,
of $\tau$. Because $\tau$ has no $j$-dependents, all free variables
that occur in the subtree $t_{j}$
directly below $\lambda \ol{x}_{j}$
are bound above $\tau$ in $t$; therefore,  the result
of applying 
{\bf T2} is still a closed
term (in normal form with the right type).
If $\tau$ is $j$-directed with respect to 
a game then each play involves
a (unique) sequence of ri intervals
which are shortest $j$-plays on $\tau$, as described in the previous section.
Game-theoretically underpinning the correctness of {\bf T2} 
is  \emph{omission} of these
inessential intervals from each play.

\begin{prop}
If ${\bf i} \in \eset{ {\bf 1}, {\bf 2}}$,
$t \,${\bf Ti}$\, t'$ and $t \models P$ 
then $t' \models P$.
\end{prop}
\proof 
This is clear in the case of  ${\bf T1}$. 
Assume $t \models P$, $\tau(\lambda \ol{x}_{1},\ldots,\lambda \ol{x}_{k})$ is 
$j$-end  and $j$-directed
with respect to the game $\G(t,P)$, $t''$ is the subtree 
at the root of $\tau$ and $t_{j}$ is the subtree directly 
beneath $\lambda \ol{x}_{j}$. Let 
$t' = t[t_{j}/t'']$.  We show that
$t' \models P$. 
We  convert each $\pi \in \G(t,P)$ into
a  play $\pi'  \in \G(t',P)$ that
ends with the same final state. 
Because $\tau$ is $j$-directed with respect to each play,
$\pi$ can be  split uniquely into the following regions
\[ \begin{array}{ccccccccccccccc}
& & t'' & & \lambda \ol{x}_{j} & & t'' & & \lambda \ol{x}_{j} & & \\
\pi(1) & \dots & \pi(i_{1}) & \ldots & \pi(m_{1}) & \ldots 
& \pi(i_{n}) & \ldots &\pi(m_{n})& 
\ldots &
\pi(|\pi|) 
\end{array} \] 
where each $\pi(i_{l},m_{l})$ is a (shortest) $j$-play 
on $\tau$ and is ri; by
definition of $j$-directed, node $t''$ cannot occur outside of these 
intervals.  Therefore, $\pi(m_{k})$ is a child of $\pi(i_{k})$
for each $k$
Consequently, the play $\pi' \in \G(t',P)$ is just
the outer intervals
$\pi(1,i_{1}-1),\pi(m_{1}+1,i_{2}-1), \ldots,\pi(m_{n}+1,|\pi|)$
(modulo the changes to the entries in the look-up tables)
because for each $l$, $\pi(m_{l})$ is a child of  $\pi(i_{l})$.
We show, that if
$s$ is a node in $\tau$ or is  $m$-below an atomic  leaf 
$\lambda \ol{x}_{m}$, $m \not= j$,
of $\tau$ then $s$ cannot occur in any outer interval of $\pi$.
If $s$ were to appear in such a position  then move C4 must have
applied: there is then a variable $y$ and a position in an outer
region  $y \in \pi(n)$ and $\theta \in \pi(n)$ and
$\theta(y) = l \xi i$ and there is a free variable $z$ in $l$ such
that $\xi(z) = s \theta' i'$. However, this is impossible.
Consider $\theta_{1} \in \pi(i_{1})$: clearly, there is no
free variable in the subtree rooted at $t''$ with this property.
When play reaches $\pi(m_{1})$ because $\tau$ is a $j$-end tile
and because the look-up
tables in $\pi(m_{1})$ extend those in  $\pi(i_{1})$ there cannot
be a free variable in the subtree $t_{j}$ with this property
either. This argument is now repeated for subsequent positions
$\pi(i_{k})$ and $\pi(m_{k})$.
\qed

\ni
The two transformations
are also \emph{reversible}: we can add gratuitous subtrees and
intersperse redundant $j$-end tiles with arbitrary
subtrees beneath their other atomic leaves in any solution term.

\begin{exa}\label{examp61}
Consider the $4$th-order problem
$x v = f a, x w = f(f a)$ where $v = \lambda y_{1}y_{2}.y_{1} y_{2}$
and $w = \lambda y_{3}y_{4}.y_{3}(y_{3} y_{4})$ from  Example~\ref{examp1}.
A solution term is in  Figure~\ref{ex1}. There are two
plays,  $\pi$ in Figure~\ref{ex2} for  the first 
equation and $\pi'$ in Figure~\ref{nex2} for the second.
\begin{figure}
\[\!\begin{array}{lllr}
(1) \, q[(w), f(fa)]\,  \theta_{1} \, \xi_{1} & & &\\
(2) \, q[ w, f(fa)] \, \theta_{2} \xi_{2} &\!
\theta_{2} = \theta_{1} \eset{w \xi_{1}1\!/\!z} 
&\! \xi_{2} = \xi_{1} &\! A3\\
(3) \, q[(y_{3}y_{4}), f(f a)] \, \theta_{3} \xi_{3}
&\! \theta_{3} = \theta_{2} &\!
\xi_{3} = \xi_{1}\eset{(3)\theta_{2}2\!/\!y_{3},(11)\theta_{2}2\!/\!y_{4}} &\! C4\\
(4) \, q[-, f(f a)] \, \theta_{4} \xi_{4} &\!
\theta_{4} = \theta_{2}\eset{(y_{3} y_{4})  \xi_{3} 3\!/\!x} &\! \xi_{4} = \xi_{3} &\! A2\\
(5) \, q[( \ ), f a] \, \theta_{5} \, \xi_{5} &\!
\theta_{5} = \theta_{4} &\! \xi_{5} = \xi_{3}
&\! B1\\
(6) \, q[w, fa] \, \theta_{6} \, \xi_{6} &\!
\theta_{6} = \theta_{4} &\! \xi_{6} = \xi_{1} &\! A3\\
(7) \, q[(y_{3}y_{4}), f a ] \, \theta_{7} \, \xi_{7} &\! \theta_{7} = \theta_{4} &\!
\xi_{7} = \xi_{1}\eset{(7)\theta_{4}6\!/\!y_{3},(9)\theta_{4}6\!/\!y_{4}} &\! C4\\
(8) \, q[y_{3} y_{4}, fa] \, \theta_{8} \, \xi_{8} &\! \theta_{8} = 
\theta_{4}\eset{y_{2} \xi_{7} 7\!/\!u} &\! \xi_{8} = \xi_{3} &\! A3\\
(3) \, q[(y_{4}), f a] \, \theta_{9} \xi_{9}
&\! \theta_{9} = \theta_{2} &\!
\xi_{9} = \xi_{3} &\! C4\\
(4) \, q[-, f a] \, \theta_{10} \xi_{10} &\!
\theta_{10} = \theta_{2}\eset{y_{4}\xi_{3}9\!/\!x} &\! \xi_{10} = \xi_{3} &\! A2\\
(5) \, q[( \ ), a] \, \theta_{11} \, \xi_{11} &\!
\theta_{11} = \theta_{10} &\! \xi_{11} = \xi_{3}
&\! B1\\
(6) \, q[w, a] \, \theta_{12} \, \xi_{12} &\!
\theta_{12} = \theta_{10} &\! \xi_{12} = \xi_{1} &\! A3\\
(7) \, q[(y_{3} y_{4}), a ] \, \theta_{13} \, \xi_{13} &\! \theta_{13} 
= \theta_{10} &\!
\xi_{13} = \xi_{1}\eset{(7)\theta_{10}12 \!/\!y_{3},(9)\theta_{10} 12\!/\!y_{4}} &\! C4\\
(8) \, q[y_{4}, a] \, \theta_{14} \, \xi_{14} &\! \theta_{14} = 
\theta_{10}\eset{y_{4} \xi_{13} 13\!/\!u} &\! \xi_{14} = \xi_{3} &\! A3\\
(11) \, q[ ( \ ), a] \, \theta_{15} \, \xi_{15} &\! \theta_{15} = \theta_{2}
&\! \xi_{15} = \xi_{3} &\! C4 \\
(12) \, q[ w, a] \, \theta_{16} \xi_{16} &\!
\theta_{16} = \theta_{2} 
&\! \xi_{16} = \xi_{1} &\! A3\\
(13) \, q[(y_{3} y_{4}), a] \, \theta_{17} \xi_{17}
&\! \theta_{17} = \theta_{2} &\!
\xi_{17} = \xi_{1}\eset{(13)\theta_{2} 16\!/\!y_{3},(19)\theta_{2}16\!/\!y_{4}} &\! C4\\
(14) \, q[w, a] \, \theta_{18} \xi_{18} &\!
\theta_{18} = \theta_{2}\eset{(y_{3}y_{4})\xi_{17}17\!/\!y} &\! \xi_{18} = \xi_{1} &\! A3\\
(15) \, q[(y_{3}y_{4}), a] \, \theta_{19} \, \xi_{19} &\!
\theta_{19} = \theta_{18} &\! \xi_{19} = \xi_{1}\eset{(15)\theta_{18} 18\!/\!y_{3},
(17)\theta_{18} 18\!/\!y_{4}} &\! C4\\
(16) \, q[y_{3}y_{4}, a] \, \theta_{20} \, \xi_{20} &\!
\theta_{20} = \theta_{18}\eset{(y_{3}y_{4})\xi_{19} 19\!/\!s} &\! \xi_{20} = \xi_{19} &\! A3\\
(15) \, q[(y_{3}y_{4}), a] \, \theta_{21} \, \xi_{21} &\!
\theta_{21} = \theta_{20} &\! \xi_{21} = \xi_{19}\eset{(15)\theta_{20} 20\!/\!y_{3},
(17)\theta_{20} 20\!/\!y_{4}} &\! C4\\

(16) \, q[y_{4}, a] \, \theta_{22} \, \xi_{22} &\!
\theta_{22} = \theta_{20}\eset{y_{4}\xi_{21} 21\!/\!s} &\! \xi_{22} = \xi_{21} &\! A3\\
(17) \, q[( \ ), a ] \, \theta_{23} \, \xi_{23} &\! \theta_{23} = \theta_{22} &\!
\xi_{23} = \xi_{21} &\! C4\\

(18) \, q[y_{3}y_{4}, a] \, \theta_{24} \, \xi_{24} &\! \theta_{24} = 
\theta_{22} &\! \xi_{24} = \xi_{17} &\! A3\\

(13) \, q[(y_{4}), a] \, \theta_{25} \xi_{25}
&\! \theta_{25} = \theta_{2} &\!
\xi_{25} = \xi_{17}  &\! C4\\
(14) \, q[w, a] \, \theta_{26} \xi_{26} &\!
\theta_{26} = \theta_{2}\eset{(y_{3} y_{4})\xi_{17}25\!/\!y} &\! \xi_{26} = \xi_{1} &\! A3\\
(15) \, q[(y_{3}y_{4}), a] \, \theta_{27} \, \xi_{27} &\!
\theta_{27} = \theta_{26} &\! \xi_{27} = \xi_{1}\eset{(15)\theta_{26} 26\!/\!y_{3},
(17)\theta_{26} 26\!/\!y_{4}} &\! C4\\
(16) \, q[y_{3}y_{4}, a] \, \theta_{28} \, \xi_{28} &\!
\theta_{28} = \theta_{26}\eset{(y_{3}y_{4})\xi_{27} 27\!/\!s} &\! \xi_{28} = \xi_{27} &\! A3\\
(15) \, q[(y_{4}), a] \, \theta_{29} \, \xi_{29} &\!
\theta_{29} = \theta_{28} &\! \xi_{29} = \xi_{27}\eset{(15)\theta_{28} 28\!/\!y_{3},
(17)\theta_{28} 28\!/\!y_{4}} &\! C4\\

(16) \, q[y_{4}, a] \, \theta_{30} \, \xi_{30} &\!
\theta_{30} = \theta_{28}\eset{y_{4}\xi_{29} 29\!/\!s} &\! \xi_{30} = \xi_{29} &\! A3\\
(17) \, q[( \ ), a ] \, \theta_{31} \, \xi_{31} &\! \theta_{31} = \theta_{30} &\!
\xi_{31} = \xi_{29} &\! C4\\

(18) \, q[y_{4}, a] \, \theta_{32} \, \xi_{32} &\! \theta_{32} = 
\theta_{30} &\! \xi_{32} = \xi_{17} &\! A3\\

(19) \, q[( \ ), a ] \, \theta_{33} \, \xi_{33} &\! \theta_{33} = \theta_{2} &\!
\xi_{33} = \xi_{17} &\! C4\\
(20) \, q[ \ \ex \ ] \, \theta_{34} \, \xi_{34} &\! \theta_{34} = 
\theta_{2} &\! \xi_{34} = \xi_{17} &\! A1\\
\end{array} \]
\caption{Another play on the tree in  Figure~\ref{ex1} from Example~\ref{examp61}}
\label{nex2}
\end{figure}
We examine applications of {\bf T2} to the term.
The simple
tile $z(\lambda u, \lambda)$ rooted at (6) is $1$-end and $1$-directed
with respect to the game:
there are three $1$-plays on it which are all ri, $\pi(6,7)$,
$\pi'(6,7)$ and $\pi'(12,13)$.
Transformation {\bf T2} allows us to remove this tile, 
so the leaf node (8) is directly beneath
node (5). The \emph{basic} tile $z(\lambda s.s,\lambda)$
rooted at (14)
is $1$-end and $1$-directed with respect to the game: 
the only plays  $\pi(12,15)$,
$\pi'(18,23)$ and $\pi'(26,31)$
are ri. A second application of {\bf T2} removes it;  therefore,  node (18)
is directly beneath node (13). Consequently, the basic tile
$z(\lambda y.y,\lambda)$ rooted at (12) is also $1$-end and $1$-directed
with respect to the ``reduced''  game; 
the plays $\pi(12,19)$ and $\pi'(16,33)$
become  the
sequences $\pi(12) \pi(13) \pi(16) \pi(19)$
and $\pi'(16) \pi'(17) \pi'(24) \pi'(25) \pi'(32) \pi'(33)$
(modulo changes to the look-up tables). 
The starting term in Figure~\ref{ex1}
is, therefore,  reduced to the smaller solution
term $\lambda z.z(\lambda x.f x) a$.
\qed
\end{exa}

Assume  $t$ is a $3$rd-order term.
If we inspect it top-down, 
starting beneath the initial lambda 
then it consists of  simple tiles,
each of  
which is  either a constant tile   or a top tile 
$y(\lambda_{1}, \ldots, \lambda_{k})$, $k \geq 0$, 
where each atomic leaf is  labelled with a dummy $\lambda$
because $y$ has order at most $2$; therefore, it is also
an end tile.

\begin{fact} \label{fact75}
If $t$ has order $3$ and $\tau$ is a simple tile in $t$
then either $\tau$ is a constant tile or an end tile which is also
a top tile.
\end{fact}

\ni
For instance, the tree in Figure~\ref{picture2} consists of
four simple top tiles $y(\lambda)$ that are also end tiles
rooted at $(2)$, $(4)$, $(10)$ and $(14)$
and the  simple constant tiles $f(\lambda x z_{1} z_{2},\lambda)$, 
$x(\lambda)$, $z_{2}$ and $a$. 
Therefore, by repeated application of
 Fact~\ref{fact57} and Proposition~\ref{end1}
each play $\pi \in \mathsf{G}(t,P)$ when $t$ is $3rd$-order
merely descends a branch of $t$ until it
reaches a final state. We now examine this 
\emph{tree model property} of plays
in more detail and show how it
leads to a very straightforward  proof of decidability of 
$3$rd-order matching.

Assume $P$ is $3rd$-order. 
We define a (unique)  partition of any play $\pi \in \G(t,P)$
in stages; at each stage we identify a simple tile, a subpart of $t$,
and the  interval at that stage.  
We call this iteratively defined notion of
partition a p-partition (a ``play partition'') to distinguish
it from Definition~\ref{def614} of the b-partition for a position.
(At $3$rd-order,  these partitions are intimately related as we shall
note; in the next section we extend 
p-partitions to all orders and its definition uses
b-partitions.)

\begin{defi} \label{def76}
Assume $P$ is $3$rd-order and $\pi \in \mathsf{G}(t,P)$. 
The \emph{p-partition} of $\pi$ is defined in stages $1 \leq k
\leq n$ for some $n$ as
$\pi(j_{0}),\pi(i_{1},j_{1}), \ldots, \pi(i_{n},j_{n})$
where $j_{0} = 1$. At each stage $k$ there is 
\begin{enumerate}[(1)]
\item the p-partition up to stage $k-1$, $\pi(1,j_{k-1}) = \pi(j_{0}),
\ldots, \pi(i_{k-1},j_{k-1})$;
\item the simple tile $\tau_{k}$ 
which occurs in $t$ directly beneath node $t' \in \pi(j_{k-1})$;
\item the composite tile  $\gamma_{k}$ of $t$ consisting
of all the nodes in the tiles $\tau_{1},\ldots,\tau_{k}$;
\item the position $\pi(i_{k})$ with $t_{k} \in \pi(i_{k})$
which is the root node of $\tau_{k}$;
\item the interval  $\pi(i_{k},j_{k})$ determined as follows:
$j_{k}$ is the least $j > i_{k}$ such that
\begin{enumerate}[$\bullet$]
\item $t'' \in \pi(j)$ is an atomic leaf of $\tau_{k}$, or
\item $j = |\pi|$.
\end{enumerate}
\end{enumerate}
\end{defi}

The idea of  a p-partition 
of $\pi \in \G(t,P)$  is to structurally relate parts of $\pi$
to  parts of $t$.
At  stage~$1$, 
$\tau_{1}$ is the simple tile directly beneath
the initial lambda of $t$
which is either a  constant or a top tile.
The subpart of $t$ at this stage, $\gamma_{1}$ is just
$\tau_{1}$.
Assume $t_{1}$ is the root of $\tau_{1}$; therefore,
$t_{1}  \in \pi(i_{1})$ (because $i_{1} =2$).
Consider the interval  $\pi(i_{1},j_{1})$:
$\pi(j_{1})$ is the first position  such that either
it is at an atomic 
leaf of $\tau_{1}$ or it is the final position of the play.
In the first case, $\pi(i_{1},j_{1})$ is a shortest play
on $\tau_{1}$; the tile $\tau_{2}$ is the simple tile
directly beneath the atomic leaf $t' \in \pi(j_{1})$ of $\tau_{1}$ 
 and $\gamma_{2}$ is $\tau_{1}$
and $\tau_{2}$. The p-partition thereby continues: at 
stage $k$, the interval
$\pi(i_{k},j_{k})$ is either a shortest play on $\tau_{k}$ or
$j_{k} = |\pi|$ and then there are no further stages. 
The following is an easy consequence  of Fact~\ref{fact57},
Proposition~\ref{prop515} and the definition of b-partition.

\begin{fact}
Assume 
the  \emph{p-partition} of $\pi \in \G(t,P)$
is 
$\pi(j_{0}),\pi(i_{1},j_{1}), \ldots, \pi(i_{n},j_{n})$
and $\tau_{k}$ is the simple tile at stage $k$.
\begin{enumerate}[(1)]
\item For $k: 1 \leq k  < n$, $\pi(i_{k},j_{k})$ is a shortest play
on $\tau_{k}$.
\item For $k: 1\leq k \leq n$, $\pi(i_{k},j_{k})$
is internal to $\tau_{k}$.
\item For $k: 1\leq k < n$ the b-partition of $\pi(j_{k}) =
\pi(j_{0}),\pi(i_{1},j_{1}), \ldots, \pi(i_{k},j_{k})$.
\end{enumerate}
\end{fact}

\begin{exa}\label{ex63}
If  $\pi$ is  the play of Example~\ref{examp42}
on the  tree of Figure~\ref{picture2}
then its p-partition is depicted below linearly.
\[\begin{array}{llllll}
y \ \ \ \lambda \ \ \ \ & y \ \ \ \lambda \ \ \ \ & f \ \ \lambda x z_{1} z_{2}
\ \ \ \ & x \ \ \ \lambda \ \ \ \ & y \ \ \ \lambda \ \ \ \ & z_{2} \\
\pi(2,3) & \pi(4,5) & \pi(6,7) & \pi(8,9) & \pi(10,11) & \pi(12,12) 
\end{array} \]
The simple tiles at each stage are; $\tau_{1} = y(\lambda)$,
$\tau_{2} = y(\lambda)$ rooted at $(4)$,
$\tau_{3} = f(\lambda x z_{1}z_{2},\lambda)$, $\tau_{4} =
x(\lambda)$, $\tau_{5} = y(\lambda)$ rooted at $(10)$ and $\tau_{6}
= z_{2}$.
For the other play $\pi'$
in this example, there is the  following p-partition.
\[ \begin{array}{lllll}
y \ \ \ \ \lambda \ \ \ \ & y \ \ \ \ \lambda \ \ \ \ & 
f \ \ \ \lambda \ \ \ \ & y \ \ \ \ \lambda \ \ \ \ 
& a \\
\pi'(2,3) & \pi'(4,5) & \pi'(6,7) & \pi'(8,9) & \pi'(10,10) 
\end{array} \]
The two plays share the first three simple tiles, but
play is at different atomic leaves of $\tau_{3}$ after stage $3$.
\qed
\end{exa}

Consider the p-partitions of all
plays in $\G(t,P)$. 
We slightly abuse notation:
we let $\pi(i_{k},j_{k})$, $\pi'(i_{k},j_{k})$
be their intervals at stage $k$  even when
they have different ranges. 
Instead of a branch  of simple tiles 
there is  a tree of simple tiles: 
as each play shares  the same simple tile $\tau_{1}$ at
stage~$1$ of its  p-partition.
The simple tile $\tau$ in $t$  is \emph{special}
if it obeys one of the following three conditions
\begin{enumerate}[$\bullet$]
\item $\tau = \tau_{k}$ for $\pi$ at stage $k$
and $\pi(i_{k},j_{k})$ is nri
(see Definition~\ref{defi48}),
\item $\tau =\tau_{k}$ for $\pi$ at stage $k$ and $j_{k} = |\pi|$,
\item $\tau =\tau_{k}$ for $\pi$, $\pi'$ at stage $k$
and $t' \not= t''$ when $t' \in \pi(j_{k})$,
$t'' \in \pi'(j_{k})$. 
\end{enumerate}
The first kind of special tile explicitly contributes to solving $P$.
The second kind identifies where a play finishes.
The third
kind separates plays;
each  p-partition after stage~$1$  that 
ends at the same atomic leaf of $\tau_{1}$ 
shares $\tau_{2}$  at stage $2$ and so on. Therefore, branching 
in the tree of simple tiles will occur at $\tau_{k}$ if there are 
plays $\pi$, $\pi'$ that end at different atomic leaves
of $\tau_{k}$  at stage $k$ 
(and agree on atomic leaves at  all earlier stages).
In Example~\ref{ex63}, $f(\lambda x z_{1} z_{2},\lambda)$
separates the plays $\pi$ and $\pi'$.  The other special tiles
in these plays are $x(\lambda)$ because of the nri interval
$\pi(8,9)$ and $z_{2}$ and $a$ as plays finish within them.

Any  simple tile in $t$ with at least one  atomic leaf
which is not special
is superfluous. Either every play avoids it (so, {\bf T1} applies)
or every play on it is ri and ends
at the same atomic leaf $\lambda \ol{x}_{j}$ for some $j$
(so, is both $j$-end and 
$j$-directed with respect to the game and {\bf T2} applies): four instance,
all  four
simple tiles  $y(\lambda)$ in Example~\ref{ex63} are both $1$-end and 
$1$-directed with respect to the game.
There is an upper bound (relative to the problem $P$)
on the number of special tiles
that can occur in a term $t$ as follows
\begin{enumerate}[$\bullet$]
\item at most $\delta$ ($=$ the right size for $P$, 
Definition~\ref{delta}) tiles that involve nri intervals;
\item at most $p$ ($=$ the number of plays\footnote{
$p$  is bounded
by the number of branches in the right terms of $P$.})
 tiles where a  play ends;
\item at most $p-1$ tiles that are play separators.
\end{enumerate}
Decidability of $3$rd-order matching, via the small model
property, is, therefore,  a simple
consequence of the p-partitions
and the identification of special simple tiles. For  Example~\ref{ex63}, 
the
term of  Figure~\ref{picture2} 
can be reduced to the smaller solution term
$\lambda y.f(\lambda x z_{1}z_{2}.x z_{2}) a$.

\begin{defi} \label{def64}
$|t|$ is the number of simple tiles with atomic leaves
in a longest  branch of $t$ and $||t||$ is the total  number
of simple tiles with atomic leaves in $t$.
\end{defi}

\begin{fact}\label{bounds}
If $t$ is a smallest
solution to $3$rd-order $P$ and $p$ is the number of plays in $\G(t,P)$,
then $||t|| \leq \delta + (2p-1)$.
\end{fact}


An alternative, and equally simple,  proof of decidability of $3$rd-order
matching that does not appeal directly to the tree model property
uses Proposition~\ref{end} 
and transformations {\bf T1} and  {\bf T2}. In a large solution term,
there must be embedded
end tiles that are redundant. 
Dowek observes that solutions with embeddings
$\lambda \ol{y}.(\ldots (y \ldots (y \ldots (y \ldots)\ldots)\ldots)
\ldots)$ can be reduced to smaller solutions $\lambda \ol{y}.
(\ldots (y\ldots)\ldots)$ in his proof of decidability of
$3$rd-order matching \cite{Dow}.
Assume $t$ is a \emph{smallest} solution 
term (with respect to $|| \cdot ||$) for  $P$.
It contains at most $\delta + p$
constant tiles with atomic leaves (otherwise, {\bf T1} would apply
and produce a smaller solution). Let $c$ be the number of constant
tiles with atomic leaves in $t_{0}$. 
Therefore, by Proposition~\ref{end}, 
$t$  contains at most 
$\lceil (\delta + p - c)/2 \rceil$ embedded simple top tiles
(otherwise, {\bf T2} would apply and produce a smaller solution).
If $\alpha$ is the arity of $P$, Definition~\ref{def3}, 
then 
there are at most $\alpha$ inequivalent top simple tiles:
as soon as a branch of $t$ contains $\alpha +1$ simple 
top tiles, there must be at least one embedded end tile.
Consequently, no branch of $t$ can contain more
than $\alpha + (\lceil (\delta+p-c)/2 \rceil)$
simple top  tiles. Because $c \leq \delta + p$,
$|t| \leq \alpha + \delta + p + 1$ and 
$3$rd-order matching is decidable.

The question is how to extend these straightforward 
proofs to all higher-orders.
With a $4$th or $5$th-order tree
there are two levels of simple non-constant tiles: top tiles
$y(\lambda \ol{x}_{1},\ldots,\lambda \ol{x}_{k})$
and end tiles $z(\lambda \ol{z}_{1},\ldots,\lambda \ol{z}_{l})$ 
where $z$ is bound by a $\lambda \ol{x}_{j}$.
The number of levels increases with order:
at $8$th or $9$th-order there are four levels.
As soon as there is more than one level, game
playing may jump around the tree
as the examples in Figures~\ref{ex2} and \ref{ex15} illustrate.
For any order, if the tree-model property holds,
so each play can be p-partitioned into internal plays on
the simple tiles from the root to a  tile where the play ends, then
decidability is assured by iteration: redundant simple end tiles are
first removed which causes further tiles to be end and so on.

\begin{rem}
Schubert defines an \emph{unsophisticated} lambda term
in \cite{Sch}: $\lambda x_{1} \ldots x_{m}.t$ is unsophisticated
if each occurrence of $x_{i} t'_{1} \ldots t'_{k}$ 
within $t$
has the property 
that no $t'_{j}$
contains a free variable $x_{l}$, $1 \leq l \leq m$.
A $5$th-order (dual) interpolation problem
is unsophisticated if in each (dis)equation 
$x v_{1} \ldots v_{m}  \approx u$ the left terms $v_{i}$
are unsophisticated. Schubert proves decidability
of $5$th-order unsophisticated dual interpolation. This restriction
implies the tree model property. When  play
is at an end tile with head variable $y$, 
the state has the form $q[\lambda \ol{z}.u,r]$
where $\lambda \ol{z}.u$ is closed: consequently, play
cannot jump back to the tile which binds $y$.
(Decidability  can 
be extended to all orders by defining \emph{hereditary}
unsophisticated terms where each
occurrence of $\lambda \ol{y}.t'$ within it is unsophisticated.)
\qed
\end{rem}

For arbitrary order, if $t$ is a smallest solution term
for $P$ and it contains $c$ simple constant
tiles with atomic leaves then $c \leq \delta + p$.
If $t$ is large
then it must contain embedded tiles. 
With $4$th-order there must be embedded top tiles
and with $5$th-order there can also be embedded end tiles.
However, Proposition~\ref{end} implies, for any order,
that there cannot be
more than $\lceil (\delta + p -c) /2 \rceil$
embedded end tiles.
Consequently, if $P$ is a $5$th-order problem and $t$
contains a bounded number $k$ of top simple tiles
then the small model property holds.

\begin{fact}\label{fact66}
If $t$ is a smallest solution to
$5$th-order $P$ that
contains at most $k$ top simple tiles,
then
$|t| \leq (k \times (\alpha +1)) + \delta + p +1$
\end{fact}
\begin{figure}
\begin{center}
\ \ \ \ \ \ \ \ \  \xymatrix{
& \ar@{-}[dl] \tau_{1}  \ar@{-}[dr] & & & &  \ar@{-}[dl] \tau_{1}
\ar@{-}[dr] & \\ 
\ar@{-}[r] & \lambda \ol{x}_{1} \ar@{.}[d]  \ar@{-}[r] & & 
&    \ar@{-}[r]  &\lambda \ol{x}_{1}  \ar@{.}[d] \ar@{-}[r]  & \\
& \ar@{-}[dl] \tau_{k}  \ar@{-}[dr] & & \Longrightarrow & &  \ar@{-}[dl] \tau_{k}
\ar@{-}[dr] & \\ 
\ar@{-}[r] & \lambda \ol{x}_{k} \ar@{.}[d]  \ar@{-}[r] & & 
&    \ar@{-}[r]  &\lambda \ol{x}_{k} \ar@{.}[d] \ar@{-}[r]  & \\
& \ar@{-}[dl] \tau_{n}  \ar@{-}[dr] & & & &  \ar@{-}[dl] \tau_{k}
\ar@{-}[dr] & \\ 
\ar@{-}[r] & \lambda \ol{x}_{n} \ar@{.}[d]  \ar@{-}[r] & & 
&    \ar@{-}[r]  & \ar@{-}[dl] \tau_{n}  \ar@{-}[dr]   \ar@{-}[r] & \\
&  & & &\ar@{-}[r]  &\lambda \ol{x}_{n} \ar@{.}[d] \ar@{-}[r]  & \\
& & & & & 
} 

\end{center}
\caption{Tile Lowering }
\label{tl}
\end{figure}

\ni
Fact~\ref{fact66} generalises  to all orders:
if $t$ is a smallest solution to $P$ that
contains at most $k$ simple tiles that are neither
end nor constant tiles then it has a bounded size. 
This result slightly extends
some  cases examined in \cite{DW,SchS} where there are
restrictions on the number of free variables
that a $\lambda \ol{y}$ can bind within a solution term.

A family of tiles in a tree  consists of  a top tile together with all
its dependents (see Definition~\ref{family}). 
The problem case  for $4$th or $5$th-order is a solution
term with arbitrary many top tiles (and, therefore, arbitrary
many families).
If a tree is large, then it  must contain  families
of tiles all of whose plays are ri and, therefore, do not contribute to solving
$P$. For even higher-orders, for the same reason,
whole subfamilies of tiles are redundant.

The difficulty is how to extract and remove redundant families
and subfamilies of tiles. They may be entangled, occurring
anywhere in a tree. What we would like to do is disentangle them
thereby  restoring, as far as possible, the tree model property.
A key ingredient is \emph{tile lowering}
where we  generalise to \emph{basic} tiles. 
Consider the left tree in  Figure~\ref{tl} and assume
$\tau_{k}$ is a basic tile and $\tau_{n}$ is a simple tile
whose head variable is bound within $\tau_{k}$ (and no variable
on the branch 
between $\lambda \ol{x}_{k}$ and $\tau_{n}$ is bound within $\tau_{k}$).
What we would like to do is to transform the left tree into
the right tree where $\tau_{k}$ 
(and all the subtrees beneath its atomic leaves other than
$\lambda \ol{x}_{k}$) is copied immediately above $\tau_{n}$
(thereby producing a larger basic tile)
with the understanding that it is the lower occurrence
of $\tau_{k}$ that binds $\tau_{n}$ and any free variables beneath
$\tau_{n}$ that are bound within  $\tau_{k}$ in the left tree.
In \cite{St2}, we introduced an explicit local transformation
that has this effect for $4$th-order matching and for
the atoms case at all orders.
The virtue of the transformation is that families
of tiles are disentangled as the lower occurrence of
$\tau_{k}$ is brought next to its dependent tile $\tau_{n}$
(and the upper 
occurrence of $\tau_{k}$ in the right tree is ``closer'' to
being an end tile, as (some) binding is lost below). 
Game-theoretically, tile lowering will be  justified in terms of
permutations, omissions and repetitions of ri plays on
tiles. 


\section{Partitioning of  plays}
\label{unfold}
The  analysis now shifts from how single tiles to
how families of tiles in a term  contribute to solving 
a problem.  We  extend the definition  
of the p-partition of a play from the previous section to all orders.
We start by examining simple properties of game playing
that involve families of tiles: the discussion uses 
the notion of when a position is  a descendent of another
position, Definition~\ref{def416},
when one tile is a dependent of another,
Definition~\ref{defdepend},  and 
when two tiles belong to the same family,
Definition~\ref{family}. 

\begin{prop} \label{C4jump}
Assume
$\tau$ with root node $t$ and  
$\gamma(\lambda \ol{x}_{1},\ldots,\lambda \ol{x}_{k})$
with root node $t'$
are simple tiles.  
\begin{enumerate}[(1)]
\item If  $t \in \pi(i)$, $t' \in \pi(j)$, $\pi(j)$
is the result of move A3  of Figure~\ref{game} and
$\pi(j)$
is a descendent of $\pi(i)$ then $\gamma$ is a dependent of
$\tau$. 
\item If $t \in \pi(i)$ 
and  $\lambda \ol{x}_{j} \in \pi(i+1)$ and $\pi(i+1)$ is the result
of move C4 then $\tau$ and $\gamma$ belong to the same family
of tiles.
\end{enumerate}
\end{prop}
\proof (1) Assume that $t \in \pi(i)$, $t' \in \pi(j)$ and $\pi(j)$
is a descendent of $\pi(i)$. Therefore, because
the descendent relation is the reflexive and transitive closure of
the child relation,
there is a subsequence
of positions $\pi(i_{1}), \pi(i_{2}), \ldots, \pi(i_{2k+1})$
with $i = i_{1}$,  $i_{2k+1} = j$, for all $n$, $\pi(i_{n+1})$ is a child of 
$\pi(i_{n})$. Assume $t_{m} \in \pi(i_{m})$ for each $m: 1 \leq m \leq 2k+1$;
so, $t = t_{1}$ and $t' = t_{2k+1}$. Because $\pi(j)$ is the result
of A3, $t'$ is labelled with a variable, that is bound by $t_{2k}$:
therefore, $\gamma$ is a dependent of the simple tile rooted
at $t_{2k-1}$. This argument is now repeated because
$\pi(i_{2k-1})$ must in turn be the result of A3 and
so the simple tile rooted at $t_{2k-1}$ is a dependent of the simple
tile rooted at $t_{2k-3}$, and so on; consequently,
as the dependency relation is transitive closed, 
$\gamma$ is a dependent of $\tau$. 
(2) Consider the following subsequences of  positions 
$\pi(j_{1}), \pi(j_{2}), \pi(j_{3}), \ldots, \pi(j_{3k-2}),\pi(j_{3k-1}),
\pi(j_{3k})$ where $i = j_{3k-1}$ and $i+1 = j_{3k}$;
each position $\pi(j_{3m})$ is the result of
C4 on $\pi(j_{3m-1})$ which is the result of $0$ or more applications
of C3 on $\pi(j_{3m-2})$, $m \geq 1$;
furthermore, for $m >1$, $\pi(j_{3m-2})$ is the result
of A3 and is a child of $\pi(j_{3m-3})$ and  $\pi(j_{1})$ is
the result of A3 and a child of $\pi(1)$.
Now by a routine induction on $k$, it follows that for every
$m : 1 \leq m \leq k$, $t_{m}\in \pi(j_{3m})$ is an atomic leaf
of some simple tile rooted at $t_{n} \in \pi(j_{3n}-2)$ for some $n \leq m$
and that all such tiles belong to the same family.
\qed

In the previous section we alluded to the notion of tile 
level.

\begin{defi} \label{level}
The \emph{level} of a non-constant tile
is defined inductively.
\begin{enumerate}[(1)]
\item If $\tau$ is a top tile then $\tau$ has level  $1$.
\item If $\tau$ is an immediate dependent of $\gamma$ 
and $\gamma$ has level $m$ then $\tau$ has level $m+1$.
\end{enumerate}
\end{defi}

\ni
The number of possible  levels increases with order.
In a  $4$th or $5$th-order term there are at most
two levels of tile: top tiles and end
tiles as illustrated in Figures~\ref{ex1} and \ref{ex14}.
With a  $8$th or $9$th-order term, there are at
most four levels of tile and so on.
The presence of dummy lambda in a term tree does not affect the notion 
of level (because a dummy lambda cannot be a binder). 

If   there is more than one level of tile in a term,
game playing  may pass repeatedly through the same 
sequences of nodes of a tree.
The following definition captures when two such intervals
are said to correspond.

\begin{defi} \label{defsimilar}
The intervals $\pi(i,j)$, $\pi(i', j')$ \emph{correspond},
written $\pi(i,j) \sim \pi(i',j')$,  provided that $j,j' < |\pi|$, 
$j-i = j'-i'$ and for all $k : 0 \leq k \leq j-i$, 
\begin{enumerate}[(1)]
\item if $t \in \pi(i+k)$ then $t \in \pi(i'+k)$.
\item if $q[l,r] \in \pi(i+k)$ then for some $r'$, $q[l,r'] \in \pi(i'+k)$
\item if $q[(l_{1},\ldots,l_{m}),r] \in \pi(i+k)$
then for some $r'$, $q[(l_{1},\ldots,l_{m}),r'] \in \pi(i'+k)$.
\item if $q[-,r] \in \pi(i+k)$ then for some $r'$,
$q[-,r'] \in \pi(i'+k)$.
\end{enumerate}
\end{defi}

\begin{fact}
The relation $\sim$ on intervals
is an equivalence relation.
\end{fact}

\ni
Although this definition abstracts from the look-up tables,
it requires  agreement on the sequences  of nodes
of the tree and on the left terms of states.
For instance, intervals that are  shortest ri plays on a top tile
correspond, as shown in the proof of
Proposition~\ref{prop515}. When  $\pi'$ is the play in  Figure~\ref{nex2}
on the tree in Figure~\ref{ex1},
the nri intervals $\pi'(4,7)$ and $\pi'(10,13)$ correspond. 

In Section~\ref{tiles}
we described some  uniformity properties of game playing  for
top and embedded tiles (especially for embedded end
tiles). 
Now our aim is to understand
uniformities of play for  tiles of arbitrary  level. 
Given a position $\pi(j)$ at a lambda node $t'$, 
there is an associated  (unique) b-partition of 
$\pi(1,j)$ into intervals that are plays on the simple tiles
between the root of the term tree and node $t'$
by Definition~\ref{def614}.
If more than one   position is  at a lambda node $t'$ then  
their associated b-partitions
must differ in  their plays at an earliest tile with the same
starting positions but different end positions.

\begin{prop} \label{prop73A}
Assume $t' \in \pi(j)$, $t' \in \pi(j')$, $j \not=j'$ and $t'$ is labelled 
$\lambda \ol{x}$ for some $\ol{x}$.  
If $\pi(j_{0}), \pi(i_{1},j_{1}), \ldots, \pi(i_{n},j_{n})$
is the b-partition for $\pi(j)$ 
and  $\pi(j'_{0}), \pi(i'_{1},j'_{1}), \ldots, \pi(i'_{n},j'_{n})$
is the b-partition for $\pi(j')$ then 
there is a $k: 1 \leq k \leq n$ such that 
$i_{k} = i'_{k}$ and $j_{k} \not= j'_{k}$
and for all $m < k$, $i_{m} = i'_{m}$ and $j_{m} = j'_{m}$. 
\end{prop}
\proof Assume that $\lambda \ol{x}_{0}, u_{1}, \lambda \ol{x}_{1},
\ldots, u_{n}, \lambda \ol{x}_{n}$ is the branch from
the root of the term tree to $t'$ labelled $\lambda \ol{x}_{n}$, 
$t' \in
\pi(j)$ and $t' \in \pi(j')$ for $j'\not= j$.
 Let $\pi(j_{0}), \pi(i_{1},j_{1}), \ldots, \pi(i_{n},j_{n})$
be the b-partition for  $\pi(j)$ 
and $\pi(j'_{0}),\pi(i'_{1},j'_{1}), \ldots, \pi(i'_{n},j'_{n})$
be  the b-partition for $\pi(j')$. 
Because $j_{0} = j'_{0} = 1$, $i_{1} = i'_{1} = 2$
and $j \not= j'$, there must be a  least
$k$ such that $i_{k} = i'_{k}$ and $j_{k} \not= j'_{k}$.
\qed

The b-partitions of Example~\ref{example512} illustrate
this proposition; node $(7)$ 
of Figure~\ref{ex14} labelled $\lambda$ belongs to both
$\pi(15)$ and $\pi(31)$ of the play in Figure~\ref{ex15}.
These b-partitions  
agree on   the play $\pi(2,3)$  on the initial tile
$z(\lambda z_{1},\lambda z_{2})$ that ends at $\lambda z_{1}$
and then they have  different $1$-plays 
$\pi(4,5)$ and $\pi(4,21)$ on
the  next tile $z(\lambda x_{1},\lambda x_{2})$.

\begin{defi} \label{defi744}
Assume $t' \in \pi(j)$, $t' \in \pi(j')$, $j \not=j'$, $t'$ is labelled 
$\lambda \ol{x}$ for some $\ol{x}$ and $\lambda \ol{x}_{0}, u_{1}, 
\lambda \ol{x}_{1}, \ldots, u_{n}, \lambda \ol{x}_{n}$ is the branch from
the root of the term tree  to $t'$.
Let  $\pi(j_{0}), \pi(i_{1},j_{1}), \ldots, \pi(i_{n},j_{n})$
be  the b-partition for $\pi(j)$ and let
$\pi(j'_{0}), \pi(i'_{1},j'_{1}), \ldots, \pi(i'_{n},j'_{n})$
be the b-partition for $\pi(j')$
and let  $\tau = u_{k}(\ldots \lambda \ol{x}_{k} \ldots)$ 
be the first simple tile for 
$k \geq 1$ 
such that 
$i_{k} = i'_{k}$ and $j_{k} \not= j'_{k}$. 
The positions $\pi(j), \pi(j')$ are then said to \emph{vary at 
$\pi(j_{k}), \pi(j'_{k})$ with (simple tile) $\tau$}.
\end{defi}

\ni
Two positions vary at $\pi(j_{k}), \pi(j'_{k})$
with $\tau$ if they are at the  same lambda node
and $\tau$ is the first simple tile in their b-partitions where there is a
difference in play; there are two distinct $m$-plays
on $\tau$ for some $m$, $\pi(i_{k},j_{k})$ and
$\pi(i_{k},j'_{k})$.
In the case of 
Example~\ref{example512} discussed previously,
the positions $\pi(15)$, $\pi(31)$ vary at $\pi(5), \pi(21)$ with 
$z(\lambda x_{1},\lambda x_{2})$.

Given two positions $\pi(j)$ and $\pi(j')$ at the same node
labelled $\lambda \ol{x}$ we are interested in 
defining when
intervals $\pi(j+1, j+m)$ and $\pi(j'+1, j'+m)$ correspond
in the sense of Definition~\ref{defsimilar}. 
The simplest case is when $\pi(j), \pi(j')$ vary at $\pi(j), \pi(j')$
with
$\tau = u_{n}(\ldots \lambda \ol{x} \ldots)$;
their b-partitions
agree except on the last simple tile in the branch from the root to 
$\lambda \ol{x}$.
The two intervals $\pi(i_{n},j)$ and $\pi(i_{n},j')$ are, therefore,  both
$k$-plays on $\tau$ for some $k$; both positions $\pi(j)$ and $\pi(j')$
are children of $\pi(i_{n})$ and, therefore,  must be the
results 
of  move C4 of Figure~\ref{game}.
This means that  
the look-up tables $\theta_{j+1} \in \pi(j+1)$  and $\theta_{j'+1}
\in \pi(j'+1)$  
only differ in their entries for the variables in $\ol{x}$:
therefore, for each $m \geq 1$ their continuations 
$\pi(j+1,j+m)$, $\pi(j'+1,j'+m)$ correspond
as long as play does not reach
children of $\pi(j)$ and $\pi(j')$; or
positions where different $\fa$ choices are  exercised;
or a position with a final state.
Consider next the general situation when 
$\pi(j), \pi(j')$ vary at $\pi(j_{k}), \pi(j'_{k})$
with $\tau = u_{k}(\ldots \lambda \ol{x}_{k} \ldots)$;
their b-partitions differ at a simple tile that is  earlier in the branch
from the root to $\lambda \ol{x}$.
What we want to capture is the following uniformity: 
if the plays $\pi(i_{m},j_{m})$
and $\pi(i'_{m},j'_{m})$ on tiles $\tau_{m} : k < m \leq n$ that 
do not involve positions that are children  of $\pi(j_{k})$ and $\pi(j'_{k})$
correspond, then the continuations from
$\pi(j+1)$ and $\pi(j'+1)$ 
will also correspond as long as the positions are not descendents 
of $\pi(j_{k})$ or $\pi(j'_{k})$ 
or the result of a  different $\fa$-choice, 
or one has a final state. 
The issue  is how to
formally capture this correspondence in terms of the relationship between
the look-up tables $\theta \in \pi(j+1)$ and $\theta' \in \pi(j'+1)$.
This is the motivation for  the following bisimulation like definition.

\begin{defi} \label{defi75}
Assume $\tau$ is a simple non-constant tile. 
\begin{enumerate}[(1)]
\item Two look-up tables $\mu$, $\mu'$ are
\emph{$n$-similar except for $\tau$}, $\mu \sim_{\tau}^{n} \mu'$,
which is defined iteratively, for $n \geq 0$. 
\begin{enumerate}[$\bullet$]
\item $\mu \sim_{\tau}^{0} \mu'$ iff $\mu = \mu'$;
\item $\theta \sim_{\tau}^{n+1} \theta'$ iff (1)  
for all $y$. $\theta(y)$ is defined iff $\theta'(y)$ is defined
and (2) if $\gamma = y(\ldots)$
is not a dependent of $\tau$ and $\theta(y) = l\xi i$ then $\theta'(y)
= l \xi' i'$ and $\xi \sim_{\tau}^{n} \xi'$;
\item $\xi \sim_{\tau}^{n+1} \xi'$ iff 
(1) for all $z$. $\xi(z)$ is defined iff $\xi'(z)$ is defined
and (2) if $\xi (z) = t \theta i$
then $\xi'(z) = t \theta' i'$ and $\theta \sim_{\tau}^{n} \theta'$.
\end{enumerate}
\item Two look-up tables $\mu$, $\mu'$ are
\emph{similar except for $\tau$}, $\mu \sim_{\tau} \mu'$, 
if there is an  $n \geq 0$ such that  $\mu \sim_{\tau}^{n}
\mu'$.
\end{enumerate}
\end{defi}

\begin{fact}
In the following assume $\mu$, $\mu'$ are look-up tables
(of the same kind).
\begin{enumerate}[(1)]
\item $\mu \sim_{\tau}^{n} \mu$.
\item If $\mu \sim_{\tau}^{n} \mu'$ then $\mu' \sim_{\tau}^{n} \mu$.
\item If $\mu \sim_{\tau}^{n} \mu'$ and $\mu' \sim_{\tau}^{n} \mu''$
then $\mu \sim_{\tau}^{n} \mu''$.
 \item If $\mu \sim^{m}_{\tau} \mu'$ and $n > m$ then $\mu \sim^{n}_{\tau} 
\mu'$.
\end{enumerate}
\end{fact}

\ni
The key point with the definition of $\theta \sim_{\tau} \theta'$ is that
the entries of $\theta$ and $\theta'$ should be very similar except
in the case that they are labels of nodes of simple tiles that are dependents 
of $\tau$.

\begin{fact} \label{fact78}
If $\pi(i,j)$ and $\pi(i,j')$ are $k$-plays on the simple 
tile $\tau$, $\theta \in \pi(j+1)$ and $\theta' \in \pi(j'+1)$
then $\theta \sim_{\tau} \theta'$. 
\end{fact}

We now come to a critical uniformity property. 

\begin{prop} \label{prop79}
Assume  $\pi(j)$,$\pi(j')$ vary at $\pi(j_{k}), \pi(j'_{k})$
with $\tau$, $\theta \in \pi(j+1)$, $\theta' \in \pi(j'+1)$
and $\theta \sim_{\tau} \theta'$.
If $\pi(j+1,j+ m)$ 
is ri, $t' \in \pi(j+m)$ is labelled $\lambda \ol{x}$ for some $\ol{x}$
and no $\pi(j+l)$, $1 \leq l \leq m$, is a descendent of $\pi(j_{k})$ 
then $\pi(j'+1,j' + m) \sim \pi(j+1,j+m)$. 
\end{prop}
\proof
Assume that $\lambda \ol{x}_{0}, u_{1}, \lambda \ol{x}_{1},
\ldots, u_{n}, \lambda \ol{x}_{n}$ is the branch from
the root of term tree $t$  to $\lambda \ol{x}_{n}$, 
$\lambda \ol{x}_{n} \in
\pi(j)$ and $\lambda \ol{x}_{n} \in \pi(j')$ for $j' \not= j$. 
Let $\pi(j_{0}), \pi(i_{1},j_{1}),\ldots, \pi(i_{n},j_{n})$
be the b-partition for $\pi(j)$ 
and $\pi(j'_{0}), \pi(i'_{1},j'_{1}), \ldots, \pi(i'_{n},j'_{n})$
be  the b-partition for $\pi(j')$. 
Assume that $\pi(j)$, $\pi(j')$
vary at $\pi(j_{k})$, $\pi(j'_{k})$ 
with $\tau = u_{k}(\ldots \lambda \ol{x}_{k}\ldots)$
and that $\theta \sim_{\tau} \theta'$ when $\theta \in \pi(j+1)$ and 
$\theta' \in \pi(j'+1)$.
Assume that $\pi(j+1,j+m)$ is ri, $t' \in \pi(j+m)$ is labelled $\lambda \ol{x}$,
no $t'' \in \pi(i+k)$ within this interval is at a dependent
of $\tau$.
By a routine induction on $m$,
$\pi(j'+1,j'+m) \sim \pi(j+1,j+m)$. 
The interval $\pi(j+1,j+m)$ only involves moves A3 and C4 of
Figure~\ref{game} because it is ri and ends at $t' \in \pi(j+m)$
by Fact~\ref{fact55}.
Initially, for some simple non-constant tile
$\tau' = y(\lambda \ol{z}_{1},\ldots,\lambda \ol{z}_{l})$,
$y \in \pi(j+1)$ and $y \in \pi(j'+1)$. Therefore, because $\pi(j+1)$
is not a descendent  of $\pi(j_{k})$,  
$\tau'$ is not a dependent of $\tau$; by A3 of Figure~\ref{game}
position $\pi(j+1) = y q[l,r] \theta \xi_{j+1}$ and
position $\pi(j'+1) = y q[l,r'] \theta' \xi_{j'+1}$ where
$\xi_{j+1} \sim_{\tau} \xi_{j'+1}$.
The next position is a result of C4: if $\xi_{j+1}(x) = t' \theta'' i$
then $\xi_{j'+1}(x) = t' \theta'''i'$ and $\theta'' \sim_{\tau} \theta'''$.
Consequently, the result follows.
\qed

\begin{rem} \label{rem811}
Proposition~\ref{prop79}
can be strengthened to include corresponding nri plays 
that involve the same  $\fa$-choices. 
For instance, in Example~\ref{example512} when $\pi$ is in Figure~\ref{ex15}
the positions $\pi(5), \pi(21)$ vary at
$\pi(5), \pi(21)$ with $z(\lambda x_{1},\lambda x_{2})$.
Position $\pi(16)$ is the first position after $\pi(5)$ that is a child
of $\pi(5)$; therefore, 
the continuations  $\pi(6,15)$ and $\pi(22,31)$ 
correspond even though they pass through the constant tile
$h(\lambda)$.
\qed
\end{rem}




We shall now extend the definition of the p-partition of a play $\pi \in
\mathsf{G}(t,P)$,
Definition~\ref{def76}, from $3$rd-order $P$ to all higher  orders.
Again, it is defined in stages using simple tiles from $t$;
so $\pi = \pi(j_{0}),\pi(i_{1},j_{1}),\ldots, \pi(i_{n},j_{n})$ for some
$n$ and there is an associated sequence of simple tiles
$\tau_{1}, \ldots, \tau_{n}$
from $t$ such that for each $m : 1 \leq m \leq n$, $t' \in \pi(i_{m})$
where $t'$ is the root of $\tau_{m}$ and $\tau_{m}$ occurs directly
below $t'' \in \pi(j_{m-1})$ in $t$. 
However, the same simple tile may occur
more than once in this sequence of tiles; so we adopt the following notation.

\begin{defi} \label{def81}
Assume that $\tau_{1},\ldots, \tau_{n}$
is a sequence of simple tiles associated with
a play $\pi \in \mathsf{G}(t,P)$.
We write $\tau^{\pi}_{k}$ to identify 
the  $k$th tile $\tau_{k}$ in this sequence
and  we use the notation 
$t'  @ \tau^{\pi}_{k}$ for node $t'$ of $\tau^{\pi}_{k}$.
\end{defi}

After stage $k$, the p-partition for $\pi$ consists of the sequence
of  tiles
$\tau^{\pi}_{1},\ldots, \tau^{\pi}_{k}$; the composite tile $\gamma_{k}$ 
consists of these tiles. However, we also provide a more graphical
representation of the composite tile $\gamma_{k}$ by including  
a labelled edge for each $m: 1 <  m \leq k$ 
of the  form $\tau^{\pi}_{m} \byre{\pi}
t' @ \tau^{\pi}_{l}$ where 
$l < m$ and  $t'$ is an atomic leaf of $\tau^{\pi}_{l}$.
The relation $\byre{\pi}$ between a tile at stage $m$ and 
an atomic leaf of a tile at an earlier stage
$l$
captures  control structure in the
p-partition that $t' @ \tau^{\pi}_{l} \in \pi(j_{m-1})$. 
This linearisation of the p-partition for each $\pi$ will be useful
in the decidability proof; we can, of course, 
reconstitute the subtree $\gamma_{k}$ 
from its linear presentation after stage $k$ 
just by viewing  $\byre{\pi}$ as the subtree relation.
We assume this extra intensionality is also
reflected in the $\xi$ look-up tables for $\pi$;
in the case of the C moves of Figure~\ref{game}
if $l = \lambda z_{1} \ldots z_{j}.w$ and 
$t_{m} @ \tau^{\pi}_{k} \downarrow_{i}
t'_{i}@\tau^{\pi}_{k}$ then $\xi_{m+1} = \xi_{m}\eset{
(t'_{1} @ \tau^{\pi}_{k}) \theta_{m} m /z_{1}.\ldots,(t'_{j}@\tau^{\pi}_{k}) \theta_{m}
m/z_{j}}$. Consequently,
at an application of move C4, if $x$ is the head variable of the left term
of the state and  $\xi(x) = (t'@ \tau^{\pi}_{k}) \theta' i$
then the next position is at  $t' @ \tau^{\pi}_{k}$.

\begin{defi} \label{partit}
Assume $\pi \in \G(t,P)$.  
The \emph{p-partition} of $\pi$ is defined in stages $1 \leq k
\leq n$ for some $n$ as
$\pi(j_{0}),\pi(i_{1},j_{1}), \ldots, \pi(i_{n},j_{n})$
where $j_{0} = 1$. At each stage $k$ there is 
\begin{enumerate}[(1)]
\item the p-partition up to stage $k-1$, $\pi(1,j_{k-1}) = \pi(j_{0}),
\ldots, \pi(i_{k-1},j_{k-1})$;
\item the composite tile  $\gamma_{k-1}$ consisting
of the tiles $\tau^{\pi}_{1},\ldots,\tau^{\pi}_{k-1}$ with  edges
$\byre{\pi}$;
\item the simple tile $\tau^{\pi}_{k}$ 
which occurs in $t$ directly beneath node 
$t'@ \tau^{\pi}_{l} \in \pi(j_{k-1})$ and $\tau^{\pi}_{k} \byre{\pi}
t'@ \tau^{\pi}_{l}$  in $\gamma_{k} = 
\gamma_{k-1} \cup \eset{\tau^{\pi}_{k}}$;
\item the position $\pi(i_{k})$ with $t'' @ \tau^{\pi}_{k} \in \pi(i_{k})$
where $t''$  is the root node of $\tau_{k}$;
\item the interval  $\pi(i_{k},j_{k})$ determined as follows:

\begin{enumerate}[$\bullet$]
\item set $j = i_{k}$
\item  $(* *)$ while $j \not= |\pi|$ and $t'@\tau^{\pi}_{l} \in \pi(j)$
is not labelled with a lambda do j = j+1;
\begin{enumerate}[$-$]
\item if $j = |\pi|$ or $t' @\tau^{\pi}_{k} \in \pi(j)$ 
then $j_{k} = j$;

\item find a largest $h \geq 0$, if there is one,
such that there is a $j' < j$,  $\pi(j), \pi(j')$ vary
at $\pi(l),\pi(l')$ with $\tau^{\pi}_{m}$ 
in the same family as $\tau^{\pi}_{k}$ and $\theta \sim_{\tau^{\pi}_{m}} 
\theta'$  where $\theta \in \pi(j+1)$, $\theta' \in \pi(j'+1)$,  and
\begin{enumerate}[$*$]
\item no position in $\pi(j', j'+h)$ is a descendent  of $\pi(l')$;
\item if $h >  0$ then  $\pi(j+1,j+h) \sim \pi(j'+1,j'+h)$; 
\end{enumerate}
\item if there is such a $h \geq 0$ set $j = j + (h +1)$ 
and goto $(* *)$ else
$j_{k} = j$.

\end{enumerate}
\end{enumerate}
\end{enumerate}
\end{defi}

As with  Definition~\ref{def76}, 
if  $\tau^{\pi}_{k}$ is a top or a constant tile then
$\pi(i_{k},j_{k})$ is internal to it and
either ends  at one of its atomic leaves, or 
a final state is reached: via clause $(* *)$, $j_{k}$
will be the least $j > i_{k}$ such that $t' \in \pi(j)$
is an atomic leaf of $\tau^{\pi}_{k}$ or $j = |\pi|$.
The new case  
is when   
$\tau^{\pi}_{k} = y(\lambda \ol{x}_{1},\ldots, \lambda
\ol{x}_{m})$ is neither  a constant tile nor a top tile.
Position $\pi(i_{k})$ is at the root of $\tau^{\pi}_{k}$;
the first position $\pi(j)$, if there is one,  after $\pi(i_{k})$
that is at a lambda node $t'$
need not be  internal to $\tau^{\pi}_{k}$; 
however $t'$  must  be an atomic leaf of some tile $\tau^{\pi}_{k'}$,
$k' \leq k$,  which belongs to
the same family as $\tau^{\pi}_{k}$
by Proposition~\ref{C4jump}. 
If $k' =k$ then $j_{k} = j$ and
$\pi(i_{k},j_{k})$ finishes at $t'$.
Otherwise  $k' \not= k$ and
$t'@\tau^{\pi}_{k'} \in \pi(j)$: 
one checks whether there are previous positions $\pi(j')$ such
that $\pi(j), \pi(j')$ vary at $\pi(l),\pi(l')$ with $\tau^{\pi}_{m}$
in the same family as $\tau^{\pi}_{k}$ (and, therefore, also,
$\tau^{\pi}_{k'}$) and whether $\theta \sim_{\tau^{\pi}_{m}} \theta'$
when $\theta \in \pi(j+1)$ and $\theta' \in \pi(j'+1)$.
If there are no such positions,
for instance if $t'$ is an atomic leaf of the composite tile
$\gamma_{k}$,  then $j_{k} = j$
and $\pi(i_{k},j_{k})$
finishes at $t'$.
Otherwise there are such positions; we then look for a longest 
continuation from $\pi(j+1)$ that corresponds to 
a previous interval from such a $\pi(j'+1)$.
If there is no such continuation, so $h = 0$, then
$\pi(j+1)$ is at a descendent of $\pi(l)$; so control is at 
the root of a tile
in $\gamma_{k}$ in the same family as $\tau^{\pi}_{k}$;
and the loop starts again.
If there is such a continuation then $h > 0$ and 
the loop starts again from $\pi(j+(h+1))$;
if $\pi(i_{k},j_{k})$ is ri 
then $\pi(j+(h+1))$ will also be at the root
of a tile in $\gamma_{k}$ in the same family
as $\tau^{\pi}_{k}$ as this position  must be  a descendent of $\pi(l)$. 
Consequently, as we shall prove, if $\pi(i_{k},j_{k})$ is ri then
it is guaranteed 
to finish at an atomic leaf
of a tile in the same family as $\tau^{\pi}_{k}$. 

\begin{exa}\label{ex713} We describe
the p-partition  of 
$\pi'$ in Figure~\ref{nex2} for the term tree in Figure~\ref{ex1}
(omitting the initial move).
We present tile $\tau^{\pi'}_{k}$, without its superscript $\pi'$,
the play $\pi'(i_{k},j_{k})$
and the edge relation $\byre{\pi'}$ in Figure~\ref{ex78}.
\begin{figure}
\begin{center}
\[ \begin{array}{lll}
\tau_{1} = z(\lambda x,\lambda) \ \ \ \ \ &  \pi'(2,3) \ \  \ \ \  & \\
\tau_{2} = f(\lambda) & \pi'(4,5) \ \  & \tau_{2} \byre{\pi'} \lambda x @
\tau_{1} \\
\tau_{3} = z(\lambda u,\lambda) & \pi'(6,7) & \tau_{3} \byre{\pi'}
\lambda @ \tau_{2}  \\
\tau_{4} = x & \pi'(8,15) & \tau_{4} \byre{\pi'} \lambda u @ \tau_{3} \\
\tau_{5} = z(\lambda y,\lambda) \ \ \ \ \ &  \pi'(16,17) \ \ \ \  \ &  
\tau_{5} \byre{\pi'} \lambda @ \tau_{1} \\
\tau_{6} = z(\lambda s,\lambda)&  \pi'(18,19) \ \ \  & \tau_{6} \byre{\pi'}
\lambda y @ \tau_{5} \\
\tau_{7} =  s & \pi'(20,23) & \tau_{7} \byre{\pi'} \lambda s @ \tau_{6} \\
\tau_{8} = y & \pi'(24,33) & \tau_{8} \byre{\pi'} \lambda @ \tau_{6} \\
\tau_{9} =  a & \pi'(34,34) & \tau_{9} \byre{\pi'} \lambda @ \tau_{5}
\end{array} \]
\end{center}
\caption{Partition of $\pi'$ from Figure~\ref{nex2} in Example~\ref{ex713}}
\label{ex78}
\end{figure}
The intervals for the first three stages are plays on top and constant
tiles.
Tile $\tau_{4} = x$: play jumps at position $\pi'(8)$
to $\lambda x$ of $\tau_{1}$;
the positions $\pi'(9)$,$\pi'(3)$ vary at $\pi'(9), \pi'(3)$ with
$\tau_{1}$ and $\theta \sim_{\tau_{1}} \theta'$
when $\theta \in \pi'(10)$ and $\theta' \in \pi'(4)$;
there are  the maximal corresponding intervals $\pi'(4,7)$ and $\pi'(10,13)$.
Play returns to $x$ and jumps to $\lambda @ \tau_{1}$ which completes
stage~$4$ as $\pi'(15)$ is the first position at this atomic leaf.
The p-partition 
continues with plays on  the top tiles $\tau_{5}$ and $\tau_{6}$.
After stage $6$, the subtree $\gamma_{6}$ of Figure~\ref{ex1}
associated with the partition consists of tiles $\tau_{1}-\tau_{6}$
plus the edges already described 
(which  has atomic leaves $(9)\lambda$, $(19)\lambda$, $(15) \lambda s$
and $(17) \lambda$). 
Tile $\tau_{7} =  s$ and $\pi'(21), \pi'(19)$
vary at $\pi'(21), \pi'(19)$ with $\tau_{6}$ and
$\theta \sim_{\tau_{6}} \theta'$ when 
$\theta \in \pi'(22)$, $\theta' \in \pi'(20)$;
however, there are no corresponding continuations because
$\pi'(22)$ and $\pi'(20)$ are children of $\pi'(21)$
and $\pi'(19)$.
The interval at stage~$7$ is, therefore,   
$\pi'(20,23)$ that finishes at $\lambda @ \tau_{6}$.
Tile $\tau_{8} = y$ and  the play
at this stage jumps to $\lambda y$ of $\tau_{5}$;
positions $\pi'(25)$, $\pi'(17)$ vary at $\pi'(25)$, $\pi'(17)$
with $\tau_{5}$ and $\theta \sim_{\tau_{5}} \theta'$ where
$\theta \in \pi'(26)$ and $\theta' \in \pi'(18)$;
$\pi'(26,31)$ corresponds to $\pi'(18,23)$
and then play returns to $\tau_{8}$ and  jumps to $\lambda @ \tau_{5}$; 
so,  $\pi'(24,33)$ is the interval at
stage $8$. 
Finally,  $\tau_{9} = a$ and stage $9$ is $\pi'(34,34)$.
\qed
\end{exa}


A reason that the same tile may be repeated in a p-partition starts with
different $\fa$-choices. For instance, consider a situation where
$\tau_{m} \byre{\pi} t'@ \tau_{k}$ and $\tau_{n} \byre{\pi} t' @ \tau_{k}$
as follows where $\tau_{k}$ is a constant tile with arity $2$, 
$f(\lambda,\lambda)$. Consider Figure~\ref{ex834}:
the p-partition for $\pi$ has a play on $\tau_{n'}$,
then $\tau_{m'}$ and then $\tau_{k}$ choosing the left branch
of $f$. Tile $\tau_{m-1}$ is a dependent of $\tau_{m'}$
and play jumps to its atomic leaf above $\tau_{k}$
and so this position with the earlier
one will vary at themselves at $\tau_{m'}$; play thus proceeds to
$\tau_{k}$; but now there is a different $\fa$-choice; so
at stage $m-1$, play finishes at the other leaf of $\tau_{k}$.
\begin{figure}
\begin{center}
\ \ \ \ \ \ \ \ \  \xymatrix{
& \tau_{n'} \ar@{.}[d] & \\
& \tau_{m'} \ar@{.}[d] & \\
& \tau_{k}  \ar@{.}[dl] \ar@{.}[dr] & \\
\tau_{m-1} & & \tau_{m} \ar@{.}[d] \\
& & \tau_{n-1}
}
\end{center}

\caption{Illustrating repeating tiles in a p-partition}
\label{ex834}
\end{figure}
Play now proceeds through $\tau_{m}$  to $\tau_{n-1}$ which is a dependent
of $\tau_{n'}$; it jumps to the atomic leaf of $\tau_{n'}$ above
$\tau_{m'}$ and again it
varies with the earlier position at themselves at $\tau_{n'}$:
instead of the corresponding play
which passes through $\tau_{k}$ twice, play
descends to $\tau_{k}$ and now the $\fa$-choice is
the right branch; therefore, play after
stage $n-1$ also finishes at $\lambda @ \tau_{k}$ of the second branch.
These different $\fa$-choices 
involve stages of a p-partition that have
nri intervals; if an interval is ri then it
is  well-behaved as Proposition~\ref{vip} shows.
 
\begin{defi} \label{def91}
Assume $\pi(j_{0}),\pi(i_{1},j_{1}),\ldots, \pi(i_{n},j_{n})$
is the p-partition of $\pi \in \mathsf{G}(t,P)$ 
and $\tau^{\pi}_{1},\ldots,\tau^{\pi}_{n}$ is the 
associated sequence of tiles.
The  \emph{$\pi$-path for $\tau^{\pi}_{k}$}
is the sequence of tiles $\tau^{\pi}_{m_{1}}, \ldots, \tau^{\pi}_{m_{l}}$ 
such that $m_{1} = 1$, $\tau^{\pi}_{m_{l}} = \tau^{\pi}_{k}$
and 
$\tau^{\pi}_{m_{j+1}} \byre{\pi} t_{j} @ \tau^{\pi}_{m_{j}}$
for $1 \leq j < l$.
\end{defi}

\begin{prop} \label{vip}
Assume $\pi(j_{0}),\pi(i_{1},j_{1}),\ldots, \pi(i_{n},j_{n})$
is the p-partition of $\pi \in \mathsf{G}(t,P)$ 
and $\tau^{\pi}_{1},\ldots,\tau^{\pi}_{n}$ is the 
associated sequence of tiles.

\begin{enumerate}[\em(1)]
\item If $m < n$ and
$\tau^{\pi}_{m}$ is a top or constant tile 
then $\tau^{\pi}_{m+1} \byre{\pi} t' @ \tau^{\pi}_{m}$ for some 
$t'$.
\item If $\pi(i_{m},j_{m})$ is ri and 
$\tau^{\pi}_{m+1} \byre{\pi} t' @ \tau^{\pi}_{k}$
then $\tau^{\pi}_{k}$ belongs to the same family as $\tau^{\pi}_{m}$.
\item If $\pi(i_{m},j_{m})$ is ri and
$\tau^{\pi}_{m}$ is an embedded tile in $t$ 
then $\tau^{\pi}_{m+1} \byre{\pi} t' @ \tau^{\pi}_{m}$ for some 
$t'$.
\item If $\pi(i_{m},j_{m})$ is ri and $\tau^{\pi}_{m+1} \byre{\pi} 
t'@\tau^{\pi}_{k}$ then $t'@\tau^{\pi}_{k}$ is not in the $\pi$-path of 
$\tau_{m}$.
\end{enumerate}
\end{prop}
\proof (1) is clear from Definition~\ref{partit}.
For (2) assume  $\tau^{\pi}_{m+1}
\byre{\pi} t' @ \tau^{\pi}_{k}$ and $\pi(i_{m},j_{m})$ is ri. 
Consider the first position $\pi(j)$ in this interval
at a lambda node.
By Proposition~\ref{C4jump}, it is an 
atomic leaf of a tile in the same family as $\tau^{\pi}_{m}$.
Either this position is $\pi(j_{m})$ and so the result follows,
or there are earlier positions $\pi(j')$ and $t' \in \pi(j')$
and $\pi(j), \pi(j')$ vary at $\pi(l), \pi(l')$ at $\tau'$
in the same family as $\tau^{\pi}_{m}$ and $\theta \sim_{\tau'}
\theta'$ when $\theta \in \pi(j+1)$ and $\theta' \in \pi(j'+1)$.
One chooses the $j'$ that allows longest
corresponding intervals $\pi(j+1,j+h)$ and $\pi(j'+1,j'+h)$ for $h\geq 0$.
Using Proposition~\ref{prop79}, it follows that
$\pi(j+(h+1))$ is a descendent of $\pi(l)$
and so is at  a tile $\tau''$ in the same family
as $\tau^{\pi}_{k}$; the same  argument is now repeated.
For (3), consider the b-partition for $\pi(j_{m-1})$;
because $\tau^{\pi}_{m}$ is an embedded tile,
there is a tile $\tau^{\pi}_{k}$ such that $\tau^{\pi}_{k} \equiv 
\tau^{\pi}_{m}$ and $\tau^{\pi}_{m}$ is below $\tau^{\pi}_{k}$
in $t$
and in the b-partition there is a play on $\tau^{\pi}_{k}$,
$\pi(i'_{k},j'_{k})$; consider the shortest interval
$\pi(i'_{k},i'_{k} + j')$ that is a play on $\tau^{\pi}_{k}$.
Using Proposition~\ref{prop516}
and Definition~\ref{partit} it follows that 
$\pi(i_{m},j_{m}) = \pi(i_{m}, i_{m}+j')$
as $t' \in \pi(i_{m}+j')$ is an atomic leaf of
$\tau^{\pi}_{m}$. In the case of $(4)$ 
assume $\pi(i_{m},j_{m})$ is ri 
and $\tau_{m+1} \byre{\pi} t'@\tau^{\pi}_{k}$
and $t' @ \tau^{\pi}_{k}$ is in the $\pi$-path of $\tau^{\pi}_{m}$;
so, $\tau^{\pi}_{k}$ and $\tau^{\pi}_{m}$ belong to
the same family by (2) because  $\pi(i_{m},j_{m})$ is ri. 
Also, there must be a tile $\tau^{\pi}_{l}$,
$k < l \leq m$ such that $\tau^{\pi}_{l} \byre{\pi} t' @ \tau^{\pi}_{k}$ and
either $\tau^{\pi}_{m} = \tau^{\pi}_{l}$ or $\tau^{\pi}_{l}$ 
is on the $\pi$-path
for $\tau^{\pi}_{m}$.
If $\tau^{\pi}_{k}$ is a top tile then consider the b-partition for
$\pi(j_{m-1})$ and its interval that is a play
on $\tau^{\pi}_{k}$ that ends at $t' @ \tau^{\pi}_{k}$  at
position $\pi(j')$; clearly, if position $\pi(j)$ in
$\pi(i_{m},j_{m})$ is at $t' @ \tau^{\pi}_{k}$ then
$\pi(j)$, $\pi(j')$ vary at $\pi(j)$, $\pi(j')$ at $\tau^{\pi}_{k}$
which means that $j < j_{m}$ and so it is impossible that
$\tau^{\pi}_{m+1} \byre{\pi} t'@\tau^{\pi}_{k}$.
Otherwise, $\tau^{\pi}_{k}$ is a dependent tile; the argument is now
similar but more general; if
position $\pi(j)$ in $\pi(i_{m},j_{m})$
is at $t''@\tau^{\pi}_{l}$ for $\tau^{\pi}_{l}$ that is
$\tau^{\pi}_{k}$ or a tile that $\tau^{\pi}_{k}$ is a dependent of
and $t''@\tau^{\pi}_{l}$ is in the $\pi$-path for $\tau^{\pi}_{m}$ 
then there is a tile $\tau^{\pi}_{m'}$ such that
$\tau^{\pi}_{m'}$ is a dependent of $\tau^{\pi}_{k}$ or of some tile 
that $\tau^{\pi}_{k}$ is a dependent of
and the b-partition for $\pi(j_{m'-1})$ which involves a play on
$\tau^{\pi}_{l}$ is such that it finishes at $t''$ at position
$\pi(j')$ and $\pi(j)$, $\pi(j')$ vary at $\pi(j_{1})$, $\pi(j'_{1})$
at $\tau^{\pi}_{l'}$ in the same family as $\tau^{\pi}_{m}$;
therefore, $\pi(j)$ cannot be a final position for
$\pi(i_{m},j_{m})$. 
\qed

\begin{exa}\label{ex715}
The p-partition of 
$\pi$ in Figure~\ref{ex15} where the term tree is in
Figure~\ref{ex14} 
is presented in Figure~\ref{ex79} (omitting the initial move);
\begin{figure}
\begin{center}
\[ \begin{array}{lll}
\tau_{1} = z(\lambda z_{1},\lambda z_{2}) \ \ \ \ &    \pi(2,3) \ \ \ \ \ \ &
 \\
\tau_{2} = z(\lambda x_{1},\lambda x_{2}) &  
 \pi(4,5) & \tau_{2} \byre{\pi} \lambda z_{1} @ \tau_{1} \\
\tau_{3} = (6)z_{1}(\lambda) & \pi(6,7) &  
\tau_{3} \byre{\pi} \lambda x_{1} @ \tau_{2} \\
\tau_{4} = h(\lambda) \ \ \ \ \ \ &  \pi(8,9) \ \ \ \ \ & 
\tau_{4} \byre{\pi}  \lambda z_{2} @ \tau_{1} \\
\tau_{5} = z_{2} & \pi(10,15) & \tau_{5} \byre{\pi} \lambda @ \tau_{4} \\
\tau_{6} =  x_{1}(\lambda) & \pi(16,17) & \tau_{6} \byre{\pi} 
\lambda @ \tau_{3}\\
\tau_{7} = g(\lambda) \ \ \ \ \ \  &  \pi(18,19) \ \ \ \ \ \ & 
\tau_{7} \byre{\pi}\lambda x_{2} @ \tau_{2} \\
\tau_{8} = x_{2} & \pi(20,33)  & \tau_{8} \byre{\pi} \lambda @ \tau_{7} \\
\tau_{9} = (10) z_{1}(\lambda)  & \pi(34,43) & 
\tau_{9} \byre{\pi} \lambda @ \tau_{6} \\
\tau_{10} = a & \pi(44,44) &  \tau_{10} \byre{\pi}\lambda @ \tau_{9}
\end{array} \]
\end{center}
\caption{Partition of $\pi$ in Figure~\ref{ex15} of Example~\ref{ex715}}
\label{ex79}
\end{figure}
where we provide the tile 
$\tau^{\pi}_{k}$, omitting $\pi$,  the interval $\pi(i_{k},j_{k})$
and the edge relation
at each stage $k$. 
Play proceeds through the top tiles $\tau_{1}$ and $\tau_{2}$.
Tile $\tau_{3} = (6)z_{1}(\lambda)$
and   play
at this stage is $\pi(6,7)$  that ends at $\lambda z_{2} @ \tau_{1}$.
The next tile is a constant tile.
Tile  $\tau_{5}$ is $z_{2}$ and the pair $\pi(11)$, $\pi(3)$
vary at $\pi(11), \pi(3)$ with $\tau_{1}$ and $\theta \sim_{\tau_{1}} \theta'$ for 
$\theta \in \pi(12)$ and $\theta' \in \pi(4)$; so 
the interval  at stage $5$ is $\pi(10,15)$
that ends at $\lambda$ of $\tau_{3}$ as  there is no earlier
position where control is at this leaf.  
The tile  $\tau_{6} = x_{1}(\lambda)$ and $\pi(16,17)$ is the 
interval at this stage.  
Play then proceeds through a   constant tile at stage $7$.
Tile $\tau_{8} =  x_{2}$ and the interval is $\pi(20,33)$;
positions $\pi(21), \pi(5)$ vary at $\pi(21), \pi(5)$ with $\tau_{2}$
and $\theta \sim_{\tau_{2}} \theta'$ for $\theta \in \pi(22)$ and 
$\theta' \in \pi(6)$; 
the intervals $\pi(6,15)$ and $\pi(22,31)$ correspond
as described in Remark~\ref{rem811}. 
Tile  $\tau_{9}$ is $(10)z_{1}(\lambda)$; 
positions $\pi(35), \pi(7)$ vary at $\pi(35),\pi(7)$ with $\tau_{1}$
and $\theta \sim_{\tau_{1}} \theta'$ when $\theta \in \pi(36)$ and
$\theta' \in \pi(8)$;
the intervals $\pi(8,10)$ and $\pi(36,38)$ correspond. 
With the next position $\lambda z @\tau_{1} \in \pi(39)$
and 
$\pi(39), \pi(11)$ vary at $\pi(39), \pi(11)$ with $\tau_{1}$
and $\theta \sim_{\tau_{1}} \theta'$ for $\theta \in \pi(40)$ and $\theta'
\in \pi(12)$. The intervals 
and $\pi(40,41)$, $\pi(12,13)$ correspond. Therefore at stage~$9$,
the interval is built from two separate subintervals. Finally,
stage~$10$ is the constant tile $a$.
\qed
\end{exa}

\section{Unfolding and the small model property}
\label{fifth}

We now prove decidability of higher-order matching at all orders, 
by showing  the
small model property; if $t \models P$ then there is a small
term $t' \models P$. The proof starts with   the  tree of tiles
that captures the   p-partitions of all plays in  a game $\G(t,P)$
and then extends it to a  tree of basic tiles. 
We then define unfolding on such trees which underpins the small
model property.

As with the $3$rd-order case in Section~\ref{third}, we examine the
p-partitions  of all plays in $\mathsf{G}(t,P)$. 
We maintain  abuse of notation: if $\pi$ and $\pi'$ are two
plays we let $\pi(i_{k},j_{k})$, $\pi'(i_{k},j_{k})$
be their intervals  at stage $k$ irrespective of their ranges.
Instead of a sequence of simple tiles
there is  a tree of simple tiles that is associated with the p-partitions
as each p-partition shares  the initial tile $\tau_{1}$ of $t$.
As a representation for this tree of tiles, we let its root  be
$\tau^{\Pi}_{1}$ when $\Pi$ is the set of all plays
in $\G(t,P)$; any other node of this tree
has the form $\tau^{\Pi'}_{k}$
which represents that for each $\pi \in \Pi'$, $\tau^{\pi}_{k}$ 
is its tile at  stage $k$ and for all earlier stages $m < k$, every play in
$\Pi'$ also shares the same tile at stage $m$.
Thus, the tree has the form depicted in  Figure~\ref{ex81}.
\begin{figure}
\begin{center}
\ \ \ \ \ \ \ \ \  \xymatrix{
& & & \tau_{n}^{\Pi'_{1}} \\
& & \tau_{k}^{\Pi'} \ar@{.}[ur] \ar@{.}[r]&  \tau_{n'}^{\Pi'_{l}} \\
\tau_{1}^{\Pi}  \ar@{.}[r]&  \tau_{j}^{\Pi} \ar@{.}[ur] \ar@{.}[dr]
\ar@{.}[r]& \ldots\\
& &\tau_{k}^{\Pi''} \ar@{.}[dr] \ar@{.}[r]& \tau_{m'}^{\Pi''_{l}}\\
& & & \tau_{m}^{\Pi''_{1}} \\
}
\end{center}

\caption{Tree of tiles in  all p-partitions}
\label{ex81}
\end{figure}
However, we also assume the induced edge relations $\byre{\pi}$ within 
this tree:  $\tau^{\Pi'}_{m} \byre{\pi} t' @ \tau^{\Pi''}_{k}$
if $\tau^{\pi}_{m} \byre{\pi} t'@ \tau^{\pi}_{k}$ and $\pi \in \Pi'\cap \Pi''$.

Let $\mathrm{T}$ be the  tree of tiles
for the p-partitions of all plays in $\G(t,P)$.
We drop the superscript $\Pi'$ from tiles
$\tau^{\Pi'}_{k}$ whenever the context allows.
We assume the definition of $\pi$-path, Definition~\ref{def91},
which picks out the sequence
of tiles in $t$ from its root to
$\tau^{\Pi'}_{k}$ when $\pi \in \Pi'$.
Also, we shall assume that Definitions~\ref{defi53}, \ref{defdepend}
and \ref{defi54}
of $j$-below,
below, immediate $j$-dependent, dependent and embedded 
apply to tiles in a tree $\mathrm{T}$
by examining  $\pi$-paths and bindings:
for instance,  $\tau_{m}$
is an immediate  $j$-dependent of $\tau_{k}$ if  
$\tau_{k}$ is in the $\pi$-path for $\tau_{m}$, 
the free variable $y$ at the head of  $\tau_{m}$ 
is bound in $\tau_{k}$
and $\tau_{m}$ is $j$-below $\tau_{k}$ relative to this  
$\pi$-path.
However, because of the linear representation 
of a p-partition there is also the idea that a tile
$\tau_{m}$ is \emph{later} than $\tau_{k}$ in
$T$ if there is a play $\pi$ and $\tau_{k} = \tau^{\pi}_{k}$
and $\tau_{m} = \tau^{\pi}_{m}$ and $m > k$; we also say that
$\tau_{k}$ is \emph{earlier} than $\tau_{m}$.

As with the $3$rd-order case in Section~\ref{third}, 
we identify  \emph{special} tiles in the tree.

\begin{defi} \label{def92}
The tile $\tau^{\Pi}_{k} \in \mathrm{T}$  
is \emph{special}
if it obeys one of the following three conditions;
\begin{enumerate}[(1)]
\item
for some $\pi \in \Pi$, $\pi(i_{k},j_{k})$ is nri,
\item for some $\pi \in \Pi$, $q \in
\pi(j_{k})$ is a final state,  
\item for some $\pi$, $\pi' \in \Pi$,
$t' \not= t''$ when $t' \in \pi(j_{k})$ and 
$t'' \in \pi'(j_{k})$. 
\end{enumerate}
A tile
is \emph{x-special} if it obeys $(1)$ or $(2)$ of these conditions.
\end{defi}

A special tile that is not x-special is a play separator.
There is the same upper bounds 
on the number of special tiles 
in $\mathrm{T}$ as in the $3$rd-order case.

\begin{fact} \label{fact92}
Assume $\mathrm{T}$ is tree of tiles associated with $\mathsf{G}(t,P)$.
Within $\mathrm{T}$ there are 
\begin{enumerate}[(1)]
\item at most $\delta$ ($=$ the right size for $P$, 
Definition~\ref{delta}) tiles that involve nri intervals;
\item at most $p$ ($=$ the number of plays in $\mathsf{G}(t,P)$) tiles
where play ends;
\item therefore, at most $\delta + p$  are x-special tiles;
\item at most $p-1$ tiles that are play separators.
\end{enumerate}
\end{fact}

The proof of the small model property for the $3$rd-order case 
is straightforward: 
use transformation ${\bf T2}$ of Section~\ref{third} to remove 
any tile that is  not special from $\mathrm{T}$ and  update edges.
With higher orders we cannot just omit a  tile
that is  not special.
It may have dependents, so its removal would result in a tree that is
no longer a closed term. Or an associated  interval  may finish
at an atomic leaf of some other tile in the tree, so
its removal may not  preserve game playing.
Instead, we introduce 
tile \emph{unfolding} as a transformation
on a  tree $\mathrm{T}$. We need to generalise
the notion of tree to that of a \emph{tree of basic} tiles. 
We will be   interested in a tile $\tau_{m}
\in \mathrm{T}$ that 
is an immediate dependent of a top or embedded tile
$\tau_{k}$ which is not x-special and which also does not have x-special
later tiles belonging to  the same family. Therefore, as we shall see,
$\tau_{m}$ can be replaced in the tree  by a basic  tile  
that is constructed from $\tau_{k}$ and 
$\tau_{m}$; this may  require
revision of later edges in the tree.
Before developing the  full account,
we shall now briefly illustrate it.

\begin{exa} \label{ex91} Consider the p-partition  in 
Figure~\ref{ex78} of Example~\ref{ex713}
for the play $\pi'$ in Figure~\ref{nex2} on  the term tree in Figure~\ref{ex1}.
The tree of tiles with the edge relation  $\byre{\pi'}$, with $\pi'$ omitted,
is pictured in the top diagram in  Figure~\ref{ex92}.
\begin{figure}
\begin{center}
\xymatrix{ \ \ \ \ \ \ \ \ \ 
\tau_{1} \ar@/^/[r] \ar@/_1pc/[rrrr]& \tau_{2} \ar@/^/[r] & \tau_{3} \ar@/^/[r]& \tau_{4} 
& \tau_{5}  \ar@/^/[r] \ar@/_1pc/[rrrr]& \tau_{6} 
\ar@/_/[rr]\ar@/^/[r] & \tau_{7}
& \tau_{8}  & \tau_{9} }

\vspace{1cm}

\xymatrix{ \ \ \ \ \ \ \ \ \ 
\tau_{1} \ar@/^/[r] \ar@/_1pc/[rrrr]& \tau_{2} \ar@/^/[r] & \tau_{3} \ar@/^/[r]& \tau_{4} 
& \tau_{5}  \ar@/^/[r]& \tau_{6} 
\ar@/_/[rr]\ar@/^/[r] & \tau_{7}
& \tau'_{8} \ar@/^/[r] & \tau_{9} }

\end{center}

\caption{Tree of tiles before and after unfolding  in Example~\ref{ex713}}
\label{ex92}
\end{figure}
There are edges, for instance,  from different atomic leaves of 
$\tau_{1}$ to 
$\tau_{2}$ and $\tau_{5}$. The special (and x-special)
tiles are $\tau_{2}$, $\tau_{4}$ and $\tau_{9}$; intervals on the first
two of these are nri and the third is where the play $\pi'$ finishes.
Transformation ${\bf T2}$  of Section~\ref{third} would allow us
to remove the tile $\tau_{3}$ with the effect that the edge from
$\tau_{2}$ would then be to  $\tau_{4}$: dynamically, in terms of play,
this means that the ri interval $\pi'(i_{3},j_{3})$, a play on $\tau_{3}$,
is omitted and
the interval $\pi'(i_{4},j_{4})$ reduced (by omitting its ri play
on $\tau_{3}$). Tile $\tau_{5}$ is a top tile 
with  immediate dependent
$\tau_{8}$; neither of these tiles is x-special.  However,
because of this binding and the fact that it has two outgoing edges
from different atomic leaves, $\tau_{5}$ cannot be omitted (like
$\tau_{3}$); nor can we remove $\tau_{8}$ because $\pi'(i_{8},j_{8})$
ends at an atomic  leaf of $\tau_{5}$. Instead
we can unfold $\tau_{5}$ at $\tau_{8}$: we introduce the basic tile
$\tau'_{8} = z(\lambda y. \tau_{8},\lambda)$ which prefaces
$\tau_{5}$ to $\tau_{8}$; the new interval 
$\pi'(i_{8},j_{8})$, in effect, includes an extra interval that corresponds to
$\pi'(i_{5},j_{5})$ as a prefix and then omits the ri plays on
$\tau_{6}$ and $\tau_{7}$ from the old interval
$\pi'(i_{8},j_{8})$. Play in the new interval $\pi'(i_{8},j_{8})$
which is still ri
now finishes at an atomic leaf
of  $\tau'_{8}$;  so edges may need to be changed;
here, $\tau'_{8}$ now has an edge to $\tau_{9}$.
The effect of this  unfold is pictured 
in the lower diagram
of Figure~\ref{ex92}. As a consequence,
both $\tau_{5}$
and $\tau'_{8}$ can now be removed by transformation
${\bf T2}$. As the reader can verify,  tile $\tau_{6}$ can also
be unfolded at $\tau_{7}$.
\qed
\end{exa}

We call the process ``unfolding'' $\tau$ at $\tau'$
(where $\tau'$ is an immediate dependent of $\tau$) 
because it is analogous  
to unfolding or unravelling a transition system 
in modal logic;  here, there is  the extra 
dimension of binding. As with unravelling, the purpose of
unfolding in $\mathrm{T}$ is also to approximate
the tree model property. Technically, from a game-theoretic point of
view,  what will justify  unfolding is 
permuting, repeating and omitting corresponding ri intervals:
for instance,  with the replacement of $\tau_{8}$
by $\tau'_{8}$ in Example~\ref{ex91}
there is a repetition and a permutation
of intervals that correspond
to the earlier ri interval  on $\tau_{5}$ within $\tau'_{8}$,
and then omission of the  ri intervals that correspond to $\pi'(i_{6},j_{6})$, 
$\pi'(i_{7},j_{7})$ within $\pi'(i_{8},j_{8})$.
Although initially unfolding increases the size of a tree, its point is
to reduce tile levels. Tile $\tau_{8}$ is level $2$
whereas its replacement $\tau'_{8}$ is a level $1$
(basic end) tile.
Unfolding is not defined as a transformation in 
the sense of Section~\ref{third} because it is not local; 
edges to later tiles
may be revised in its application.

First, we extend the notion of a tree  to that of a tree 
of basic tiles with associated plays that are p-partitioned.

\begin{defi} \label{defitree}
The tree of basic tiles $\mathrm{T}$ has associated plays
$\Pi$ if $\Pi$ is the set of plays down the branches of
$\mathrm{T}$ and for each 
$\pi \in \Pi$, $\pi = \pi(j_{0}),\pi(i_{1},j_{1}),\ldots, \pi(i_{n},j_{n})$
for some $n$ such that
\begin{enumerate}[(1)]
\item if $m \leq n$ then  $\pi(i_{m},j_{m})$ is an interval
on the tiles $\tau^{\pi}_{1},\ldots, \tau^{\pi}_{m}$ in $\mathrm{T}$
that starts at the root of $\tau_{m}$,

\item if $m < n$ and $t'@\tau^{\pi}_{k} \in \pi(j_{m})$
then $t'$ is an atomic leaf of $\tau^{\pi}_{k}$ and $\tau^{\pi}_{m+1}
\byre{\pi} t'@ \tau^{\pi}_{k}$,

\item if $m < n$ and
$\tau^{\pi}_{m}$ is a top or constant tile 
then $\tau^{\pi}_{m+1} \byre{\pi} t' @ \tau^{\pi}_{m}$ for some 
$t'$,

\item if $\pi(i_{m},j_{m})$ is ri and 
$\tau^{\pi}_{m+1} \byre{\pi} t' @ \tau^{\pi}_{k}$
then $\tau^{\pi}_{k}$ belongs to the same family as $\tau^{\pi}_{m}$,
\item if $\pi(i_{m},j_{m})$ is ri and
$\tau^{\pi}_{m}$ is an embedded tile in $t$ 
then $\tau^{\pi}_{m+1} \byre{\pi} t' @ \tau^{\pi}_{m}$ for some 
$t'$,
\item if $\pi(i_{m},j_{m})$ is ri
and $\tau_{m+1} \byre{\pi} t' @ \tau_{k}$
then $t' @ \tau_{k}$ is not in the $\pi$-path of $\tau_{m}$.
\end{enumerate}
\end{defi} 

\ni
Initially, when  $\mathrm{T}$ is the tree of simple tiles 
constructed from the p-partitions of
the plays in $\mathsf{G}(t,P)$ then $\mathrm{T}$ has associated plays
$\mathrm{G}(t,P)$; parts (1) and  (2) of Definition~\ref{defitree}
follow from  Definition~\ref{partit} of p-partition,
the remainder from 
Proposition~\ref{vip}. We assume that the definitions of b-partition,
dependent, special,
x-special, $\pi$-path  and so on are extended to basic 
tiles in a tree of basic tiles.

\begin{defi} \label{def93}
Assume  $\mathrm{T}$ is a tree of basic tiles
with associated plays. Tile $\tau_{k}$ \emph{is unfoldable at}
$\tau_{m}$  if
\begin{enumerate}[(1)]
\item $\tau_{k}$ is a top or embedded tile,
\item $\tau_{m}$  is the first tile in $\tau_{k+1}, \ldots, \tau_{m}$ 
that is a dependent
of $\tau_{k}$, 
\item $\tau_{k}$ and no later  tile that is in the same family as $\tau_{k}$
is x-special. 
\end{enumerate}
\end{defi}

If $\tau_{k}$ is unfoldable at $\tau_{m}$ then we define the unfolding of 
$\tau_{k}$ at $\tau_{m}$ in $\mathrm{T}$ as the tree $\mathrm{T'}$
with the same nodes as $\mathrm{T}$ except that $\tau_{m}$ is replaced by
a basic tile $\tau'_{m}$ that is a composition of $\tau_{k}$ and $\tau_{m}$.
(Edges in $\mathrm{T}$ may also be changed in $\mathrm{T'}$.)
In the following we define the associated plays on $\mathrm{T'}$
from those on $\mathrm{T}$: the definition uses the notion of corresponding
positions as defined (for intervals) in 
Definition~\ref{defsimilar}. However, because $\mathrm{T'}$
is different from $\mathrm{T}$, the notion of correspondence
is slightly weakened in specific circumstances to allow that a sequence
of positions may correspond to a single position.

\begin{defi} \label{def94}
Assume $\tau_{k} = y(\ldots\lambda \ol{x}\ldots)$ is unfoldable
at $\tau_{m}= \tau^{\Pi'}_{m}$ in $\mathrm{T}$  and  for each $\pi \in \Pi'$,
$\tau_{k+1} \byre{\pi} \lambda
\ol{x}@\tau_{k}$.
The \emph{unfolding of $\tau_{k}$ at $\tau_{m}$}
in $\mathrm{T}$ is  the tree $\mathrm{T'}$ where 
$\mathrm{T'}$ has the same tiles as $\mathrm{T}$
except that $\tau_{m}$ is replaced by 
$\tau'_{m}  =  y(\ldots\lambda \ol{x}.\tau_{m}\ldots)$.
For each play
$\pi = \pi(j_{0}),\pi(i_{1},j_{1}),\ldots, \pi(i_{n},j_{n})$
on $\mathrm{T}$ there is a play 
$\sigma = \sigma(j_{0}),\sigma(i_{1},j_{1}),\ldots, \sigma(i_{n},j_{n})$
on $\mathrm{T'}$, defined as follows in stages and top down:
\begin{enumerate}[(1)]
\item if $\pi \not\in \Pi'$ 
or $\pi \in \Pi'$ and $l < m$ then $\sigma(i_{l},j_{l}) =
\pi(i_{l},j_{l})$ and $\tau_{l+1} \byre{\sigma} t' @ \tau_{k'}$ 
iff $\tau_{l+1} \byre{\pi} t' @ \tau_{k'}$;
\item if $\pi \in \Pi'$, $l \geq m$, 
$t' @ \tau_{j} \in \pi(j_{m-1})$ 
and $t' @ \tau_{j} \in \sigma(j_{m-1})$
then 
$\sigma(i_{l},j_{l})$ is the continuation 
from the head of $\tau_{l}$ in $\gamma_{l}
= \tau_{1},\ldots, \tau_{l}$ with edges $\byre{\sigma}$.
Any position 
$\sigma(i')$ in $\sigma(i_{l},j_{l})$ corresponds to  $\pi(i')$
in $\pi(i_{l},j_{l})$ in the sense of Definition~\ref{defsimilar}
except in the following circumstances
where the notion of correspondence is weakened:
\begin{enumerate}[$\bullet$]
\item \emph{the positions are at $\tau'_{m}$ and $\tau_{m}$}: if
$t'@\tau_{j} \in \sigma(i'-1)$ and $t' @ \tau_{j} \in \pi(i'-1)$
then $\sigma'\sigma(i')$ corresponds to   $\pi(i')$  where
$\sigma'$ is a shortest play  on $\tau_{k}$ in $\tau'_{m}$ that ends 
at $\lambda \ol{x}@\tau'_{m}$,
\item \emph{there is a jump into $\tau'_{m}$ and $\tau_{k}$}:
as a result of move C4 of Figure~\ref{game}, 
$t''@\tau_{k} \in \pi(i')$ and $t'' @ \tau'_{m} \in \sigma(i')$;
then $\sigma(i',i'+i'')$ corresponds to $\pi(i',i'+i'')$ if these intervals
are internal to $\tau'_{m}$ and $\tau_{k}$, 
\item \emph{the positions are at $\lambda \ol{x} @ \tau'_{m}$
and $\lambda \ol{x} @ \tau_{k}$}: $\lambda \ol{x} @ \tau'_{m} \in \sigma(i')$,
$\lambda \ol{x} @ \tau_{k} \in 
\pi(i')$ and $\pi(i'')$ is the first later position such that 
$t'@\tau_{j}  \in \pi(i'')$ and $\pi(l,i')$ is the play on $\tau_{k}$
in the b-partition for $\pi(i'')$; then
$\sigma(i')$ corresponds to  $\pi(i',i'')$. 
\end{enumerate}
For the edges: if $\tau_{l}$ is $\tau_{m}$ or is below $\tau_{m}$ 
and $\tau_{l+1} \byre{\pi} t''@\tau_{k}$ then $\tau_{l+1} \byre{\sigma}
t''@\tau'_{m}$; otherwise, $\tau_{l+1} \byre{\sigma} t''@\tau_{p}$
if $\tau_{l+1} \byre{\pi} t''@\tau_{p}$. 
\end{enumerate}
We say that $\sigma$ on $\mathrm{T'}$ is \emph{the companion} of $\pi$
on $\mathrm{T}$ and that $\mathrm{T'}$ is \emph{an unfolding} of $\mathrm{T}$.
\end{defi}

In Example~\ref{ex91},
$\tau_{5}$ is unfolded at $\tau_{8}$; $\mathrm{T}$ is the 
upper and $\mathrm{T'}$ the lower tree in Figure~\ref{ex92}.
The companion play $\sigma'$ on $\mathrm{T'}$ of $\pi'$ on $\mathrm{T}$
has the same intervals up to and including stage $7$.
In defining the weakened correspondence, play
in $\sigma'$ is at $\tau'_{8}$ in $\mathrm{T'}$ and
in $\pi'$ at $\tau_{8}$ in $\mathrm{T}$: so $\sigma'' \sigma'(i_{8})$
where $\sigma''$ is the initial play on $\tau_{5}$ in $\tau'_{8}$
now corresponds to $\pi'(i_{8})$; next there is a jump into
$\tau'_{8}$ and $\tau_{5}$ by move C4 which is to $\lambda y@\tau'_{8}$
and $\lambda y@\tau_{5}$ and so this $\sigma'$ position corresponds
to the interval that is from $\lambda y @ \tau_{5}$ to
the lambda node directly above $\tau_{8}$; so then the next positions
will again correspond. In both these cases,  where  an interval corresponds to
a position, the interval must be ri.

\begin{prop} \label{prop95}
Assume $\mathrm{T}$ has associated plays $\Pi$. 
If $\mathrm{T'}$ is an  unfolding of $\mathrm{T}$ then 
$\mathrm{T'}$ has associated plays
$\Sigma = \eset{\sigma \, | \,  \mbox{for some } \pi \in \Pi,
\sigma \mbox{ is a companion of } \pi}$.
\end{prop}
\proof
Assume that $\mathrm{T}$ has associated plays $\Pi$ and
$\mathrm{T'}$ is the unfolding of $\tau_{k}$ at $\tau^{\Pi'}_{m}$
in $\mathrm{T}$. By definition, if $\pi \not\in \Pi'$ and $\sigma$
is its companion then $\sigma = \pi$ is a play on
a branch of $\mathrm{T'}$ as required.
Otherwise,
assume $\tau_{k} = y(\ldots\lambda \ol{x}\ldots)$,
$\tau_{k+1} \byre{\pi} \lambda
\ol{x}@\tau_{k}$ for each $\pi \in \Pi'$,
$t'@\tau_{j} \in \pi(j_{m-1})$
and 
$\tau'_{m}  =  y(\ldots\lambda \ol{x}.\tau_{m}\ldots)$.
We now show that we can find corresponding positions
$\sigma(i')$ in $\sigma(i_{l},j_{l})$ and
$\pi(i')$ in $\pi(i_{l},j_{l})$ as described in
Definition~\ref{def94} for each $l \geq 1$. 
For $l < m$, this holds because $\sigma(i_{l},j_{l}) = \pi(i_{l},j_{l})$
and $\tau_{l+1} \byre{\sigma} t''@ \tau_{k'}$ iff $\tau_{l+1} 
\byre{\pi} t''@\tau_{k'}$.
Consider next  the case that corresponding  positions are at $\tau'_{m}$
and $\tau_{m}$: $t' @ \tau_{j} \in \sigma(i'-1)$ and
$t'@\tau_{j} \in \pi(i'-1)$. 
We examine  the b-partitions for $\sigma(i'-1)$ and $\pi(i'-1)$ and their
component plays $\sigma'$, $\pi'$ on $\tau_{k}$;
we show that $\sigma'$ and $\pi'$ correspond, that they are ri
and that they are shortest plays on $\tau_{k}$ (that end
at the leaf $\lambda \ol{x}@\tau_{k}$). 
We prove this by induction on corresponding  positions $\sigma(i'-1)$,
$\pi(i'-1)$, $i' \geq i_{m}$. The base case
is when $i' = i_{m}$.
At this point the b-partitions for $\sigma(i'-1)$ and $\pi(i'-1)$
are the same. Let $\sigma'$ be the component interval in this
b-partition on $\tau_{k}$. That $\sigma'$ is ri and is a shortest
play on $\tau_{k}$ that ends at
$\lambda \ol{x}$ follows from the
fact that $\tau_{m}$ is the first 
tile that is a dependent of $\tau_{k}$ in the sequence
$\tau_{k+1}, \ldots, \tau_{m}$, that $\tau_{k}$ is a top
or embedded tile  and that $\tau_{k}$ and all
later tiles in the same family are not x-special; consequently, 
any tile between $\tau_{k}$ and $\tau_{m}$ in the same family as $\tau_{k}$
has an associated ri interval and, therefore, if play in such an interval
were at a different atomic leaf
of $\tau_{k}$ than $\lambda \ol{x}$, then 
$\tau_{m}$ would not be a
dependent of $\tau_{k}$. For the inductive 
step the argument is similar after noting the following property:
if corresponding  positions are at $\tau'_{m}$ and $\tau_{m}$ then
it is not possible that play was in a dependent tile $\tau_{m'}$ of
$\tau_{k}$ that is below $\tau_{m}$ before jumping back into
$\tau_{k}$,  then to $\lambda \ol{x} @\tau_{k}$ and then proceeding to
$\tau_{m}$ because in $\sigma$ the simulating position would be at
$\tau_{m}$ in $\tau'_{m}$ (because in $\mathrm{T'}$, $\tau_{m'}$
is a dependent of $\tau'_{m}$ and edges are updated in $\mathrm{T'}$).
So this property holds. 
Therefore, returning to the main argument, 
the continuation
from $\sigma(i'-1)$ consists first of a sequence of moves $\sigma''$
that corresponds to $\sigma'$ except it is on $\tau_{k}$ within
$\tau'_{m}$; therefore, $\sigma'' \sigma(i')$ weakly corresponds to  $\pi(i')$.
Consider next corresponding positions such that at the next positions
they jump into $\tau'_{m}$ and $\tau_{k}$; $t'' @\tau_{k} \in \pi(i')$
and $t'' @ \tau'_{m} \in \sigma(i')$. Then the intervals 
$\pi(i',i'+i'')$, $\sigma(i',i'+i'')$ that are internal to these
tiles correspond except that  they take place in $\tau_{k}$
and $\tau'_{m}$.
The final circumstance to examine is that corresponding
positions are $\sigma(i')$, $\pi(i')$ such that 
$\lambda \ol{x} @ \tau'_{m} \in  \sigma(i')$
and $\lambda \ol{x} @ \tau_{k} \in \pi(i')$.
We show that there is a first position such that 
$t' @ \tau_{j} \in \pi(i'')$ and for some $l$, $\pi(l,i')$ 
is the play on $\tau_{k}$ in the b-partition for $\pi(i'')$
and the b-partition for $\sigma(i')$ contains the ri interval
$\sigma(i'',j'')$ which corresponds to $\pi(i',i'')$.
Positions $\sigma(i')$ and $\pi(i')$  must be the result
of a C4 move. However, the look-up table where this entry
is defined must be at a  position within a tile $\tau_{m'}$
that is a dependent of $\tau_{k}$ in $\mathrm{T}$ and of $\tau'_{m}$
in $\mathrm{T'}$ that is, below $\tau_{m}$.
Now the result follows as $\tau_{k}$ and all later tiles in
the same family are not x-special. 
Clearly, $\mathrm{T'}$ has no other plays than the companions of
$\Pi$. Moreover, each companion play  $\sigma$  obeys the six conditions
in Definition~\ref{defitree} given that they
hold for each $\pi$ on $\mathrm{T}$.
\qed

To prove the small model property, 
assume a smallest term $t$ such that $t\models P$
and let $\mathrm{T}$ be its  tree of
simple tiles that captures the 
p-partitions of every  $\pi \in \mathsf{G}(t,P)$.
First, we describe the proof for a particular case of $\mathrm{T}$,  
a \emph{general atoms case},
that obeys the following condition:  
if $\tau_{k}$ 
is not a constant tile  
then it is not x-special.
What this means is that every top and embedded tile
is unfoldable at a first dependent\footnote{This feature,
that every top and embedded tile is unfoldable at a first  dependent, 
is true in 
the atoms case where  $\delta = 0$ even though some of these tiles
may be x-special.}.
The decidability proof now reduces to the
$3$rd-order case as the tree-model property is regained.

Initially, restricting further, assume  
$P$ is $5$th-order: there are, therefore,
only two levels of non-constant tiles, top
and end tiles. Starting top down with $\mathrm{T}_{0} = \mathrm{T}$,
at each stage $\mathrm{T}_{i}$, a top tile 
$\tau$ that is closest to the root and that has
dependents
is unfolded at a first  dependent $\tau'$.
The construction finishes at some stage $n$, when $\mathrm{T}_{n}$
has no unfoldable tiles; that is, when it has no top tiles
with dependents.  This means that $\mathrm{T}_{n}$ only
consists of constant tiles and basic top tiles that are also end tiles.
Once the tree is in this form,
only the special tiles (those that are
constant tiles and play separators) need to be kept:
the remainder are redundant using transformation {\bf T2}. 
To obtain a small term, 
the initial $\lambda \ol{y}$ of $t$ is placed at its top and 
the constant $d : \Nil$ is placed below every leaf $\lambda \ol{z}$.
The bound on the size of $t$ is  larger than
in Fact~\ref{bounds} because the units are now  \emph{basic} tiles
instead of simple tiles.
In the worst case, each whole tile consists of one simple top
tile and at most $\alpha$ simple end tiles in any branch
(where $\alpha$ is the arity of $P$).
Therefore, using this construction,
we obtain the following bound where the measures
are all from $P$:  $|t| \leq (\alpha + 1) \times (\delta + (2p -1))$.

For higher-orders, 
the bounds become  even larger.
Initially, all top and embedded tiles with dependents are unfoldable.
Now unfolding is iterated. At each stage $\mathrm{T}_{i}$,
a top or embedded tile  $\tau$ with dependents which has the \emph{greatest}
level is unfolded; if there is more than one such tile then
one that is closest to the root is chosen to be $\tau$
and it is unfolded at a first dependent to give $\mathrm{T}_{i+1}$.
Consider what may happen when  $P$ is $7$th-order. There are
now three levels of non-constant  tiles: top,  middle and end.
Unfolding reconciles embedded middle tiles with their
immediate end dependents, which may in turn create
larger  embedded middle tiles or end tiles. 
For instance, assume the following branch
of tiles in $\mathrm{T}_{0}$ 
\[ \tau \ \ \ \ \tau_{1} \ \ \ \ \tau_{2} \ \ \ \tau_{21} \ \ \  \ \tau_{3} \ \ \tau_{31}
\]
where $\tau$ is a top tile, $\tau_{1}$, $\tau_{2}$, $\tau_{3}$ are dependents
of $\tau$, $\tau_{2}$ and  $\tau_{3}$ are both embedded middle
tiles because of $\tau_{1}$ and $\tau_{21}$ is a dependent of
$\tau_{2}$ and $\tau_{31}$ of $\tau_{3}$.
First, $\tau_{2}$ is unfolded at $\tau_{21}$ and then $\tau_{3}$ is unfolded
at $\tau_{31}$ to give the following sequence.
\[ \tau \ \ \ \ \tau_{1} \ \ \ \ \tau_{2} \ \ \ \tau_{2}\tau_{21} \ \ \  \ 
\tau_{3} \ \ \ \tau_{3}\tau_{31}
\]
The situation has reduced to the $5$th-order case, as there are now
only two levels of tiles. So, 
the complete  unfolding is  the following sequence of top tiles that
are also end tiles. 
\[ 
\tau \ \ \ \tau \tau_{1} \ \ \ \tau \tau_{1} \tau_{2} \ \ \ \tau 
\tau_{1} \tau_{2} (\tau_{2} \tau_{21})
\ \ \ \tau \tau_{1} \tau_{2} (\tau_{2} \tau_{21}) \tau_{3} \ \ \ 
\tau  \tau_{1} \tau_{2} (\tau_{2} \tau_{21}) \tau_{3} (\tau_{3} \tau_{31}) \]
Therefore,
using {\bf T2} we can remove any non-special
tiles from the unfolded tree and any embedded (basic) end tiles
within special tiles.
The following is a very crude size bound on a smallest term that solves
the problem $P$ of order $2n +1$ in this restricted case: 
$|t| \leq g(n) \times (\delta + (2p-1))$
where $g(1) = 1$ and $g(k+1) = (\alpha + 1)^{g(k)}$. 

\begin{rem}
In the general  atoms case 
there is a bounded size solution term with 
a simple  form that is  a   \emph{transferring}
term \cite{Pad1}. For instance, consider the definition of transferring
in \cite{Lo1};
for every subterm of $\lambda x_{1} \ldots x_{n}.t$
of the form $x_{i} s_{1} \ldots s_{k}$ 
the free variables of  any $s_{j}$ belong
to $\eset{x_{1},\ldots,x_{n}}$.
\qed
\end{rem}

Let us return to the unrestricted case.
We show that there is a
bounded size term that \emph{almost} has the 
tree model property. Not all top or embedded tiles with
dependents
can be unfolded because of their contribution to solving $P$;
for instance, in Example~\ref{ex91}, tile $\tau_{1}$ cannot be
unfolded at $\tau_{4}$.
There is also a further issue 
that  does not occur with the general atoms case. 
After unfolding, 
we need 
to extract a term from the  unfolded tree.
The intention is that the edge relation $\byre{\pi}$
should be the subtree
relation.
However, there can  be multiple edges of the form 
$\tau^{\pi}_{m} \byre{\pi} t' @ \tau_{k}$ 
and $\tau^{\pi'}_{n} \byre{\pi'} t' @ \tau_{k}$:
so, we need to guarantee 
that the ``subterms'' rooted at
$\tau^{\pi}_{m}$ and $\tau^{\pi'}_{n}$ are compatible. 

\begin{defi} \label{defi816}
The tree $\mathrm{T}$ has the \emph{subterm property}
if there is a smallest  equivalence relation $\cong$
on its tiles 
such that whenever $\tau_{m} \cong \tau_{n}$, 
\begin{enumerate}[(1)]
\item they are syntactically the same tile,
\item if $\tau_{k} \byre{\pi} t''@ \tau_{m}$ and 
$\tau_{l}
\byre{\pi'} t'' @ \tau_{n}$ then $\tau_{k} \cong
\tau_{l}$.
\end{enumerate}
\end{defi}

If $\mathrm{T}$ 
has the subterm property
then  its \emph{extraction} is  defined top-down.
For any edge  $\tau_{m} \byre{\pi} t' @ \tau_{k}$ 
the tile $\tau_{m}$ is moved
to be directly below $t'@ \tau_{k}$.
The initial $\lambda \ol{y}$ of $t$ is placed at the
top and 
the constant $d : \Nil$ is placed below every leaf $\lambda \ol{z}$.
Next we compute the smallest  equivalence relation 
$\cong$ starting with  the identity, for each tile
$\tau_{m} \cong \tau_{m}$ and then closing up inductively
under the conditions
in Definition~\ref{defi816}:
tiles $\tau_{m} \cong \tau_{n}$ are identified.

\begin{fact} \label{lem818}
If $\mathrm{T}$ has  
has the subterm property
then  its extraction is a term.
\end{fact}

To obtain the small model property, we show that given 
the  initial tree there
is a way of  unfolding such that the resulting
tree has the subterm property. 
We then examine the extraction  and  apply the transformation
{\bf T2}
to remove any redundant tiles.
The upper bound is very crude.

\begin{thm} \label{thm93}
If $t$ is a smallest solution
of $P$ of order at most $2n + 1$, then 
$|t| \leq g(n) \times ((p^{2} \times \delta \times N(n)) + p-1)$
where $\alpha$ is the arity of $P$,
$N(n) = p \times \sum \eset{ \alpha^{i}
\, : \, 1 \leq i \leq g(n)}$, $g(1) = 1$ and $g(k+1) = (\alpha + 1)^{g(k)}$.
\end{thm}
\proof
Assume $t$ is a smallest solution to $P$ of order $2n+1$.
Let $\mathrm{T}_{0}$ be the tree of simple tiles
for the p-partitions of each play in  $\mathsf{G}(t,P)$.
By Fact~\ref{fact92} there are at most $\delta + (2p-1)$ special
tiles 
in the tree.
The issue is to define a largest sequence of unfolds 
such that afterwards the resulting tree 
has the subterm property.
If every top and embedded tile is unfoldable then 
at each stage $i+1$ those  top or embedded tiles  with the largest level
that have dependents are chosen; one of them that is closest
to the root is then unfolded at a first dependent; so $\mathrm{T}_{i+1}$
is an 
unfolding of  $\mathrm{T}_{i}$.
Unfolding is continued until $\mathrm{T}_{n}$ for some $n$
when there are no more unfoldable tiles; an easy argument shows that
there is such an $n$ as the unfolding reduces tile level.
By Proposition~\ref{prop95}, it follows that
companion plays preserve final states. 
After unfolding, every tile in the resulting
tree is a basic top  tile or a constant tile
and, consequently, obeys the subterm property.
So, its extraction $t'$ is a term.
Except for the special tiles in $t'$,
the remainder are redundant using transformation {\bf T2}.
Therefore, this produces a term $t''$ where 
$|t''| \leq g(n) \times (\delta + (2p -1))$. Because $t$ is a smallest
term that solves $P$, $t$ obeys this bound.
Otherwise, some top or embedded tiles must be excluded from
being unfolded.  
In the worst case there are  at most $\delta + p$ x-special
tiles in $\mathrm{T}_{0}$ and, therefore, at most $(\delta + p) \times N(n)$
distinct tiles in $\mathrm{T}_{0}$ that are  x-special
or  dependents of  x-special tiles  and
that are also not embedded tiles.
Moreover, there are at most a further $p - 1$ tiles that are special.
We now proceed with a sequence of  unfoldings as follows:
find a highest level unfoldable tile that is closest to
the root and unfold it at a first  dependent; and keep repeating
this until there are no more unfoldable tiles. 
If the resulting tree $\mathrm{T}_{n}$ fails to have the subterm property then 
at some earlier stage $j$ there are
tiles $\tau_{k}$ and $\tau_{l}$ in $\mathrm{T}_{j}$ that are 
syntactically the same and one of them, say 
$\tau_{k}$ is later unfolded whereas  $\tau_{l}$
is not and both have edges to the same
atomic leaf: a failure
of condition (2) of Definition~\ref{defi816}. 
Both these tiles are dependents of the same tile
in $\mathrm{T}_{j}$.
Therefore, we exclude 
the tile that $\tau_{k}$ is a dependent of from
being unfolded and then examine a complete sequence of   unfoldings without it.
By repeating this argument, after
the  sequence of  unfoldings the resulting   tree has  the subterm property;
the number of distinct tiles that
are excluded  from being unfolded is at most $(\delta + p) \times N(n)$. 
By Proposition~\ref{prop95} 
its extraction $t'$ solves $P$.
To begin with there are $((\delta + p)\times N(n)) + p-1$ tiles
that are either special or dependents of x-special tiles.
Again, we wish to apply the transformation {\bf T2}
to remove redundant end tiles from $t'$. 
The question
is how many extra unfolded tiles are also special
because they have become separators.
This depends on the number of plays $p$. 
If $p = 1$ then no extra tiles are needed. 
If $p > 1$ then
for each of the  $((\delta + p) \times N(n))$ tiles,
there could be at most 
$p-1$ new play separators in  the full unfolding.
Therefore, the result follows as each tile has bounded length
$g(n)$. 
\qed

\section{Conclusion}
\label{conclusion}
Although we have shown that higher-order matching is decidable, the
upper size bound is very coarse and more work needs to
be done to make it more accurate. Although our complexity analysis is
in terms of term size,
it crudely agrees with the known 
non-elementary complexity lower bound based on
\cite{Stat}. 
As order of a problem increases, so does its level and, therefore, 
the size of a smallest solution term according to our analysis
increases (exponentially). 
Implicit in the analysis are positive
sensible algorithms for solving  dual
interpolation problems. 
The game-theoretic characterisation of dual interpolation
allows us to examine incomplete terms. As a first step, given a problem $P$
its component set of simple tiles can be defined from the subtypes
and the constants (including forbidden constants and the new constant
$d$ of Section~\ref{treecheck}).
From these,  we can then define
varieties of basic tiles that have no embedded end tiles. 
We can then proceed to construct a term tree in stages,
first by seeing  if there are 
basic top tiles that separate plays, and, thereby, continue recursively.
Otherwise, we need to check if there must be
non-top basic tiles, and so on. In the worse case, we need to examine all
possible terms whose size is bounded by Theorem~\ref{thm93}.
It may be worth investing effort to implement a tool that builds such terms.


An   open question is  whether
the set of all solutions of an interpolation problem 
is  independently characterisable.
For instance, Comon and Jurski define tree automata that characterise
all solutions to a $4$th-order problem \cite{CJ}. 
In Section 8 of their paper they
describe  two problems with extending their 
automata beyond order $4$. The first is that
states of an automaton are constructed
out of the observational equivalence classes of terms. 
Up to a 4th-order problem, one only needs to consider
finitely many terms.
With $5$th and higher orders, this is no longer true 
and  one needs to quotient
the potentially infinite terms into their respective observational
equivalence classes in order to define
only finitely many states: however as Padovani shows this procedure is,
in fact, equivalent to the matching problem itself \cite{Pad2}. 
The second problem is the term trees that the automata recognise.
For a $4$th-order problem,
there is an automaton that recognises its full set of solutions
(up to $\alpha$-equivalence) even though the syntax may be infinite.
Comon and Jurski define a special kind of
automata, $\Box$-automata, to achieve this \cite{CJ}. The occurrence
of a leaf $\Box$ in a term tree represents any (syntactically correct) subtree.
A $\Box$ cannot contribute to the solution of a matching
problem. 
This is no longer true at $5$th-order,
as illustrated by their  example
$x \, \lambda y z.y(\lambda z'.z z') = a$.
Solutions of this problem include the terms
\[ \lambda x_{1}.x_{1}(\lambda y_{1}.x_{1}(\ldots x_{1}(\lambda y_{n}.y_{i_{1}}
(y_{i_{2}}(\ldots (y_{i_{k}} a )\ldots))) u_{n} \ldots ) u_{2} ) u_{1} \]
where for some $m \leq k$, every $u_{i_{j}}$, $j \leq m$, is the identity
and $u_{i_{m+1}}$ is the constant function $a$. 
Because $k$ is arbitrarily large,
one  cannot use a bounded number of variables
to capture all these terms \cite{CJ}.
In the general case, by iterated unfolding, any solution term
can be transformed into another  solution term 
that only uses (and reuses) boundedly many
variables. Whether
this is capturable using some kind of automaton (such as a transducer) is open.

In subsequent work \cite{St3},
we are able to overcome  the first problem of Comon and Jurski
at higher orders but not the second.
We provide  a tree-automata  
characterisation 
relative to a \emph{finite} alphabet: 
given a problem $P$, a finite set of variables
and constants  the (possibly infinite) set of terms
that are built from those components and that solve $P$
is regular.
The states of the automaton are  built from abstractions of
sequences of moves in a (variant version of the)  game
which works for all  orders.
Although there is  active research
extending automata on words and trees to infinite alphabets which preserve
``good'' properties, such as decidability of non-emptiness, 
see \cite{luc} for a recent survey, the results  do not yet  
apply to the case caused by  higher-order binding.

As we briefly mentioned in Section~\ref{treecheck}, Ong has shown that
the tree-checking game can be presented using game-semantics
and innocent strategies. Another   question is whether this
framework provides an alternative basis for understanding higher-order
matching.

\section*{Acknowledgement} I am indebted to Luke Ong for many 
extremely helpful discussions
about matching. I would also like to thank
the  LMCS referees  who engaged with the paper and 
proposed  productive improvements.

\end{document}